\documentclass{jfm}

\usepackage{graphicx,natbib,empheq,relsize} 

\ifCUPmtlplainloaded \else
\checkfont{eurm10}
\iffontfound
\IfFileExists{upmath.sty}
{\typeout{^^JFound AMS Euler Roman fonts on the system,
		using the 'upmath' package.^^J}%
	\usepackage{upmath}}
{\typeout{^^JFound AMS Euler Roman fonts on the system, but you
		dont seem to have the}%
	\typeout{'upmath' package installed. JFM.cls can take advantage
		of these fonts,^^Jif you use 'upmath' package.^^J}%
}
\else
\fi
\fi

\ifCUPmtlplainloaded \else
\checkfont{msam10}
\iffontfound
\IfFileExists{amssymb.sty}
{\typeout{^^JFound AMS Symbol fonts on the system, using the
		'amssymb' package.^^J}%
	\usepackage{amssymb}%
	  \let\leq=\leqslant
	  \let\geq=\geqslant
}{}
\fi
\fi

\ifCUPmtlplainloaded \else
\IfFileExists{amsbsy.sty}
{\typeout{^^JFound the 'amsbsy' package on the system, using it.^^J}%
	\usepackage{amsbsy}}
{\providecommand\boldsymbol[1]{\mbox{\boldmath $##1$}}}
\fi

\providecommand\bcdot{\boldsymbol{\cdot}} 

\newcommand\Elas{\mbox{\textit{El}}}  
\newcommand\Nzero{N^2}

\usepackage{mathrsfs} 
\usepackage{amsmath,mathtools}
\usepackage[usenames,dvipsnames]{color} 

\newcommand{\ddtwo}[2]{{ \frac{\text{d}^2 {#1}}{\text{d} {#2}^2}   }} 
\newcommand{\ddo}[2]{{ \frac{\text{d} {#1}}{\text{d} {#2}}   }}

\newcommand{\bld}[1]{{\boldsymbol{ #1 }} } 
\DeclareMathAlphabet{\mathscrbf}{OMS}{mdugm}{b}{n} 
\newcommand\Wie{\mbox{\textit{Wi}}}  

\def\lline{\vrule width12pt height2.5pt depth -2pt}
\def\mline{\vrule width4pt height2.5pt depth -2pt}

\def\bdot{\raise.2em\hbox to .15em{.}}
\def\dashed{\mline\hskip4.5pt\mline\hskip4.5pt\mline\thinspace}

\def\dotted{\bdot\ \bdot\ \bdot\ \bdot\thinspace}

\definecolor{gray}{gray}{0.5}

\def\bdotblack{\raise.25em\hbox to .15em{.}}

\def\sldsh{\rule[0.2\baselineskip]{0.075in}{0.65pt}}
\def\sldot{\rule[0.2\baselineskip]{0.025in}{0.65pt}}
\def\bldot{\hspace{0.025in}}
\def\linesolids{\rule[0.2\baselineskip]{0.275in}{0.65pt}}
\def\linedashed{\sldsh\bldot\sldsh\bldot\sldsh} 
\def\linedshdot{\sldsh\bldot\sldot\bldot\sldsh\bldot\sldot} 
\def\linedotted{\sldot\bldot\sldot\bldot\sldot\bldot\sldot\bldot\sldot\bldot\sldot}

\title[Scaling of amplification in viscoelastic flows]{Scaling of energy amplification in viscoelastic channel and Couette flow}

\author[Ismail Hameduddin, Tamer A. Zaki, Dennice F. Gayme]%
{Ismail Hameduddin%
	\thanks{Email address for correspondence: ismailh@jhu.edu},\ns
	Tamer A. Zaki 
	and Dennice F.  Gayme}

\affiliation{Department of Mechanical Engineering, The Johns Hopkins University,
	Baltimore, MD 21218, USA}

\begin{document}

\maketitle

\begin{abstract} 
	
The linear amplification of disturbances is critical in setting up transition scenarios in viscoelastic channel and Couette flow, and may also play an important role when such flows are fully turbulent.
As such, it is of interest to assess how this amplification, defined as the steady-state variance maintained under Gaussian white noise forcing, scales with the main nondimensional parameters: the Reynolds ($\Rey$) and Weissenberg ($\Wie$) numbers.
This scaling is derived analytically in the two limits of strong and weak elasticity for when the forcing is streamwise-constant.
The latter is the relevant forcing for capturing the overall behaviour because it was previously shown to have the dominant contribution to amplification.
The final expressions show that for weak elasticity the scaling retains a form similar to the well-known $\mathcal{O}(\Rey^3)$ relationship with an added elastic correction. 
For strong elasticity, however, the scaling is $\mathcal{O}(\Wie^3)$ with a viscous correction. 
The key factor leading to such a mirroring in the scaling is the introduction of forcing in the polymer stress.
The results demonstrate that energy amplification in a viscoelastic flow can be very sensitive to the model parameters even at low $\Rey$.
They also suggest that energy amplification can be significantly increased by forcing the polymer stress, thereby opening up possibilities such as flow control using systematically designed polymer stress perturbations.
\end{abstract}

\begin{keywords} 
\end{keywords} 

\section{Introduction}  
Elastic effects introduced by dissolving polymers in a Newtonian solvent have the potential for dramatically changing the dynamics of parallel shear flows.
For example, such flows have been found to have a `reverse Orr' mechanism that enables them to extract energy from the mean even while disturbance wavefronts are aligned with the mean shear \citep{Page2015}.
Similarly, the introduction of new time scales due to elasticity has been suggested  as a cause for transition to turbulence at low Reynolds numbers, $\Rey$ \citep{Meulenbroek2004,Morozov2005}.
We explore the dynamics of such viscoelastic flows by studying the input-output amplification, which is defined as the steady-state variance maintained in the output of a linear system under white noise input forcing.
Thus, when the output is velocity, the amplification is the steady-state kinetic energy.
This type of input-output, or energy, amplification has been used to study a variety of flows and regime such as Newtonian laminar \citep{Farrell1993,Jovanovic2005} and turbulent \citep{delAlamo2006} channel flows,  midlatitude atmospheric jets \citep{Farrell1995}, as well as the laminar viscoelastic channel and Couette flows \citep{Hoda2008} for which we will derive scalings in the present work.

As with their Newtonian counterparts, viscoelastic parallel shear flows can exhibit bypass transition to turbulence under favorable conditions \citep{Atalik2002,Agarwal2014}. 
Moreover, in marked contrast to purely viscous flows, transition to a turbulent-like state (elastic turbulence) has been observed even in creeping flow \citep{Pan2013}.
It has been suggested that this latter type of `elastic' transition occurs due to a mean flow distortion caused by the large amplification of particular initial disturbances that decay over a large timescale \citep{Meulenbroek2004,Morozov2005}. In the complete absence of inertia, \citet{Jovanovic2010} showed that such initial disturbances can evolve over timescales on the order of the Weissenberg number, $\Wie$. 
 Energy amplification can be helpful towards understanding such flows since large amplification implies that small input disturbances can lead to large output variance that can then push  the system out of the basin of attraction of a stable equilibrium and setup a bypass transition scenario.

In the present work, we focus on the streamwise constant ($k_x = 0$) component  of the flow because it was previously shown by \citet{Hoda2008} and also \citet{Zhang2013} to have the predominant amplification.
\citet{Hoda2008}  also found that the relative contribution of the amplification due to the streamwise constant components increases with increasing elasticity $\Elas = \Wie/\Rey$.
These findings are consistent with the low $\Rey$ regime  experiments of \citet{Qin2017} and the direct numerical simulations of \citet{Agarwal2014} in the subcritical high $\Rey$ regime.
In these papers, the authors found that prior to bypass transition the flow was dominated by its streamwise constant component.  

The energy amplification of the $k_x = 0$ component in Newtonian channel and Couette flow is known to scale as $f_N\Rey + g_N\Rey^3$,  where the coefficients $f_N$ and $g_N$ are functions that depend on the spanwise wavenumber \citep{Bamieh2001,Jovanovic2005}.
This scaling can be directly extended to viscoelastic channel and Couette flow when the velocity is subjected to stochastic forcing and is also the output under consideration \citep{Hoda2009}.
The coefficients in the viscoelastic scaling then also depend on the elasticity number, $\Elas$, because for sufficiently large $\Elas$, the elasticity can introduce dynamically important, $\Elas$-dependent timescales that appear as peaks in the frequency response \citep{Hoda2009}.
\citet{Hoda2009}  numerically calculated the $\Elas$-dependent  coefficients in the energy amplification and found that for strongly elastic flows ($\Elas\rightarrow \infty$), the coefficient of $\Rey$ becomes independent of $\Elas$ and the coefficient of $\Rey^3$ scales linearly with $\Elas$.

\citet{Jovanovic2011} further specialized the study of the energy amplification at the $k_x = 0$ component to the case of strong elasticity, i.e., $1/\Elas = \Rey/\Wie \rightarrow 0$.
The authors used singular perturbation methods and the frequency response approach used by \citet{Hoda2009} to derive the scaling of the amplification of the velocity field and the polymer stresses with $\Wie$ and $\Rey$, correct up to $\mathcal{O}(1/\Elas)$.
In that work, the authors considered stochastic forcing only in the velocity field and found that the variance in the field scales as $\Rey f_{\textrm{JK}} + \Wie \Rey^2  g_{\textrm{JK}} + \mathcal{O}(\Elas^{-1})$, where the coefficients $f_{\textrm{JK}}$ and $g_{\textrm{JK}}$ are functions of the spanwise wavenumber and viscosity ratio. 
This scaling is consistent with the numerical findings of \citet{Hoda2009}.
\citet{Jovanovic2011} found that the variance in the polymer stresses scales as $(\Rey^2/\Wie)a_{\textrm{JK}} + \Wie \Rey^2 b_{\textrm{JK}} + \Wie^3 c_{\textrm{JK}} + \mathcal{O}(\Elas^{-1})$, where the coefficients $a_{\textrm{JK}}$, $b_{\textrm{JK}}$, and $c_{\textrm{JK}}$ are functions of the spanwise wavenumber and viscosity ratio. 
In this scaling, the term $c_{\textrm{JK}}$ is purely the contribution of the variance in the streamwise normal stress perturbations.
In addition to the scaling, \citet{Jovanovic2011} discovered a physical mechanism due to elasticity, analogous to the vortex tilting mechanism in Newtonian flows, that generates large energy amplification even when $\Elas \rightarrow 0$ (weakly inertial regime):  streamwise vorticity and coupling by the base flow leads to large $\mathcal{O}(\Wie)$ cross-plane polymer stresses which then produce large fluctuations in streamwise polymer shear stresses, whose gradients amplify the streamwise velocity. 
The main amplification mechanism in this process, intermediate amplification by the polymer stresses, does not depend on inertia associated with the perturbations but yet  arises due to coupling via the base flow shear much like in purely viscous flows. 

The literature reviewed above only considered stochastic forcing of the velocity field, and did not consider the possibility of input forcing in the polymer stresses.
Polymer stress forcing can be a useful tool to grossly represent  unmodeled phenomena such as chain scission, molecular aggregation, polydispersity, thermal fluctuation effects, and others. 
These unmodeled phenomena may have a significant effect on the macroscale dynamics such as drag reduction, which is known to be sensitive to how the polymers are introduced into the flow  \citet{Warholic1999}.
The polymer stress forcing can also represent general uncertainties that appear due to the assumptions  in the modeling process such as parametric uncertainty, statistical closures and assumptions on the forces associated with the polymer chains. 
In general modeling of non-Newtonian flows and deriving the associated constitutive relations is a large and intensive area of research. 
Although the rich tapestry of constitutive relations available allows one to qualitatively replicate most experimentally observed macroscale phenomena, the physical relevance of these models is still a matter of speculation \citep{Leal1990,Nieuwstadt2001}.
Sensitivity of the output to Gaussian white noise polymer stress forcing is an important indicator of the robustness of these models to slight discrepancies. 

To the authors' best knowledge, the problem of velocity and polymer stress amplification in laminar channel/Couette flow due to stochastic polymer stress forcing has not been previously studied.
In a closely related study, \citet{Jovanovic2010} examined the impulse response of the polymer stresses in the case with no inertia, i.e $\Rey \rightarrow 0$.
Such an impulse response can be trivially recast as the frequency response of the polymer stress under stochastic forcing of the same.
The authors found significant transient growth of the polymer stresses even without inertia, suggesting that purely elastic flows may exhibit transient turbulence-like behaviour under appropriate conditions.

\begin{figure}
	\includegraphics[scale=1.0]{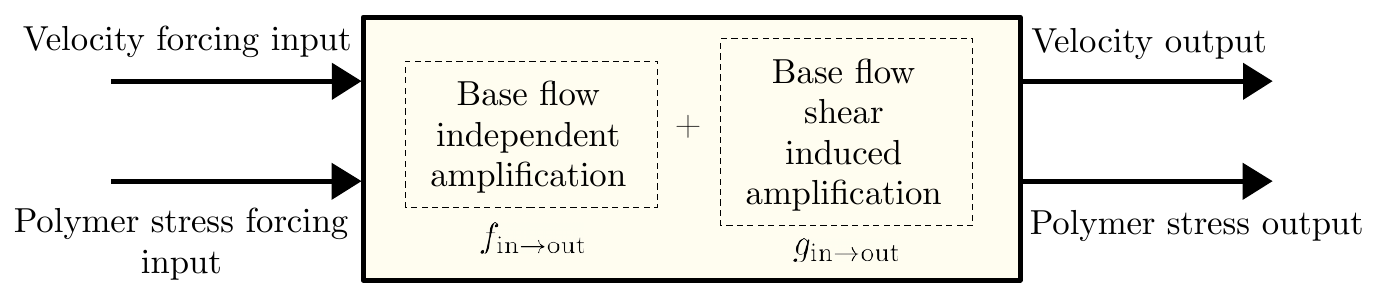}
	\caption{The input-output system.}
	\label{fig:block}
\end{figure}

In the present work, we do not assume  $\Rey \rightarrow 0$ and consider the response of both the velocity and polymer stresses under stochastic forcing of both, as shown in figure \ref{fig:block}.
We thus have four distinct components, each associated with an input-output pair  shown in figure \ref{fig:block}.
We derive the scaling  of each of these components of the amplification  with $\Rey$ and $\Wie$ in the limits of strong ($\Elas \rightarrow \infty$) and also weak ($\Elas \rightarrow 0$) elasticity, and are also able to separate the contribution to the amplification that arises due to the base flow shear.
We validate the theoretically predicted scalings with numerical calculations.

An important limitation of the present work is that we do not derive the $\Elas$ scaling of the contribution of the streamwise normal stress perturbations to the polymer stress amplification.
\citet{Hoda2009} showed that when $k_x=0$, this stress is only one-way coupled with the remaining state variables so that it does not affect the velocity and the remaining polymer stress components. 
However, this component itself can be significantly amplified, with a contribution to the energy amplification that can be shown to scale as $\Wie^p \Rey^{q}$, where $p+q=5$ and $p=3$ in the strongly elastic limit \citep{Jovanovic2011}, i.e.  $\Rey/\Wie \rightarrow 0$.
In the appendix we show that $p=0$  for fixed $\Elas$.

In the limit of strong elasticity $\Elas \rightarrow \infty$, our scaling result reads
 \begin{multline}
\mathcal{E} =  
(f_{\mathsfbi{T} \rightarrow \bld{u}}  +
f_{\mathsfbi{T} \rightarrow \mathsfbi{T}})\Wie   +
(g_{\mathsfbi{T} \rightarrow \bld{u}}+
g_{\mathsfbi{T} \rightarrow \mathsfbi{T}})\Wie^3\\
+ f_{\bld{u} \rightarrow \bld{u}}\Rey +
f_{\bld{u} \rightarrow \mathsfbi{T}}   \frac{\Rey^2}{\Wie}+
\left(g_{\bld{u} \rightarrow \bld{u}} +
g_{\bld{u} \rightarrow \mathsfbi{T}} \right)\Wie\Rey^2  
\end{multline}
where $\mathcal{E}$ is the total input/output amplification.
Here $f_{a\rightarrow b}$ ($g_{a\rightarrow b}$) represents the normalized base-flow independent (dependent) contribution to the amplification in $b$ due to forcing in $a$, where $a,b\in \{\bld{u},\mathsfbi{T}\}$ and $\bld{u}$ and $\mathsfbi{T}$ represent the velocity and polymer stresses, respectively.
In this expression we retain the leading order term in each $f_{a\rightarrow b}$ and $g_{a\rightarrow b}$ in order to describe the scaling associated with each input/output pair and how the base-flow affects this scaling, even if a particular term is subdominant with respect to the remaining leading order terms.
As a result, each  $f_{a\rightarrow b}$ and $g_{a\rightarrow b}$ are correct up to $\mathcal{O}(\Elas^{-1})$ but the truncation error in the overall expression is $\mathcal{O}\left(\Rey^2/\Wie,\Rey^4/\Wie\right)$.
Our expression for $\mathcal{E}$ is consistent with that in \citet{Jovanovic2011}, i.e. when only the velocity field is stochastically forced, modulo the $\Wie^3 \Rey^2 c_{\textrm{JK}}$ contribution from the variance in the streamwise normal stress described previously and which we do not consider here.
The energy amplification expression derived by \citet{Jovanovic2011} appears as a `viscous correction' to the dominant scaling, $(f_{\mathsfbi{T} \rightarrow \bld{u}}  +
f_{\mathsfbi{T} \rightarrow \mathsfbi{T}})\Wie   +
(g_{\mathsfbi{T} \rightarrow \bld{u}}+
g_{\mathsfbi{T} \rightarrow \mathsfbi{T}})\Wie^3$, which results from stochastic forcing in the polymer stresses.

In the limit of weak elasticity $\Elas \rightarrow 0$, the scaling reads
\begin{multline}
\mathcal{E} = 
(f_{\bld{u} \rightarrow \bld{u}}+
f_{\bld{u} \rightarrow \mathsfbi{T}} )\Rey +
(g_{\bld{u} \rightarrow \bld{u}} +
g_{\bld{u} \rightarrow \mathsfbi{T}} )\Rey^3 \\
+  f_{\mathsfbi{T} \rightarrow \mathsfbi{T}} \Wie  +
f_{\mathsfbi{T} \rightarrow \bld{u}} \frac{\Wie^2}{\Rey}  +
\left(g_{\mathsfbi{T} \rightarrow \bld{u}}+
g_{\mathsfbi{T} \rightarrow \mathsfbi{T}} \right)\Rey\Wie^2.   
\end{multline}
Both  $f_{a\rightarrow b}$ and $g_{a\rightarrow b}$ are correct up to $\mathcal{O}(\Elas)$ but the truncation error now reads $\mathcal{O}(\Wie,\Rey^2\Wie)$, which means we require an additional constraint that $\Wie \rightarrow 0$.
In analogy with the strongly elastic case, an `elastic correction' supplements the primary contribution to the amplification, which arises due to forcing in the velocity field.

The coefficients $f_{a\rightarrow b}$ and $g_{a\rightarrow b}$ are, in principal, independent of the model parameters in both strong and weak elasticity and thus can be used to explore the change in behaviour between weak and strong elasticity.

We present the mathematical problem in section \ref{sec:background}. In section \ref{sec:ReScaling} we present and extend some previous results with discussion. We derive the $\Rey$, $\Wie$ scaling of energy amplification in the two elastic limits in section \ref{sec:main_results}. We validate these results against numerical computations in section \ref{sec:discussion} and discuss their implications. We present our conclusions and directions for future work in section \ref{sec:conclusions}. 
 
\section{Mathematical Formulation} \label{sec:background} 

\subsection{Problem Setup}
\begin{figure}
\centering
\includegraphics[scale=1.0]{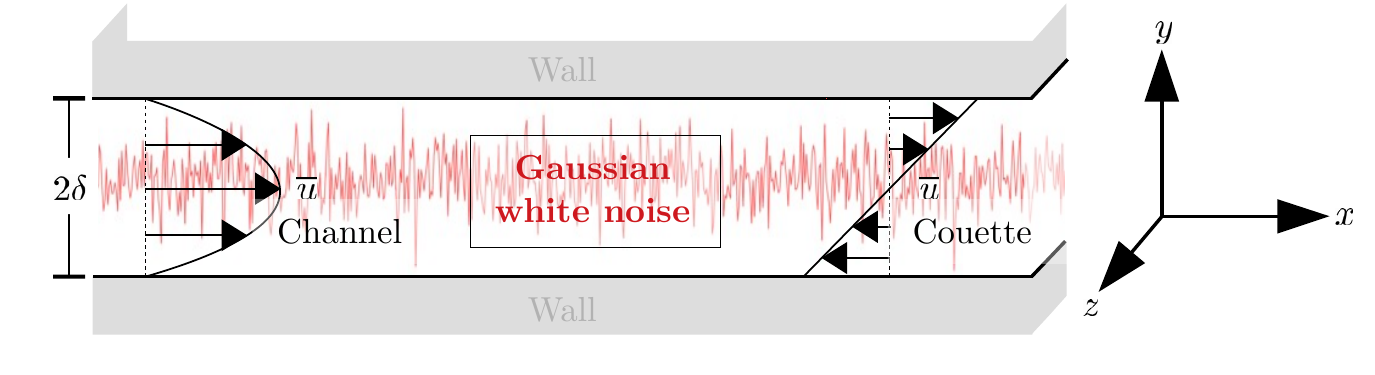}
\caption{Schematic for velocity profile for laminar channel and Couette flows, with superimposed Gaussian white noise forcing.}
\label{figSchematic}
\end{figure}

We are interested in the linear evolution of small perturbations in laminar viscoelastic  channel and Couette flow. A schematic of the physical problem is given in Fig. \ref{figSchematic}. The relevant parameters for the flow are the Reynolds number $\Rey$, the elasticity number ${\Elas}$ and the Weissenberg number $\Wie$. In terms of the three timescales of the problem (convective, viscous and polymer relaxation), these parameters are given by
\begin{align}
\Rey = \frac{(\delta^2/\nu) }{(\delta/U)}, \quad 
{\Elas} = \frac{\lambda_{\text{p}}  }{ (\delta^2/\nu)}, \quad
\Wie = \frac{\lambda_{\text{p}} }{(\delta/U)}=\Elas\Rey \label{params}
\end{align}
where $\delta$ is the channel half-height, $U$ is the centreline velocity for channel flow and the wall speed in Couette flow, $\lambda_{\text{p}}$ is the polymer relaxation time, and $\nu$ is the   mixture kinematic viscosity. 

We decompose the state variables, i.e. the velocity $\bld{u} = \begin{bmatrix} u & v & w \end{bmatrix}^{\mathsf{T}}$ and the symmetric polymer stress tensor $\mathsfbi{T}$, as follows
\begin{align}
\bld{u} = \bld{\overline{u}} + \bld{u}', \qquad
\mathsfbi{T} = \mathsfbi{\overline{T}} + \mathsfbi{T}'
\end{align}
where the primed quantities are perturbations about the laminar base flow, which is indicated by the overlines. The laminar velocity $\bld{\overline{u}}$ and polymer stress $\mathsfbi{\overline{T}}$ are given by
\begin{align}
\bld{\overline{u}} = \begin{bmatrix}
\overline{u}(y) \\ 0 \\ 0
\end{bmatrix}, \qquad
\mathsfbi{\overline{T}} =
\begin{bmatrix}
2\Wie (\ddo{\overline{u}}{y})^2 & \ddo{\overline{u}}{y} & 0 \\
\ddo{\overline{u}}{y} & 0 & 0 \\
0 & 0 & 0
\end{bmatrix}\label{baseState}
\end{align}
where $\overline{u}(y)=1-y^2$ and $\overline{u}(y)=y$ for channel and Couette flow, respectively.

The linearized Oldroyd-B equations \citep{Bird1987}, which describe the evolution of  $\bld{u}'$ and $\mathsfbi{T}'$ around the base flow, are given by
\begin{align}
\nabla \bcdot \bld{u}' &= 0 \label{LOB1}\\
\p_t \bld{u}' + \bld{u}'\bcdot \nabla \bld{\overline{u}} + \bld{\overline{u}}\bcdot\nabla \bld{u}' &= -\nabla p' + \frac{\beta}{\Rey}\Delta \bld{u}' + \frac{1-\beta}{\Rey}\nabla \bcdot \mathsfbi{T}' + \bld{d}_{\bld{u}'} \label{LOB2}\\
\p_t \mathsfbi{T}' + \bld{u}'\bcdot \nabla \mathsfbi{\overline{T}}  + \bld{\overline{u}}\bcdot\nabla \mathsfbi{T}'& -\left[ \mathsfbi{T}' \bcdot \nabla \bld{\overline{u}} + \mathsfbi{\overline{T}}\bcdot \nabla \bld{u}'
+
(\mathsfbi{T}' \bcdot \nabla \bld{\overline{u}})^{\mathsf{T}} + (\mathsfbi{\overline{T}}\bcdot \nabla \bld{u}')^{\mathsf{T}}\right] \nonumber\\
&\hspace{0.5in}=  \frac{1}{\Wie}\left[ \nabla\bld{u}' + (\nabla \bld{u}')^{\mathsf{T}} - \mathsfbi{T}'\right] + \mathsfbi{D}_{\mathsfbi{T}}\label{LOB3}
\end{align}
where $p'$ is the pressure perturbation and the viscosity ratio is given by $\beta = \nu_s/\nu$ where $\nu_s$ is the kinematic viscosity of the Newtonian solvent. 
The velocity satisfies the no-slip condition $\bld{u}'(x,y=\pm 1,z) = 0$ and the polymer stresses do not have boundary conditions.
The   vector $\bld{d}_{\bld{u}}$ and the symmetric tensor $\mathsfbi{D}_{\mathsfbi{T}}$  represent the input disturbances, and are defined as
\begin{align}
\bld{d}_{\bld{u}} =  \begin{bmatrix} d_u \\ d_v \\ d_w \end{bmatrix}, \qquad
\mathsfbi{D}_{\mathsfbi{T}} = \begin{bmatrix}
d_{xx} & d_{xy} & d_{xz} \\
d_{yx} & d_{yy} & d_{yz} \\
d_{zx} & d_{zy} & d_{zz} \\
\end{bmatrix} .\label{inpdefn}
\end{align} 
We also define the vector equivalent of $\mathsfbi{D}_{\mathsfbi{T}}$ as
\begin{align}
\bld{d}_{\mathsfbi{T}} \equiv \begin{bmatrix} d_{xx} & d_{xy}& d_{xz}& d_{yy}& d_{yz}& d_{zz} \end{bmatrix}^{\mathsf{T}}.
\end{align}   

\subsection{The streamwise constant equations} 
The largest amplification of disturbances in an Oldroyd-B fluid occurs in the streamwise constant component, similar to the situation in Newtonian flow \citep{Hoda2008,Farrell1993}. 
The equations for this component  take on a particularly convenient block chain form that was exploited in previous studies \citep{Hoda2009,Jovanovic2011,Page2014}. 
We introduce this form of the state-space representation of the streamwise constant component of the linearized Oldroyd-B equation (\ref{LOB1})--(\ref{LOB2}) below.

Taking the Fourier transform of (\ref{LOB1})--(\ref{LOB3}) in the homogenous $x$ and $z$ directions such that for any quantity, say $\phi = \phi(x,y,z,t)$ with $\phi'=\phi -\overline{\phi}$, we have
\begin{align}
\phi'(x,y,z,t) = \hat{\phi}(y,t)\,\text{e}^{\text{i}(k_x x + k_z z)}.
\end{align}
Using the standard approach to eliminate $p$ from the governing equations, reduces the state variables to the wall-normal velocity $\hat{v}$, the wall-normal vorticity $\hat{\eta}$ and the individual polymer stresses $\hat{T}_{ij}$ \citep{Hoda2009}.
The state vector can then be written as the Cartesian product $\bld{\varphi}_1 \times \bld{\varphi}_2  \times \bld{\varphi}_3$, where
\begin{align}
\bld{\varphi}_1 &\equiv \begin{bmatrix} \hat{v }&\hat{T}_{yy} &\hat{T}_{yz} & \hat{T}_{zz} \end{bmatrix}^{\mathsf{T}},\quad
\bld{\varphi}_2 \equiv \begin{bmatrix} \hat{\eta} & \hat{T}_{xy} & \hat{T}_{xz} \end{bmatrix}^{\mathsf{T}},\quad
\bld{\varphi}_3 \equiv \hat{T}_{xx}.
\end{align}

In the case when $k_x=0$, by incompressibility we have $-\text{i}k_z\hat{w} = \p_y \hat{v}$ and thus $\bld{\varphi}_1$ represents all the velocity fluctuations in the cross-plane, i.e. the $(y,z)$, plane.
The polymer stress components in $\bld{\varphi}_1$ can be used to form a stress tensor  
\begin{align}
\begin{bmatrix}
\hat{T}_{yy} & \hat{T}_{yz} \\
\hat{T}_{yz} & \hat{T}_{zz}
\end{bmatrix}
\end{align}
which arises due to shearing and volumetric deformation of the polymer in the cross-plane.
The wall-normal vorticity, $\hat{\eta}$, is included in $\bld{\varphi}_2$ and is well-known to be linearly coupled to the cross-plane velocity fluctuation due to the tilting of the streamwise vorticity perturbation by the base-flow \citep{Farrell1993}.
In an analogous manner, we group into $\bld{\varphi}_2$ the stresses that couple the cross-plane polymer perturbations to the purely streamwise perturbation $\hat{T}_{xx}=\bld{\varphi}_3$.

Setting $k_x = 0$ in the linearized, Fourier transformed equations, we retrieve the streamwise constant  equations in terms of the variables $\bld{\varphi}_i$, 
\begin{align}
\p_t\begin{bmatrix}
\bld{\varphi}_1  \\
\bld{\varphi}_2 \\
\bld{\varphi}_3
\end{bmatrix} &=
\begin{bmatrix}
\frac{1}{\Rey}\bld{\mathcal{L}} & \bld{0} & \bld{0} \\
\bld{\mathcal{C}} & \frac{1}{\Rey}\bld{\mathcal{S}} & \bld{0} \\
\bld{0} &\bld{\mathcal{A}} & -\frac{1}{\Wie}
\end{bmatrix}\begin{bmatrix}
\bld{\varphi}_1  \\
\bld{\varphi}_2 \\
\bld{\varphi}_3
\end{bmatrix} + 
 \begin{bmatrix}
\bld{\mathcal{B}}_{\bld{1}}  \\
\bld{\mathcal{B}}_{\bld{2}} \\
\bld{\mathcal{B}}_{\bld{3}} 
\end{bmatrix}  
\begin{bmatrix}
\bld{\hat{d}}_{\bld{u}} \\  \bld{\hat{d}}_{\mathsfbi{T}}
\end{bmatrix}. \label{swconstsys}
\end{align}
 The no-slip boundary conditions lead to the boundary conditions $\hat{v}(y = \pm 1,t) = \p_y\hat{v}(y = \pm 1,t) = 0$ on the wall-normal velocity and $\hat{\eta}(y = \pm 1,t) = 0$ on the wall-normal vorticity. The polymer stresses are free on the boundaries.  The operators $\bld{\mathcal{L}}$, $\bld{\mathcal{C}}$, $\bld{\mathcal{S}}$ and $\bld{\mathcal{A}}$ are matrix-valued and are given by
\begin{align}
\bld{\mathcal{L}} &\equiv \begin{bmatrix}  
\beta \mathcal{L}_{11}   & \alpha \mathcal{L}_{12}						 	& \alpha \mathcal{L}_{13}  										 &  \alpha \mathcal{L}_{14}\\
\frac{1}{{\Elas}}\mathcal{L}_{21}	  & -\frac{1}{{\Elas}}       & 0 	 & 0 \\
\frac{1}{{\Elas}}\mathcal{L}_{31}  & 0 	 & -\frac{1}{{\Elas}}								 & 0 \\
\frac{1}{{\Elas}}\mathcal{L}_{41}	  & 0 					  				       & 0 															 & -\frac{1}{{\Elas}} \\
\end{bmatrix}, \quad
\bld{\mathcal{A}} \equiv 
\begin{bmatrix}
\mathcal{A}_{11} & \mathcal{A}_{12} & 0 
\end{bmatrix}\nonumber\\
\bld{\mathcal{S}} &\equiv 
\begin{bmatrix}  
\beta \mathcal{S}_{11}   & \alpha \mathcal{S}_{12}						 	& \alpha \mathcal{S}_{13}  										\\
\frac{1}{{\Elas}}\mathcal{S}_{21}	  & -\frac{1}{{\Elas}} 				       & 0  \\
\frac{1}{{\Elas}}\mathcal{S}_{31}  & 0 									       & -\frac{1}{{\Elas}}
\end{bmatrix},  \quad
 \bld{\mathcal{C}}  \equiv 
\begin{bmatrix}
\mathcal{C}_{11} & 0 & 0 & 0\\
\mathcal{C}_{21} & \mathcal{C}_{22} & 0 & 0\\
\mathcal{C}_{31} & 0 & \mathcal{C}_{33} & 0
\end{bmatrix} \label{Opsdefn}
\end{align}
where we let $\alpha \equiv 1-\beta$ for notational convenience and
\begin{align}
\mathcal{L}_{11} &\equiv    \Delta^{-1}\Delta^2, \quad
\mathcal{L}_{12}  \equiv - \Delta^{-1} k_z^2\p_y, \quad
\mathcal{L}_{13}  \equiv - \text{i}k_z\Delta^{-1}(\p_{yy} + k_z^2), \nonumber\\
\mathcal{L}_{14} &\equiv  \Delta^{-1}k_z^2\p_y  \quad
\mathcal{L}_{21}  \equiv 2\p_y, \quad
\mathcal{L}_{31}  \equiv  - \frac{(\p_{yy} + k_z^2)}{\text{i}k_z}, \,\quad
\mathcal{L}_{41}  \equiv -2\p_y \nonumber\\
\mathcal{S}_{11} &\equiv \Delta, \quad
\mathcal{S}_{12}  \equiv \text{i}k_z\p_y, \quad
\mathcal{S}_{13}  \equiv -  k_z^2 , \quad
\mathcal{S}_{21}  \equiv \frac{1}{\text{i}k_z} \p_y, \quad 
\mathcal{S}_{31}  \equiv 1  \label{OpsOpsdefn} \\
\mathcal{C}_{11} &\equiv -\text{i}k_z \ddo{\overline{u}}{y}, \quad
\mathcal{C}_{21}  \equiv \ddo{\overline{u}}{y}\p_y - \ddtwo{\overline{u}'}{y} \nonumber\\
\mathcal{C}_{31}  &\equiv -\frac{1}{\text{i}k_z}\ddo{\overline{u}}{y}\p_{yy}, \quad
\mathcal{C}_{22} = \mathcal{C}_{33} \equiv \ddo{\overline{u}}{y},\quad
\mathcal{A}_{11} \equiv \frac{2}{\text{i}k_z}\ddo{\overline{u}}{y}\p_y, \quad
\mathcal{A}_{12} \equiv 2\ddo{\overline{u}}{y}. \nonumber
\end{align}
Finally the disturbance coefficient matrices $\bld{\mathcal{B}}_{\bld{1}}$, $\bld{\mathcal{B}}_{\bld{2}}$ and $\bld{\mathcal{B}}_{\bld{3}}$ are defined as
\begin{align}
\bld{\mathcal{B}}_{\bld{1}} &\equiv  
 \begin{bmatrix}
 0 & \mathcal{B}_{v,2} & \mathcal{B}_{v,3} & \bld{0} & \bld{0} \\
\bld{0} & \bld{0} & \bld{0} & \bld{0} & \text{diag}\left(\mathcal{B}_{yy},\mathcal{B}_{yz},\mathcal{B}_{zz}\right)   
\end{bmatrix}  \nonumber\\
\bld{\mathcal{B}}_{\bld{2}} &\equiv 
\begin{bmatrix} 
\text{diag}\left(\mathcal{B}_{\eta},0,0 \right) 
& \text{diag}\left(0, \mathcal{B}_{xy}, \mathcal{B}_{xz} \right) & \bld{0} 
\end{bmatrix}  \label{Bdefns} \\
\bld{\mathcal{B}}_{\bld{3}} &\equiv \begin{bmatrix} \bld{0}_{1\times 3} & \mathcal{B}_{xx} & \bld{0}   \end{bmatrix} \nonumber
\end{align}
where the disturbance coefficients $\mathcal{B}_{(\cdot)}$ are defined in the appendix and the notation $\text{diag}(\cdot)$ refers to a diagonal matrix with diagonal elements given in the argument. Alternatively by setting a disturbance coefficient to zero we can eliminate the forcing in the corresponding equation. For example, setting $\mathcal{B}_{\eta} = 0$ eliminates the forcing in the wall-normal vorticity $\hat{\eta}$. Similarly setting $\mathcal{B}_{v,i} = 0$ for $i \in \{2,3\}$ and $\mathcal{B}_{qr} = 0$ for $q,r \in \{x,y,z\}$ completely eliminates the forcing in the wall-normal velocity $\hat{v}$ and the polymer stress $\hat{T}_{qr}$, respectively. Note that
\begin{align}
\bld{\mathcal{B}}_{\bld{1}} \bld{\mathcal{B}}_{\bld{1}} ^* &= 
 \text{diag}\left( \mathcal{B}_{v,2}\mathcal{B}_{v,2}^* + \mathcal{B}_{v,3}\mathcal{B}_{v,3}^*, \mathcal{B}_{yy}\mathcal{B}_{yy}^*, \mathcal{B}_{yz}\mathcal{B}_{yz}^*,\mathcal{B}_{zz}\mathcal{B}_{zz}^*\right) \\
\bld{\mathcal{B}}_{\bld{2}} \bld{\mathcal{B}}_{\bld{2}}^* &= \text{diag}\left( \mathcal{B}_{\eta}\mathcal{B}_{\eta}^*, \mathcal{B}_{xy}\mathcal{B}_{xy}^*, \mathcal{B}_{xz}\mathcal{B}_{xz}^*\right).
\end{align}

%

We now consider  general system outputs that can be expressed as a linear combination of $\bld{\varphi}_1$ and $\bld{\varphi}_2$.
As discussed previously, although we allow forcing in $\bld{\varphi}_3 = \hat{T}_{xx}$, we do not consider  $\bld{\varphi}_3 = \hat{T}_{xx}$ as an output in the present work.
The output $\bld{\psi} = \bld{\psi} (y,k_z,t)$ is then 
\begin{align}
\bld{\psi}  =
  \bld{\mathcal{H}} 
\begin{bmatrix}
\bld{\varphi}_1  \\
\bld{\varphi}_2 
\end{bmatrix}  
=
\left(\bld{\mathcal{H}}_{\bld{u}}+ \bld{\mathcal{H}}_{\mathsfbi{T}}\right)
\begin{bmatrix}
\bld{\varphi}_1  \\
\bld{\varphi}_2 
\end{bmatrix} 
\label{outputEqn}
\end{align}
where $\bld{\mathcal{H}}_{\bld{u}}$ and $\bld{\mathcal{H}}_{\mathsfbi{T}}$ are given by 
\begin{align}
\bld{\mathcal{H}}_{\bld{u}}
&= 
\begin{bmatrix}  
0 & \bld{0}  & \mathcal{H}_{u} & \bld{0}  \\
\mathcal{H}_{v} & \bld{0}  & 0 & \bld{0}    \\
\mathcal{H}_{w} & \bld{0}& 0 & \bld{0} \\
\bld{0} &\bld{0} &\bld{0} &\bld{0} 
\end{bmatrix}, \quad
\bld{\mathcal{H}}_{\mathsfbi{T}} = 
\begin{bmatrix}
\bld{0} &\bld{0} &\bld{0} &\bld{0} \\
\bld{0}  & \bld{0} & \bld{0}  & \text{diag}\left(\mathcal{H}_{xy},\mathcal{H}_{xz} \right) \\
\bld{0}  & 
\text{diag}\left(\mathcal{H}_{yy},\mathcal{H}_{yz},\mathcal{H}_{zz}\right) &
\bld{0} & \bld{0}  
\end{bmatrix} \label{Hdefns}
\end{align}
and the matrix elements $\mathcal{H}_{u}$ , $\mathcal{H}_{v}$, and $\mathcal{H}_{w}$ are nonzero only if the output contains the quantity indicated in the subscript. 
Similarly  $\mathcal{H}_{ij}$ is nonzero only if the output contains $\hat{T}_{ij}$.
In this case, the matrix elements are defined in the appendix.
In particular, if we set $\bld{\mathcal{H}}_{\mathsfbi{T}} = \bld{0}$ or $\bld{\mathcal{H}}_{\bld{u}} = \bld{0}$, we have
\begin{align}
\bld{\psi}  &=
\begin{bmatrix} 
\hat{u} & \hat{v} &\hat{w} & \bld{0}
\end{bmatrix}^{\mathsf{T}}
\equiv \bld{\Psi}_{\bld{u}}, 
\quad
\bld{\psi} =
\begin{bmatrix}
\bld{0} &  
\hat{T}_{xy} & \hat{T}_{xz} &
\hat{T}_{yy} & \hat{T}_{yz} &
\hat{T}_{zz}
\end{bmatrix}^{\mathsf{T}}
\equiv \bld{\Psi}_{\mathsfbi{T}}, 
\end{align}
respectively.


The energy amplification $\mathcal{E}(k_z) $ is the ensemble average steady-state energy density (in physical space) maintained in the output $\bld{\psi}$ under spatiotemporal Gaussian white noise forcing and is given by
\begin{align}
\mathcal{E}(k_z) \equiv \lim_{t\rightarrow \infty} \frac{1}{2}\int_{-1}^{1}  
\Big\langle
\bld{\psi}^*\bld{\psi}(y,k_z,t)   \Big\rangle \,dy.
\end{align}
where $\langle \cdot \rangle$ denotes the ensemble average. This quantity, $\mathcal{E}$, is referred to as energy since the streamwise, wall-normal averaged fluid kinetic energy is $(1/2)\bld{\psi}^*\bld{\psi}(y,k_z,t)$ with $\bld{\mathcal{H}}_{\mathsfbi{T}} = 0$. For convenience, we refer to $\bld{\psi}^*\bld{\psi}(y,k_z,t)$ with $\bld{\mathcal{H}}_{\bld{u}} = 0$ but $\bld{\mathcal{H}}_{\bld{u}} \neq 0$ as polymer stress energy although this differs from the polymer strain energy. 


\section{Reynolds number scaling of energy amplification} \label{sec:ReScaling} 

We use the Reynolds number scaling  of the energy amplification  derived by  \citet{Hoda2009} as a starting point to derive the main scaling results in this paper.
In this section, we briefly discuss this preliminary Reynolds number scaling and in the appendix provide an extension that incorporates the contribution from $\bld{\varphi}_3$. 
We also examine the role of the base flow shear in the overall amplification.

If we assume the disturbances $\bld{\hat{d}}_{\bld{u}}$ and $\bld{\hat{d}}_{\mathsfbi{T}}$ are $\delta$-correlated Gaussian white-noise processes, then the ensemble average energy density for the system (\ref{swconstsys}) with output (\ref{outputEqn}) is given by \citep{Farrell1993}
\begin{align}
\mathcal{E}(k_z)= \text{tr}( \bld{\mathcal{H}} ^* \bld{\mathcal{H}}  \bld{\mathcal{P}}) \label{Esoln}
\end{align}
 where $\bld{\mathcal{P}}$ is the solution to the Lyapunov equation
\begin{align}
\begin{bmatrix}
\frac{1}{\Rey}\bld{\mathcal{L}} & 0  \\
\bld{\mathcal{C}} & \frac{1}{\Rey}\bld{\mathcal{S}}
\end{bmatrix}\bld{\mathcal{P}}
+
\bld{\mathcal{P}}
\begin{bmatrix}
\frac{1}{\Rey}\bld{\mathcal{L}}^*  & \bld{\mathcal{C}}^*   \\
0 & \frac{1}{\Rey}\bld{\mathcal{S}}^* 
\end{bmatrix}=
- 
\begin{bmatrix} 
\bld{\mathcal{B}}_{\bld{1}}\bld{\mathcal{B}}_{\bld{1}}^* & \bld{0} \\
\bld{0} & \bld{\mathcal{B}}_{\bld{2}}\bld{\mathcal{B}}_{\bld{2}}^*
\end{bmatrix} . \label{fullLyap}
\end{align}
where from (\ref{Bdefns}) we
\begin{align}
\bld{\mathcal{B}}_{\bld{1}} \bld{\mathcal{B}}_{\bld{1}} ^* &= \text{diag}\left( \mathcal{B}_{v,2}\mathcal{B}_{v,2}^* + \mathcal{B}_{v,3}\mathcal{B}_{v,3}^*, \mathcal{B}_{yy}\mathcal{B}_{yy}^*, \mathcal{B}_{yz}\mathcal{B}_{yz}^*,\mathcal{B}_{zz}\mathcal{B}_{zz}^*\right) \nonumber\\
\bld{\mathcal{B}}_{\bld{2}} \bld{\mathcal{B}}_{\bld{2}} ^*& = \text{diag}\left( \mathcal{B}_{\eta}\mathcal{B}_{\eta}^*, \mathcal{B}_{xy}\mathcal{B}_{xy}^*, \mathcal{B}_{xz}\mathcal{B}_{xz}^*\right) \label{B1B1tB2B2t}
\end{align} 
and we note that $\bld{\mathcal{B}}_{\bld{1}}\bld{\mathcal{B}}_{\bld{2}}^* = \bld{0}$. With the definitions in (\ref{Hdefns}), we can also write the expression $\bld{\mathcal{H}}^*\bld{\mathcal{H}}$ in (\ref{Esoln}) is as a diagonal matrix
\begin{align}
\bld{\mathcal{H}}^*\bld{\mathcal{H}} &=
\text{diag}\left(
\bld{\mathcal{H}}_{\bld{\varphi}_1},
\bld{\mathcal{H}}_{\bld{\varphi}_2} \right)   \label{DDt-HtH}
\end{align}
where
\begin{align}   
\bld{\mathcal{H}}_{\bld{\varphi}_1} &\equiv
 \text{diag}( \mathcal{H}_v^* \mathcal{H}_v + 
\mathcal{H}_w^* \mathcal{H}_w,0,0,0)
+
 \text{diag}(0,\mathcal{H}_{yy}^* \mathcal{H}_{yy},
\mathcal{H}_{yz}^* \mathcal{H}_{yz},
\mathcal{H}_{zz}^* \mathcal{H}_{zz})  \nonumber \\ 
\bld{\mathcal{H}}_{\bld{\varphi}_2} &\equiv
 \text{diag}( 
\mathcal{H}_u^* \mathcal{H}_u,0,0) +
 \text{diag}(0,\mathcal{H}_{xy}^* \mathcal{H}_{xy},
\mathcal{H}_{xz}^* \mathcal{H}_{xz})  \label{Hfpdef}
\end{align}
and the notation $\bld{\mathcal{H}}_{\bld{\varphi}_i}$ with $i \in \{ 1, 2\}$ implies that $\bld{\mathcal{H}}_{\bld{\varphi}_i}$ only contains quantities that are output coefficients of the state vector $\bld{\varphi}_i$ in the output equation (\ref{outputEqn}).
Using (\ref{DDt-HtH}) in (\ref{Esoln}) and  (\ref{fullLyap}) we obtain the following exact expression for $\mathcal{E}\left(k_z;\beta,{\Elas}\right)  $
\begin{align}
\mathcal{E}\left(k_z;\beta,{\Elas}\right)  = ( f_{1}\left(k_z;\beta,{\Elas}\right)  + f_{2}\left(k_z;\beta,{\Elas}\right)  )\,\Rey + g\left(k_z;\beta,{\Elas}\right)  \,\Rey^3 \label{Eexpr}
\end{align}
where
\begin{align} 
f_{1}\left(k_z;\beta,{\Elas}\right)  &= \text{tr}\left(\bld{\mathcal{H}}_{\bld{\varphi}_1}   \bld{\mathcal{X}} \right)  \label{f1defn}\\
 f_{2}\left(k_z;\beta,{\Elas}\right)&=  \text{tr}\left(\bld{\mathcal{H}}_{\bld{\varphi}_2}  \bld{\mathcal{W}} \right)\label{f2defn}\\
  g\left(k_z;\beta,{\Elas}\right)  &= \text{tr}\left(\bld{\mathcal{H}}_{\bld{\varphi}_2} \bld{\mathcal{Z}} \right)\label{f3defn}
\end{align}
and $\bld{\mathcal{X}}$, $\bld{\mathcal{W}}$ and $\bld{\mathcal{Z}}$ are solutions to the following Lyapunov/Sylvester equations 
\begin{align}
 \bld{\mathcal{S}} \bld{\mathcal{W}} + \bld{\mathcal{W}}  \bld{\mathcal{S}}^*  &= - \bld{\mathcal{B}}_{\bld{2}}\bld{\mathcal{B}}_{\bld{2}}^* \qquad
 \label{WLyap} \\ 
\bld{\mathcal{L}} \bld{\mathcal{X}} + \bld{\mathcal{X}}  \bld{\mathcal{L}}^*  &= -\bld{\mathcal{B}}_{\bld{1}}\bld{\mathcal{B}}_{\bld{1}}^*  \label{XLyap}\\
 \bld{\mathcal{S}} \bld{\mathcal{Y}} + \bld{\mathcal{Y}}  \bld{\mathcal{L}}^*  &= - \bld{\mathcal{Q}}, \quad
 \bld{\mathcal{Q}} = \bld{\mathcal{C}}\bld{\mathcal{X}}
 \label{YSylv} \\
 \bld{\mathcal{S}} \bld{\mathcal{Z}} + \bld{\mathcal{Z}}  \bld{\mathcal{S}}^*  &= -\bld{\mathcal{R}}, \quad
 \bld{\mathcal{R}} =  \bld{\mathcal{C}}\bld{\mathcal{Y}}^*  + \bld{\mathcal{Y}}\bld{\mathcal{C}}^* .
 \label{ZLyap} 
\end{align} 
where we note that $\bld{\mathcal{W}}, \bld{\mathcal{X}}$ and $\bld{\mathcal{Z}}$ are Hermitian since $\bld{\mathcal{B}}_{\bld{1}}\bld{\mathcal{B}}_{\bld{1}}^*$,  $\bld{\mathcal{B}}_{\bld{2}}\bld{\mathcal{B}}_{\bld{2}}^*$ and $\bld{\mathcal{R}}$ are Hermitian.  A derivation for the expression (\ref{Eexpr}) is provided in the appendix and was first reported by \citet{Hoda2009}. 

Defining the new operators $\bld{\mathcal{\tilde{L}}} = \Elas \bld{\mathcal{L}} $ and $\bld{\mathcal{\tilde{S}}} = \Elas \bld{\mathcal{S}}$, it is easy to show $\bld{\mathcal{P}}$ also solves the Lyapunov equation
\begin{align}
\begin{bmatrix}
\frac{1}{\Wie}\bld{\mathcal{\tilde{L}}} & 0  \\
\bld{\mathcal{C}} & \frac{1}{\Wie}\bld{\mathcal{\tilde{S}}}
\end{bmatrix}\bld{\mathcal{P}}
+
\bld{\mathcal{P}}
\begin{bmatrix}
\frac{1}{\Wie}\bld{\mathcal{\tilde{L}}}^*  & \bld{\mathcal{C}}^*   \\
0 & \frac{1}{\Wie}\bld{\mathcal{\tilde{S}}}^* 
\end{bmatrix}=
- 
\begin{bmatrix} 
\bld{\mathcal{B}}_{\bld{1}}\bld{\mathcal{B}}_{\bld{1}}^* & \bld{0} \\
\bld{0} & \bld{\mathcal{B}}_{\bld{2}}\bld{\mathcal{B}}_{\bld{2}}^*
\end{bmatrix}. \label{fullWieLyap}
\end{align}
and thus the $\Wie$ scaling of the energy amplification is given by
\begin{align}
\mathcal{E}\left(k_z;\beta,{\Elas}\right)  = ( \tilde{f}_{1}\left(k_z;\beta,{\Elas}\right)  + \tilde{f}_{2}\left(k_z;\beta,{\Elas}\right)  )\,\Wie + \tilde{g}\left(k_z;\beta,{\Elas}\right)  \,\Wie^3 \label{EWieexpr}
\end{align}
where $\tilde{f}_{1}(k_z;\beta,{\Elas})  = \text{tr}(\bld{\mathcal{H}}_{\bld{\varphi}_1}\bld{\mathcal{\tilde{X}}})$,  $\tilde{f}_{2}(k_z;\beta,{\Elas})=  \text{tr}(\bld{\mathcal{H}}_{\bld{\varphi}_2}  \bld{\mathcal{\tilde{W}}} )$, and
$\tilde{g}(k_z;\beta,{\Elas}) = \text{tr}(\bld{\mathcal{H}}_{\bld{\varphi}_2} \bld{\mathcal{\tilde{Z}}})$ and the operators $\bld{\mathcal{\tilde{W}}}$, $\bld{\mathcal{\tilde{X}}}$, and $\bld{\mathcal{\tilde{Z}}}$ are obtained from a set Lyapunov/Sylvester equations similar to (\ref{WLyap})--(\ref{ZLyap}), with $\bld{\mathcal{L}}$ and $\bld{\mathcal{S}}$ replaced by $\bld{\mathcal{\tilde{L}}}$ and $\bld{\mathcal{\tilde{S}}}$, respectively. 

The contribution to $\mathcal{E}$ from the cross-stream kinetic energy and from the polymer stress energy associated with cross-stream polymer deformation, i.e. $|\hat{T}_{yy}|^2$, $|\hat{T}_{yz}|^2$ and $|\hat{T}_{zz}|^2$, is contained in $f_1$.  This can be seen most clearly by substituting (\ref{Hfpdef}) into (\ref{f1defn})  
\begin{multline} 
f_{1}\left(k_z;\beta,{\Elas}\right)  = \text{tr}\Big[
\underbrace{( \mathcal{H}_v^* \mathcal{H}_v + 
\mathcal{H}_w^* \mathcal{H}_w)\mathcal{X}_{11}}_{\text{streamwise vortices}} \\
+
\underbrace{\mathcal{H}_{yy}^* \mathcal{H}_{yy}\mathcal{X}_{22}  +
\mathcal{H}_{yz}^* \mathcal{H}_{yz}\mathcal{X}_{33}  +
\mathcal{H}_{zz}^* \mathcal{H}_{zz}\mathcal{X}_{44} }_{\text{cross-stream deformation of polymers}}
\Big]   \label{f1full} 
\end{multline}
where $\mathcal{X}_{ij}$ is the $(i,j)$ element of the $4\times 4$ matrix $\bld{\mathcal{X}}$. Thus $\mathcal{X}_{11}$ is proportional to the contribution of the cross-stream kinetic energy to $\mathcal{E}$. This contribution is dominated by streamwise vortices \citep{Jovanovic2011} that eventually lead to streaks of alternating low/high streamwise velocity $\hat{u}$. The remaining elements $\mathcal{X}_{22}$, $\mathcal{X}_{33}$ and $\mathcal{X}_{44}$ form the polymer stress energy contribution to $\mathcal{E}$ associated with polymer deformation in the cross-stream plane. The elements $\mathcal{X}_{ij}$ for $i\neq j$ are the cross-correlations between $\hat{v}$, $\hat{T}_{yy}$, $\hat{T}_{yz}$ and $\hat{T}_{zz}$.
The input that leads to $f_1$ is the forcing in $\bld{\varphi}_1$, which includes forcing in the cross-stream velocity components as well as the cross-stream polymer stresses.

The contribution to $\mathcal{E}$ from the cofficients  $f_2$ and $g$ comes from the streamwise kinetic energy $|\hat{u} |^2$ and from the energy associated  with the stresses due to shear or isochoric polymer deformation in the streamwise planes, i.e. $|\hat{T}_{xy}|^2$ and $|\hat{T}_{xz}|^2$.  We see this by substituting (\ref{Hfpdef}) into (\ref{f2defn}) and (\ref{f3defn})  
\begin{align}
 f_{2}\left(k_z;\beta,{\Elas}\right)&= 
 \text{tr}\big(
\underbrace{\mathcal{H}_u^* \mathcal{H}_u \mathcal{W}_{11} }_{\text{I}} +
\underbrace{\mathcal{H}_{xy}^* \mathcal{H}_{xy}\mathcal{W}_{22}  +
\mathcal{H}_{xz}^* \mathcal{H}_{xz}\mathcal{W}_{33}    }_{\text{II}}
\big)  \label{f2full}
\end{align}
 and
\begin{align}
  g\left(k_z;\beta,{\Elas}\right)  &=  \text{tr}\big(
\underbrace{ \mathcal{H}_u^* \mathcal{H}_u \mathcal{Z}_{11}}_{\text{I'}    }  +
 \underbrace{\mathcal{H}_{xy}^* \mathcal{H}_{xy}\mathcal{Z}_{22}  +
\mathcal{H}_{xz}^* \mathcal{H}_{xz}\mathcal{Z}_{33}    }_{\text{II}'}
\big).  \label{f3full}
\end{align}
where $\mathcal{W}_{ij}$ and $\mathcal{Z}_{ij}$ refer to the $(i,j)$ elements of $\bld{\mathcal{W}}$ and $\bld{\mathcal{Z}}$, respectively. 
The contribution to $|\hat{u}|^2$ is contained in I and I' of (\ref{f2full}) and (\ref{f3full}) while II and II' contain the contribution towards  $|\hat{T}_{xy}|^2$ and $|\hat{T}_{xz}|^2$.  The elements $\mathcal{W}_{ij}$ and $\mathcal{Z}_{ij}$ for $i\neq j$ do not appear in  (\ref{f2full}) and (\ref{f3full}) because their direct contribution is to the cross-correlations between  $\hat{\eta}$,  $\hat{T}_{xy}$ and $\hat{T}_{xz}$.

The distinction between $f_2$ and $g$ (and hence $\bld{\mathcal{W}}$ and $\bld{\mathcal{Z}}$ or I, II and I', II') and their different Reynolds number scaling is that $g$ arises due to a coupling brought about by the base-flow shear, $\ddo{\overline{u}}{y}$. 
Since the right-hand side of  (\ref{WLyap})  only consists of $\bld{\mathcal{B}}_{\bld{2}}$, $f_2$ arises because of amplification of the white-noise forcing in $\bld{\varphi}_2$.
However, from (\ref{ZLyap}) we see that $\bld{\mathcal{Z}}$ is a function of $\bld{\mathcal{X}}$ (via $\bld{\mathcal{Y}}$). 
Therefore $g$ is the contribution due to white-noise forcing in $\bld{\varphi}_1$
feeding into $\bld{\varphi}_2$ via the coupling term $\bld{\mathcal{C}}\bld{\varphi}_1$ in (\ref{swconstsys}). 

The coupling coefficient $\bld{\mathcal{C}}$ is independent of $\beta$ and $\Elas$ but is strongly dependent on the background shear $\ddo{\overline{u}}{y}$ as can be seen from (\ref{Opsdefn}) and (\ref{OpsOpsdefn}). 
In Newtonian flows, the coupling mechanism is referred to as the `lift-up effect' and is represented by the term I' in our analysis.  
In the viscoelastic case, we now have the term II' which is the amplification in $\hat{T}_{xy}$ and $\hat{T}_{xz}$.
By employing singular perturbation methods with $\bld{d}_{\bld{u}} \neq 0$ but $\bld{d}_{\mathsfbi{T}} = 0$, \citet{Jovanovic2011} found that the `viscoelastic lift-up effect' implied by II' is significant even in weakly inertial flows. 
We discuss the critical role of $\ddo{\overline{u}}{y}$  in generating the amplification that leads to I' and II' in the next section.

\subsection{Dependence of the energy amplification on the background shear}
\label{sec:sheardep}
We can illustrate the contribution of the $\ddo{\overline{u}}{y}$ dependent `viscoelastic lift-up' mechanism  to the energy amplification $\mathcal{E}$  by considering the special case of Couette flow, i.e. where $\ddtwo{\overline{u}}{y} = 0$. 
Note that unless we choose a different normalization, $\ddo{\overline{u}}{y} = 1$.

We can then write the coupling coefficient $\bld{\mathcal{C}}$ given in (\ref{Opsdefn}) as 
\begin{align}
\bld{\mathcal{C}} = \ddo{\overline{u}}{y}\bld{\mathcal{C}}_{0}
\end{align}
where $\bld{\mathcal{C}}_{0}$ is independent of  $\ddo{\overline{u}}{y}$. The remaining operators, $\bld{\mathcal{L}}$ and $\bld{\mathcal{S}}$ are also independent of $\ddo{\overline{u}}{y}$. Hence the solutions  of (\ref{WLyap}) and (\ref{XLyap}), $\bld{\mathcal{W}}$ and $\bld{\mathcal{X}}$, remain unaffected by $\ddo{\overline{u}}{y}$. 
On the other hand, the solution $\bld{\mathcal{Y}}$ to (\ref{YSylv}) is then given by $\bld{\mathcal{Y}} = \ddo{\overline{u}}{y}\bld{\mathcal{Y}}_{0}$ where $\bld{\mathcal{Y}}_{0}$ is the background shear independent solution to
\begin{align}
 \bld{\mathcal{S}} \bld{\mathcal{Y}}_0 + \bld{\mathcal{Y}}_0  \bld{\mathcal{L}}^*  &= -   \bld{\mathcal{C}}_0\bld{\mathcal{X}}. \nonumber
\end{align}
Similarly, the solution $\bld{\mathcal{Z}}$ to (\ref{ZLyap}) is given by $\bld{\mathcal{Z}} = (\ddo{\overline{u}}{y})^2\bld{\mathcal{Z}}_0$ where $\bld{\mathcal{Z}}_0$ is the background shear independent solution to
\begin{align}
 \bld{\mathcal{S}} \bld{\mathcal{Z}}_0 + \bld{\mathcal{Z}}_0  \bld{\mathcal{S}}^*  &= -( \bld{\mathcal{C}}_0\bld{\mathcal{Y}}_0^*  + \bld{\mathcal{Y}}_0\bld{\mathcal{C}}_0^*) . \label{Z0Lyap}
\end{align}
From (\ref{Eexpr}) and (\ref{f3defn}) we then have
\begin{align}
\mathcal{E}(k_z;\beta,\Elas) =  \left[ f_1(k_z;\beta,\Elas) + f_2(k_z;\beta,\Elas)   +\left(\Rey \ddo{\overline{u}}{y}\right)^2 f_{3'}(k_z;\beta,\Elas) \right] \,\Rey \label{shearEexpr}
\end{align}
where $g = (\ddo{\overline{u}}{y})^2 f_{3'} $ is as defined in (\ref{f3defn}) for the particular case of Couette flow and $\bld{\mathcal{Z}}_0$ is the solution to (\ref{Z0Lyap}). 
The functions $f_1$, $f_2$ and $f_{3'}$ do not depend on $\ddo{\overline{u}}{y}$.
A result similar to (\ref{shearEexpr}) can be shown to hold in Newtonian Couette flow as well.
In addition, previous results suggest that the $(\ddo{\overline{u}}{y})^2$ part of the scaling of kinetic energy also appears in Poiseuille flow  \citep{Landahl1980}. The scaling $\left(\ddo{\overline{u}}{y}\right)^2$ immediately follows from the $\ddo{\overline{u}}{y}$ proportional coupling term in the Squire equation. 

An interesting implication of (\ref{shearEexpr}) is that there is a $\ddo{\overline{u}}{y}$ dependent contribution to $\mathcal{E}$ that is not kinetic energy. 
Specifically, from (\ref{f3full}) this contribution arises because of the polymer stresses $\hat{T}_{xy}$ and $\hat{T}_{xz}$  and is given by
\begin{align}
\Rey^3\, \text{tr}(\mathcal{H}_{xy}^* \mathcal{H}_{xy}\mathcal{Z}_{22}  + \mathcal{H}_{xz}^* \mathcal{H}_{xz}\mathcal{Z}_{33}). \nonumber
\end{align} 
The dependence of this contribution on $\ddo{\overline{u}}{y}$ arises not only due to a direct coupling via the base velocity $\bld{\overline{u}}$ but also due to coupling via the $\ddo{\overline{u}}{y}$ dependent base state polymer stress $\mathsfbi{\overline{T}}$. 
 From (\ref{swconstsys}) and (\ref{Opsdefn}), we see that the $\mathcal{C}_{21} =  \ddo{\overline{u}}{y}\p_y - \ddtwo{\overline{u}}{y}$ and $\mathcal{C}_{31} = -(\ddo{\overline{u}}{y}/\text{i}k_z)\p_{yy}$ components of $\bld{\mathcal{C}}$ provide direct one-way coupling via the base-flow from $\hat{v}$ to $\hat{T}_{xy}$ and $\hat{T}_{xz}$.  In the case of Couette flow, this coupling is effective only due to gradients in $\hat{v}$. 

There is a similar base-flow dependent one-way coupling acting between the polymer stresses; $\mathcal{C}_{22} = \mathcal{C}_{33}  = \ddo{\overline{u}}{y}$   providing a coupling from $\hat{T}_{yy}$, $\hat{T}_{yz}$ and $\hat{T}_{zz}$ to $\hat{T}_{xy}$ and $\hat{T}_{xz}$ . This polymer stress-polymer stress coupling can become significant even in the weakly inertial limit and does \emph{not} require spatial gradients in the polymer stresses.
This cross-stream polymer stress driven growth in streamwise polymer stress via coupling by $\ddo{\overline{u}}{y}$ does not require forcing in the velocity, i.e. $\bld{d}_{\bld{u}} =0$ with $\bld{d}_{\mathsfbi{T}} \neq 0$ is sufficient.

In the next section we derive the dependence of $f_i(k_z;\beta,{\Elas})$ on the elasticity number ${\Elas}$ in the limits of high/low $\Elas$.

\section{Derivation of $\Elas$ scaling of amplification for $\Elas \rightarrow 0$ and $\Elas \rightarrow \infty$ } \label{sec:main_results}
In this section we derive $\Elas$ explicit expressions for $f_1$, $f_2$ and $g$ in the weak and strong elastic regimes ($\Elas \rightarrow \infty$ and $\Elas \rightarrow 0$, respectively) by solving the algebraic equations (\ref{WLyap})--(\ref{ZLyap}) in these limits.
The expressions separation of the contribution to $\mathcal{E}$ from a particular input-output pair.
 The procedure is as follows:
\begin{enumerate}
\item Rewrite each of (\ref{WLyap})--(\ref{ZLyap}) as a linear system in the form $\mathbf{Ax} = \mathbf{b}$, where $\mathbf{A} = \mathbf{A}_0 + \Elas \mathbf{A}_1$ for some $\mathbf{A}_0$, $\mathbf{A}_1$ operators independent of $\Elas$. Each of (\ref{WLyap})--(\ref{ZLyap})  reduces to a linear system with the same structure -- the \emph{generic reduced form}.
\item Let $\Elas^{-1}$ for $\Elas \rightarrow \infty$ and $\Elas$ for $\Elas \rightarrow 0$ be small parameters in the problem. Then $\mathbf{A}^{-1}$ can be written as a convergent series because $\mathbf{A}^{-1} = \Elas^{-1}(\Elas^{-1}\mathbf{A}_0 + \mathbf{A}_1)^{-1}= (\mathbf{A}_0 + \Elas \mathbf{A}_1)^{-1}$. This gives us $\mathbf{x} = \mathbf{A}^{-1} \mathbf{b}$ explicit in $\Elas$.
\end{enumerate}


In the following we replace all infinite-dimensional operators with their $N\times N$ regular  finite-dimensional representations and, by  an abuse of notation, retain the same symbols.
The $N \times N$ block elements of a matrix $\bld{\mathcal{A}}$ are denoted $\mathcal{A}_{ij}$. Finally, $\mathsfbi{I} = \mathsfbi{I}_{N}$, the $N\times N$ identity matrix.



We cast each of (\ref{WLyap})--(\ref{ZLyap})  into the generic reduced form by algebraically reducing the equations and using the vectorize operator $\mathcal{V}:\mathbb{C}^{n\times m} \rightarrow \mathbb{C}^{nm}$ .
This operator is  defined for some given matrix $\mathbf{A}\in\mathbb{C}^{n\times m}$ as
\begin{align}
\mathcal{V}\left( \mathbf{A} \right) 
=
\mathcal{V}\left( \begin{bmatrix}
\mathbf{a}_{\mathbf{1}} & \mathbf{a}_2 & \hdots & \mathbf{a}_{m}
\end{bmatrix} \right)
\equiv \begin{bmatrix}
\mathbf{a}_{\mathbf{1}}  \\
\vdots  \\
\mathbf{a}_m 
\end{bmatrix} \label{vecdefn}
\end{align}
where $\mathbf{a}_i \in \mathbb{C}^n$ (for $i = 1,\hdots,m$) are the columns of $ \mathbf{A}$. With the help of the useful identity  $\mathcal{V}(\mathbf{AXB}) = (\mathbf{B}^\mathsf{T}\otimes \mathbf{A})\mathcal{V}(\mathbf{X})$  for any compatible matrices $\mathbf{A}$, $\mathbf{X}$ and $\mathbf{B}$, vectorization changes each $N\times N$ matrix unknown in  (\ref{WLyap})--(\ref{ZLyap}) to a $\Nzero \times 1$ vector unknown. For example, the $N\times N$ matrix unknowns $\mathcal{X}_{ij}$ in (\ref{XLyap}) are vectorized to $\Nzero \times 1$ vector unknowns $\mathcal{V}(\mathcal{X}_{ij})$. 

Suppose the unknowns in the generic reduced form are given in $\bld{a}_{\bld{u}}  \in \mathbb{C}^{{\Nzero}\times 1}$ and $\bld{a}_{\bld{1}}  \in \mathbb{C}^{{\Nzero}M\times 1}$ for $M \in \mathbb{N}$. In the context of (\ref{WLyap})--(\ref{ZLyap}), $\bld{a}_{\bld{u}}$ and $\bld{a}_{\bld{1}}$ contain a subset of the vectorized unknowns $\mathcal{W}_{ij}$, $\mathcal{X}_{ij}$, $\mathcal{Y}_{ij}$ or $\mathcal{Z}_{ij}$.  With $M = 4$, $6$, $4$, and $5$ we thus have
\begin{align}
 \bld{w}_{\bld{u}}   &\equiv 
\mathcal{V}(\mathcal{W}_{11}), \quad
 \bld{w}_{\bld{1}}
 \equiv 
\mathcal{V}\big(\begin{bmatrix}
\mathcal{V}(\mathcal{W}_{12})  &
\mathcal{V}(\mathcal{W}_{13})  & 
\mathcal{V}(\mathcal{W}_{12}^* ) &
\mathcal{V}(\mathcal{W}_{13}^* ) 
\end{bmatrix}\big)   \nonumber\\
 \bld{x}_{\bld{u}}   &\equiv 
\mathcal{V}(\mathcal{X}_{11}), \quad  
\bld{x}_{\bld{1}}
 \equiv 
\mathcal{V}\big(\begin{bmatrix}
\mathcal{V}(\mathcal{X}_{12})  &
\mathcal{V}(\mathcal{X}_{13})  &
\mathcal{V}(\mathcal{X}_{14})  &
\mathcal{V}(\mathcal{X}_{12}^* ) &
\mathcal{V}(\mathcal{X}_{13}^* ) &
\mathcal{V}(\mathcal{X}_{14}^* )  
\end{bmatrix}\big)\nonumber\\
\bld{z}_{\bld{u}} &\equiv   \mathcal{V}(\mathcal{Z}_{11}), \quad\bld{z}_{\bld{1}}  \equiv 
\mathcal{V}\big(
\begin{bmatrix} 
\mathcal{V}(\mathcal{Z}_{12})  & 
\mathcal{V}(\mathcal{Z}_{13})  & 
\mathcal{V}(\mathcal{Z}_{12}^*)  & 
\mathcal{V}(\mathcal{Z}_{13}^*) 
\end{bmatrix}\big)   \nonumber \\ 
\bld{y}_{\bld{0}} &\equiv   \mathcal{V}(\mathcal{Y}_{11}), \quad
\bld{y}_{\bld{1}} \equiv 
\mathcal{V}\big(\begin{bmatrix}  
\mathcal{V}(\mathcal{Y}_{12}) & 
\mathcal{V}(\mathcal{Y}_{13}) & 
\mathcal{V}(\mathcal{Y}_{14}) & 
\mathcal{V}(\mathcal{Y}_{21}) & 
\mathcal{V}(\mathcal{Y}_{31})
\end{bmatrix}\big) \label{vecUnknowns}
\end{align} 
The subscript $\bld{u}$ indicates that the quantity is used to compute the (fluid) kinetic energy contribution in the expressions (\ref{f1full})--(\ref{f3full}). The elements $\mathcal{X}_{1j}$, $\mathcal{Z}_{1j}$ and $\mathcal{W}_{1j}$ for $j \neq 1$ contain contributions to the velocity-polymer stress cross-correlations. The $y$-dependent correlations $\langle \hat{v}^* \hat{T}_{yy}\rangle $, $\langle \hat{v}^* \hat{T}_{yz}\rangle $, and $\langle \hat{v}^* \hat{T}_{zz}\rangle$ are given by $\Rey \mathcal{X}_{12}$, $\Rey \mathcal{X}_{13}$ and $\Rey \mathcal{X}_{14}$. Similarly $\langle \hat{\eta}^* \hat{T}_{xy}\rangle = \text{i}k_z\langle \hat{u}^* \hat{T}_{xy}\rangle $ and $\langle \hat{\eta}^* \hat{T}_{xz}\rangle = \text{i}k_z\langle \hat{u}^* \hat{T}_{xz}\rangle $ are given by $\Rey \mathcal{W}_{12} + \Rey^3 \mathcal{Z}_{12}$ and $\Rey \mathcal{W}_{13} + \Rey^3 \mathcal{Z}_{13}$. The cross-correlations are of general interest but we do not consider them further.

For the generic unknowns $\bld{a}_{\bld{u}}$ and $\bld{a}_{\bld{1}}$, the generic reduced form is given by
\begin{align} 
\beta \bld{\Gamma}_{\bld{a}} \bld{a}_{\bld{u}} + \alpha\bld{\Gamma}_{\bld{b}} \bld{a}_{\bld{1}}   =
-  \bld{b}_{\bld{u}}  \nonumber \\
{\Elas}^{-1}  \bld{\Gamma}_{\bld{c}}(\bld{1}_{M\times 1} \otimes \bld{a}_{\bld{u}} )
+ \bld{\Gamma}_{\bld{d}}  (\beta,{\Elas}) \bld{a}_{\bld{1}}  =
-    \bld{b}_{\mathsfbi{T}}  \label{presysA}
\end{align}
where $\bld{1}_{M\times 1}$ is the $M\times 1$ vector with 1 in its entries, $\bld{\Gamma}_{\bld{c}}  \in \mathbb{C}^{{\Nzero}M\times {\Nzero}M}$ is a diagonal matrix, $\bld{\Gamma}_{\bld{a}} \in \mathbb{C}^{{\Nzero}\times \Nzero}$, $\bld{\Gamma}_{\bld{b}} \in \mathbb{C}^{{\Nzero}\times {\Nzero}M}$  and $\bld{\Gamma}_{\bld{d}}: [0,1] \times (0,\infty)\mapsto  \mathbb{C}^{{\Nzero}M\times {\Nzero}M}$ is given by
\begin{align}
\bld{\Gamma}_{\bld{d}}  (\beta,{\Elas}) &\equiv  \bld{\gamma}(\beta) - \frac{1}{{\Elas}}\mathsfbi{I}_{{\Nzero}M}, \quad
%
\bld{\gamma}(\beta)  \equiv (1-\beta) \bld{\gamma}_\alpha + \beta \bld{\gamma}_\beta \label{pregammadefn} 
\end{align}
where $\bld{\gamma}_\alpha,\bld{\gamma}_\beta \in \mathbb{C}^{{\Nzero}M\times {\Nzero}M}$. All the matrices defined are independent of $\Elas$ except for $\bld{\Gamma}_{\bld{d}}$ and all are defined in terms ${\Nzero}\times{\Nzero}$ block matrix elements. Similarly, all the matrices defined are also independent of $\beta$ (hence also $\alpha$) except for $\bld{\Gamma}_{\bld{d}}$ and $\bld{\gamma}$. The vectors $\bld{b}_{\bld{u}} \in \mathbb{C}^{{\Nzero}\times 1}$ and $\bld{b}_{\mathsfbi{T}} \in \mathbb{C}^{{\Nzero}M\times 1}$ are known vectors. In the context of the matrix equations (\ref{WLyap})--(\ref{ZLyap})  the vectors $\bld{b}_{\bld{u}}$ and $\bld{b}_{\mathsfbi{T}}$ are related to the right-hand sides of these equation. Thus, for example the $\bld{b}_{\bld{u}}$ vector associated with (\ref{WLyap}) is related to the first diagonal element of $-\bld{\mathcal{B}}_{\bld{2}}\bld{\mathcal{B}}_{\bld{2}}^*$ which by (\ref{B1B1tB2B2t}) is $-\mathcal{B}_{\eta}\mathcal{B}_{\eta}^*$ while the remaining diagonal elements $-\bld{\mathcal{B}}_{\bld{2}}\bld{\mathcal{B}}_{\bld{2}}^*$ are contained in $\bld{b}_{\mathsfbi{T}}$. 

The solution to the generic linear system (\ref{presysA})  for the two cases $\Elas \rightarrow \infty$ and $\Elas \rightarrow 0$ (strongly and weakly elastic) can be derived with the aid of the Neumann series. This solution is given in the appendix along with sufficient bounds on $\Elas$ for the two regimes. We leverage this solution to find the $\Elas$ scaling of $f_i(k_z;\beta,\Elas)$ in the two limits of elasticity. 
The particular reduced forms, for each of (\ref{WLyap})--(\ref{ZLyap}), are given in the appendix.

We now write the expressions for $f_1$, $f_2$ and $g$ given in (\ref{f1defn})--(\ref{f3defn}) in terms of the vectorized unknowns. Defining the vectors
\begin{align}
\bld{x}_{\mathsfbi{T}} 
&\equiv
\mathcal{V}\big(\begin{bmatrix}
\mathcal{V}(\mathcal{X}_{22}) & 
\mathcal{V}(\mathcal{X}_{33}) & 
\mathcal{V}(\mathcal{X}_{44}) 
\end{bmatrix}\big) \label{xpdefn}\\
\bld{w}_{\mathsfbi{T}} &\equiv
\mathcal{V}\big(\begin{bmatrix}
\mathcal{V}(\mathcal{W}_{22}) & 
\mathcal{V}(\mathcal{W}_{33}) 
 \end{bmatrix}\big)  \label{wpdefn}\\ 
\bld{z}_{\mathsfbi{T}} &\equiv 
\mathcal{V}\big(\begin{bmatrix}
\mathcal{V}(\mathcal{Z}_{22})  &
\mathcal{V}(\mathcal{Z}_{33})
\end{bmatrix}\big). \label{zpdefn}
\end{align}
we have from (\ref{f1defn})--(\ref{f3defn}) an equivalent expression for $f_1$ as
\begin{align}
f_{1}\left(k_z;\beta,{\Elas}\right)  &= \text{tr}_{\mathcal{V}}\left(\bld{\mathscr{H}}_{\bld{u},\bld{\varphi}_1} \bld{x}_{\bld{u}} \right) 
+   \text{tr}_{\mathcal{V}}\left(\bld{\mathscr{H}}_{\mathsfbi{T},\bld{\varphi}_1} \bld{x}_{\mathsfbi{T}} 
\right)  \label{vecf1} \\
%
  f_{2}\left(k_z;\beta,{\Elas}\right)  
%
&=  \text{tr}_{\mathcal{V}}\left(\bld{\mathscr{H}}_{\bld{u},\bld{\varphi}_2} \bld{w}_{\bld{u}} \right) 
+  \text{tr}_{\mathcal{V}}\left(\bld{\mathscr{H}}_{\mathsfbi{T},\bld{\varphi}_2} \bld{w}_{\mathsfbi{T}} 
\right)  \label{vecf2} \\
  g\left(k_z;\beta,{\Elas}\right)  
%
&=  \text{tr}_{\mathcal{V}}\left(\bld{\mathscr{H}}_{\bld{u},\bld{\varphi}_2} \bld{z}_{\bld{u}} \right) 
+ \text{tr}_{\mathcal{V}}\left(\bld{\mathscr{H}}_{\mathsfbi{T},\bld{\varphi}_2} \bld{z}_{\mathsfbi{T}}  
\right) \label{vecf3}  
\end{align} 
where we define the operator $\text{tr}_{\mathcal{V}}: \mathbb{C}^{n^2 \times 1} \rightarrow \mathbb{C}$   with $n \in \mathbb{N}$  
\begin{align}
\text{tr}_{\mathcal{V}}( \mathcal{V}(\bld{X}) ) = \text{tr}(\bld{X})
\label{trVdefn}
\end{align} 
and the operators $\bld{\mathscr{H}}_{\bld{u},\bld{\varphi}_i}$  and $\bld{\mathscr{H}}_{\mathsfbi{T},\bld{\varphi}_i}$ by
\begin{align}
\bld{\mathscr{H}}_{\bld{u},\bld{\varphi}_1} &\equiv 
\mathsfbi{I} \otimes (\mathcal{H}_v^* \mathcal{H}_v + 
\mathcal{H}_w^* \mathcal{H}_w) \nonumber \\
\bld{\mathscr{H}}_{\mathsfbi{T},\bld{\varphi}_1} &\equiv 
 \begin{bmatrix}
\mathsfbi{I} \otimes \mathcal{H}_{yy}^* \mathcal{H}_{yy} &
\mathsfbi{I} \otimes \mathcal{H}_{yz}^* \mathcal{H}_{yz} &
\mathsfbi{I} \otimes \mathcal{H}_{zz}^* \mathcal{H}_{zz}
\end{bmatrix}.
\label{vecHfp1defn}  \\
\bld{\mathscr{H}}_{\bld{u},\bld{\varphi}_2} &\equiv \mathsfbi{I} \otimes \mathcal{H}_u^* \mathcal{H}_u, \quad
\bld{\mathscr{H}}_{\mathsfbi{T},\bld{\varphi}_2}  \equiv  \begin{bmatrix}
\mathsfbi{I} \otimes \mathcal{H}_{xy}^* \mathcal{H}_{xy} &
\mathsfbi{I} \otimes \mathcal{H}_{xz}^* \mathcal{H}_{xz}  
\end{bmatrix}.\label{vecHfp2defn}
\end{align}

From (\ref{vecf1}) and (\ref{vecHfp1defn}) we see that $\bld{x}_{\bld{u}}$ contains the contribution of cross stream velocities to $f_1$ and $\bld{x}_{\mathsfbi{T}}$ contains the contribution of $\hat{T}_{yy}$, $\hat{T}_{yz}$, $\hat{T}_{zz}$ to $f_1$. Similarly from (\ref{vecf2})--(\ref{vecHfp2defn}) we see that $\bld{w}_{\bld{u}}$ and $\bld{z}_{\bld{u}}$ are the contributions of the streamwise velocity to $f_2$ and $g$, respectively while  $\bld{w}_{\mathsfbi{T}}$ and $\bld{z}_{\mathsfbi{T}}$ are the contributions of $\hat{T}_{xy}$, $\hat{T}_{xz}$ to $f_2$ and $g$, respectively.


The origin of the forcing that leads to a certain contribution in $\mathcal{E}$  can be identified using the vectorized diagonal elements of (\ref{B1B1tB2B2t}) 
\begin{align}
\bld{\mathscr{D}}_{\bld{u},\bld{\varphi}_1}
&\equiv \mathcal{V}(\mathcal{B}_{v,2} \mathcal{B}_{v,2}^* + \mathcal{B}_{v,3} \mathcal{B}_{v,3}^*), \quad   
\bld{\mathscr{D}}_{\mathsfbi{T},\bld{\varphi}_1}
\equiv
\mathcal{V}\big(\begin{bmatrix}
 \mathcal{V}(\mathcal{B}_{yy} \mathcal{B}_{yy}^*)  &
 \mathcal{V}(\mathcal{B}_{yz}\mathcal{B}_{yz}^*)  &
 \mathcal{V}(\mathcal{B}_{zz} \mathcal{B}_{zz}^*) 
\end{bmatrix}\big)\nonumber \\
\bld{\mathscr{D}}_{\bld{u},\bld{\varphi}_2}
&\equiv \mathcal{V}(\mathcal{B}_{\eta} \mathcal{B}_{\eta}^*), \quad  
\bld{\mathscr{D}}_{\mathsfbi{T},\bld{\varphi}_2}
\equiv
\mathcal{V}\big(\begin{bmatrix}
 \mathcal{V}(\mathcal{B}_{xy} \mathcal{B}_{xy}^*)  &
 \mathcal{V}(\mathcal{B}_{xz}\mathcal{B}_{xz}^*) 
\end{bmatrix}\big)
\label{vecBfpdefn}
\end{align}
where $\bld{\mathscr{D}}_{\bld{u},\bld{\varphi}_i} = 0$ implies no stochastic forcing in the fluid component of $\bld{\varphi}_i$ and similarly $\bld{\mathscr{D}}_{\mathsfbi{T},\bld{\varphi}_i} = 0$ implies no stochastic forcing in polymer stress component of $\bld{\varphi}_i$.

\begin{table}
\centering
\begin{tabular}{c p{12cm} }
Symbol & Interpretation  \\ 
\\
$\mathcal{E}$ & Total variance $\langle \hat{u}_k \hat{u}_k'\rangle + \langle \hat{T}_{k\ell} \hat{T}_{k\ell} \rangle$, with separable contributions from $\bld{\hat{u}}$ and $\mathsfbi{\hat{T}}$. \\
 $f_{a \rightarrow b}$ & Base-flow independent coefficient contributing to variance in $\hat{a}$ due to stochastic forcing in $\hat{b}$, independent of $\Rey$, and only weakly$^\dagger$ dependent on $\Elas$.\\
  $g_{a \rightarrow b}$ &  Base-flow dependent coefficient contributing to variance in $\hat{a}$ due to stochastic forcing in $\hat{b}$, independent of $\Rey$, and only weakly$^\dagger$ dependent on $\Elas$.\\\\
\end{tabular} 
\caption{Notation for the coefficients, with $a,b \in \{ \bld{u}, \mathsfbi{T} \}$.  $^\dagger$Independent up to zero-th order for $\Elas \rightarrow 0$ and $\Elas \rightarrow \infty$.}
\label{tab:notation}
\end{table} 

Applying the weak and strong elasticity solutions of the generic reduced (\ref{presysA}) provided in the appendix to each of (\ref{WLyap})--(\ref{ZLyap}), we then obtain the following expressions 

\begin{align}
f_1 + f_2 &=  
\begin{cases} 
f_{\bld{u} \rightarrow \bld{u}}+
\Elas^2 f_{\mathsfbi{T} \rightarrow \bld{u}}  +
f_{\bld{u} \rightarrow \mathsfbi{T}} +
\Elas f_{\mathsfbi{T} \rightarrow \mathsfbi{T}}  & \Elas \rightarrow 0 \\ 
f_{\bld{u} \rightarrow \bld{u}}+
\Elas f_{\mathsfbi{T} \rightarrow \bld{u}}  +
\Elas^{-1} f_{\bld{u} \rightarrow \mathsfbi{T}} +
\Elas f_{\mathsfbi{T} \rightarrow \mathsfbi{T}}  & \Elas \rightarrow \infty 
\end{cases}
\label{f1f2_components} \\
g &=  
\begin{cases}
g_{\bld{u} \rightarrow \bld{u}} +
\Elas^2 g_{\mathsfbi{T} \rightarrow \bld{u}}+
g_{\bld{u} \rightarrow \mathsfbi{T}} +
\Elas^2 g_{\mathsfbi{T} \rightarrow \mathsfbi{T}} & \Elas \rightarrow 0 \\
\Elas g_{\bld{u} \rightarrow \bld{u}} +
\Elas^3 g_{\mathsfbi{T} \rightarrow \bld{u}}+
\Elas g_{\bld{u} \rightarrow \mathsfbi{T}} +
\Elas^3 g_{\mathsfbi{T} \rightarrow \mathsfbi{T}} & \Elas \rightarrow \infty
\end{cases}  \label{f3_components}
\end{align}
where  $f_1 + f_2$ is the base-flow independent part with coefficients given by
\begin{align}
f_{\bld{u} \rightarrow \bld{u}}&= \text{tr}_{\mathcal{V}}\left(\bld{\mathscr{H}}_{\bld{u},\bld{\varphi}_1} \bld{\mathscr{X}}_{\bld{u} \rightarrow \bld{u}}\bld{\mathscr{D}}_{\bld{u},\bld{\varphi}_1}  + \bld{\mathscr{H}}_{\bld{u},\bld{\varphi}_2} \bld{\mathscr{W}}_{\bld{u} \rightarrow \bld{u}}\bld{\mathscr{D}}_{\bld{u},\bld{\varphi}_2}\right)  + \mathcal{O}(\Elas^p)\label{chi_uu} \\
f_{\mathsfbi{T} \rightarrow \bld{u}} &=  \text{tr}_{\mathcal{V}}\left(
 \bld{\mathscr{H}}_{\bld{u},\bld{\varphi}_1}\bld{\mathscr{X}}_{\mathsfbi{T} \rightarrow \bld{u}}  \bld{\mathscr{D}}_{\mathsfbi{T},\bld{\varphi}_1} 
 +  \bld{\mathscr{H}}_{\bld{u},\bld{\varphi}_2}\bld{\mathscr{W}}_{\mathsfbi{T} \rightarrow \bld{u}}  \bld{\mathscr{D}}_{\mathsfbi{T},\bld{\varphi}_2}   
\right) + \mathcal{O}(\Elas^p)\label{chi_tauu}\\
f_{\bld{u} \rightarrow \mathsfbi{T}} &=   \text{tr}_{\mathcal{V}}
\left(
 \bld{\mathscr{H}}_{\mathsfbi{T},\bld{\varphi}_1} \bld{\mathscr{X}}_{\bld{u} \rightarrow \mathsfbi{T}}\bld{\mathscr{D}}_{\bld{u},\bld{\varphi}_1} 
+  \bld{\mathscr{H}}_{\mathsfbi{T},\bld{\varphi}_2} \bld{\mathscr{W}}_{\bld{u} \rightarrow \mathsfbi{T}}\bld{\mathscr{D}}_{\bld{u},\bld{\varphi}_2} \right)+ \mathcal{O}(\Elas^p)\label{chi_utau}\\
f_{\mathsfbi{T} \rightarrow \mathsfbi{T}} &=  
 \text{tr}_{\mathcal{V}}\left(
  \bld{\mathscr{H}}_{\mathsfbi{T},\bld{\varphi}_1} \bld{\mathscr{X}}_{\mathsfbi{T} \rightarrow \mathsfbi{T}}   \bld{\mathscr{D}}_{\mathsfbi{T},\bld{\varphi}_1} 
 +   \bld{\mathscr{H}}_{\mathsfbi{T},\bld{\varphi}_2} \bld{\mathscr{W}}_{\mathsfbi{T} \rightarrow \mathsfbi{T}}  \bld{\mathscr{D}}_{\mathsfbi{T},\bld{\varphi}_2}
\right)+ \mathcal{O}(\Elas^p)  \label{chi_tautau}
\end{align}  
and   $g$ is the base-flow dependent part with coefficients
\begin{align}
g_{\bld{u} \rightarrow \bld{u}} &= \text{tr}_{\mathcal{V}}\left(\bld{\mathscr{H}}_{\bld{u},\bld{\varphi}_2} \bld{\mathscr{Z}}_{\bld{u}\rightarrow \bld{u}}\bld{\mathscr{D}}_{\bld{u},\bld{\varphi}_1}  \right) + \mathcal{O}(\Elas^p)  \label{rho_uu}\\
g_{\mathsfbi{T} \rightarrow \bld{u}} &=  \text{tr}_{\mathcal{V}}\left(   \bld{\mathscr{H}}_{\bld{u},\bld{\varphi}_2} \bld{\mathscr{Z}}_{\mathsfbi{T}\rightarrow \bld{u}}  
\bld{\mathscr{D}}_{\mathsfbi{T},\bld{\varphi}_1}  \right)  + \mathcal{O}(\Elas^p)\label{rho_tauu}\\
g_{\bld{u} \rightarrow \mathsfbi{T}}  &= \text{tr}_{\mathcal{V}}\left( \bld{\mathscr{H}}_{\mathsfbi{T},\bld{\varphi}_2}\bld{\mathscr{Z}}_{\bld{u}\rightarrow \mathsfbi{T}} \bld{\mathscr{D}}_{\bld{u},\bld{\varphi}_1}  \right)+ \mathcal{O}(\Elas^p) \label{rho_utau} \\
g_{\mathsfbi{T} \rightarrow \mathsfbi{T}}  &=  \text{tr}_{\mathcal{V}}\left(\bld{\mathscr{H}}_{\mathsfbi{T},\bld{\varphi}_2}\bld{\mathscr{Z}}_{\mathsfbi{T}\rightarrow \mathsfbi{T}} \bld{\mathscr{D}}_{\mathsfbi{T},\bld{\varphi}_1}  \right) + \mathcal{O}(\Elas^p), \label{rho_tautau}
\end{align}
where $p=1$ for $\Elas \rightarrow 0$ and $p=-1$ for $\Elas \rightarrow \infty$.
The physical interpretation associated with $f_{a\rightarrow b}$ and $g_{a \rightarrow b}$ is summarized in Table \ref{tab:notation}.

The quantities $f_{a\rightarrow b}$ and $g_{a \rightarrow b}$  are obtained using a zero-th order truncation of the operators $\bld{\mathscr{X}}_{a\rightarrow b}$, $\bld{\mathscr{W}}_{a\rightarrow b}$ and $\bld{\mathscr{Z}}_{a \rightarrow b}$ for $a,b \in \{\bld{u},\mathsfbi{T}\}$.
These operators can be expressed in terms of series expressions in $\Elas$ and are defined in the appendix up to zero-th order.
As a result,  $f_1+f_2$ and $g$ in (\ref{chi_uu})--(\ref{chi_tautau}) are correct $\mathcal{O}(\Elas^{-1})$ and $\mathcal{O}(\Elas)$ for $\Elas \rightarrow \infty$ and $\Elas \rightarrow 0$, respectively.
Each of the operators $\bld{\mathscr{X}}_{a\rightarrow b}$, $\bld{\mathscr{W}}_{a\rightarrow b}$ and $\bld{\mathscr{Z}}_{a \rightarrow b}$ is associated with an input-output pair which can be determined from the coefficients involved in the term. 
For example from (\ref{vecf1}) and (\ref{vecBfpdefn}) we see that $\text{tr}_{\mathcal{V}}(\bld{\mathscr{H}}_{\bld{u},\bld{\varphi}_1}\bld{\mathscr{X}}_{\mathsfbi{T} \rightarrow \bld{u}}  \bld{\mathscr{D}}_{\mathsfbi{T},\bld{\varphi}_1})$ is the contribution towards the cross-stream kinetic energy amplification due to forcing in the polymer stresses in $\bld{\varphi}_1$. We refer the reader to section \ref{sec:ReScaling} for a more detailed discussion on the input-output pairs in the present work. 

The expressions in  (\ref{f1f2_components}) and (\ref{f3_components}) for $f_1 + f_2$ and $g$ are explicit in their dependence on $\Elas$. 
In each case, this dependence takes the form of a geometric series' in either $\Elas$ for $\Elas \rightarrow 0$ or $\Elas^{-1}$ for $\Elas \rightarrow \infty$. 
Therefore, the leading order term in the geometric series' dominates but only in the limits of $\Elas \rightarrow 0$ or $\Elas \rightarrow \infty$ and the scaling of $f_i(k_z;\beta,\Elas)$ with $\Elas$ in these limits can be captured by the leading order terms.  
Higher-order terms may be significant at finite $\Elas$.
Figures \ref{fig:f1f2_ratio} and \ref{fig:f3_ratio} show the ratio of each coefficient $f_{a \rightarrow b}$ and $g_{a \rightarrow b}$ for $a,b\in\{\bld{u},\mathsfbi{T}\}$ evaluated between two $\Elas$ selected to represent the asymptotic regime under consideration and separated by atleast three orders of magnitude. 
A value of $1$ implies no dependence on $\Elas$.
We see a weak dependence on $\Elas$ throughout even though certain coefficients show much stronger dependence on higher order terms than others, for e.g., $g_{\bld{u} \rightarrow \bld{u}}$ in the strongly elastic regime.
We verify the expressions (\ref{f1f2_components}) and (\ref{f3_components}) more fully in the next section, where we also outline the numerical method.

\section{Discussion and verification of $\Wie$ and $\Rey$ scaling of amplification}  \label{sec:discussion}
\subsection{Numerical Methods}

We generate our numerical results using a standard pseudospectral method described in \citep{Weideman2000}. 
We discretize each of the operators in (\ref{swconstsys}) and (\ref{outputEqn}) using 60 collocation points  in the wall-normal direction and consider 60 spanwise wavenumbers $k_z$ logarithmically spaced in $[0.1,10]$. 
We use Gauss-Legendre quadrature to average variance in the wall-normal direction. 
We verified our computations for convergence and also compared them to results generated using a Galerkin scheme based on expansions in Chebyshev polynomials \citep{Boyd2001}.
Using the numerical results, we also examine the $k_z$ dependence for each case and discuss in particular the energy amplification due to stochastically forcing the polymer stresses.

\begin{figure}
\centering
\includegraphics[scale=1.0]{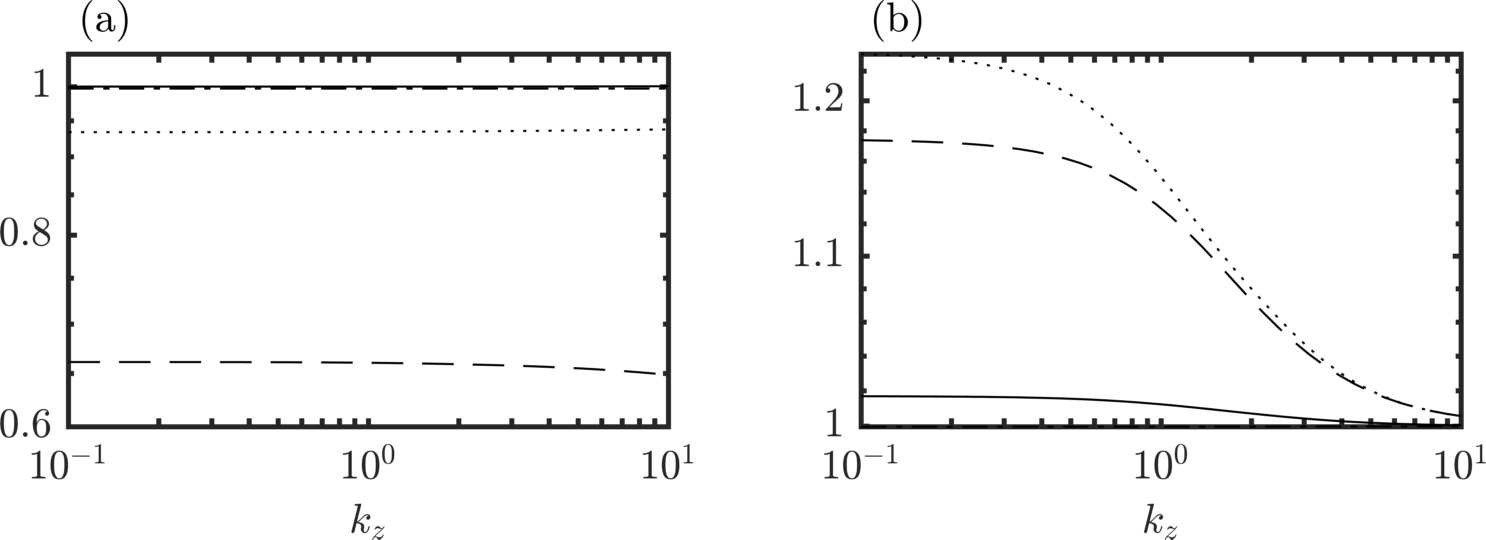}
\caption{Ratio $ f_{a \rightarrow b}(k_z;\Elas = \Elas_{\max},\beta) \big/f_{a \rightarrow b}(k_z;\Elas = \Elas_{\min},\beta) $. (a) Weakly elastic regime, $\Elas_{\min} =  10^{-8}$, $\Elas_{\max} = 10^{-5}$ and (b) Strongly elastic regime, $\Elas_{\min} = 10^0$, $\Elas_{\max} = 10^4$.
In both panels: solid lines (\linesolids) are $f_{\bld{u} \rightarrow \bld{u}}$,
dashed lines (\linedashed) are	$f_{\mathsfbi{T} \rightarrow \bld{u}}$,
dotted lines (\linedotted) are	$f_{\bld{u} \rightarrow \mathsfbi{T}}$, and
dash-dot lines (\linedshdot) are	$f_{\mathsfbi{T} \rightarrow \mathsfbi{T}}$.
Results are for both Couette and channel flow since $f_1 + f_2$ does not depend on the base flow. 
}
\label{fig:f1f2_ratio}
\end{figure} 

\begin{figure}
\centering
\includegraphics[scale=1.0]{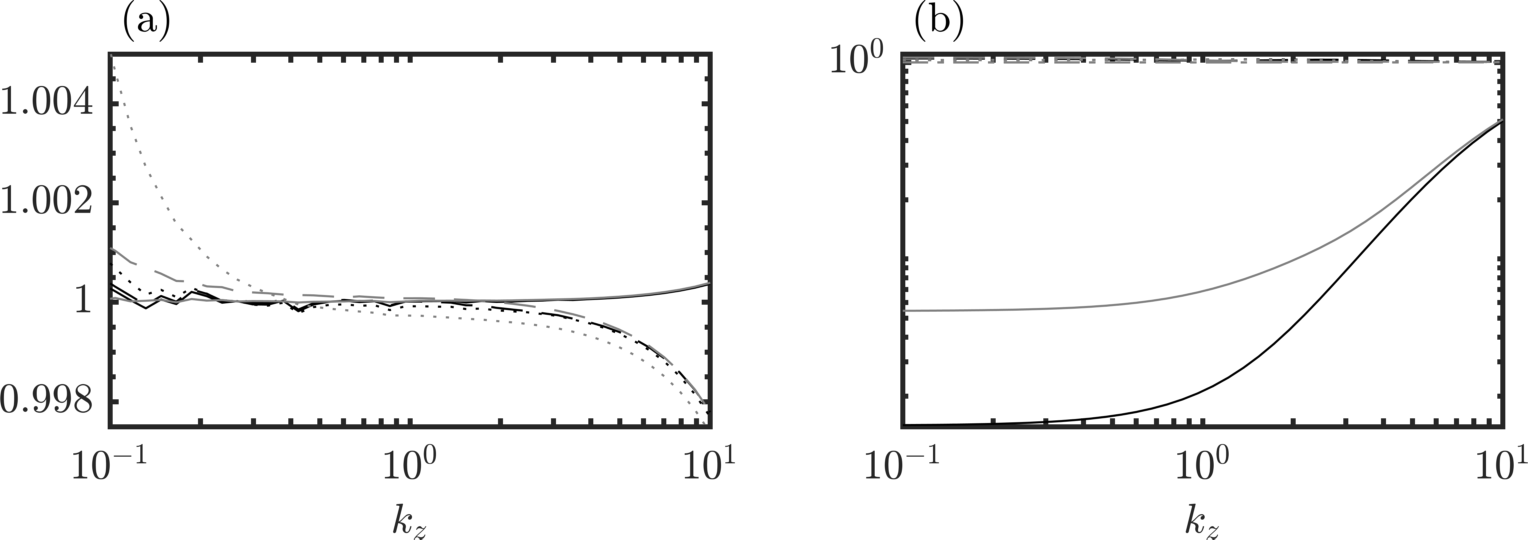}
\caption{Ratio $ g_{a \rightarrow b}(k_z;\Elas = \Elas_{\max},\beta) \big/g_{a \rightarrow b}(k_z;\Elas = \Elas_{\min},\beta) $. (a) Weakly elastic regime, $\Elas_{\min} =  10^{-8}$, $\Elas_{\max} = 10^{-5}$ and (b) Strongly elastic regime, $\Elas_{\min} = 10^0$, $\Elas_{\max} = 10^4$. 
In both panels: solid lines (\linesolids , {\color{gray}\linesolids}) are $g_{\bld{u} \rightarrow \bld{u}}$,
dashed lines (\linedashed, {\color{gray} \linedashed}) are	$g_{\mathsfbi{T} \rightarrow \bld{u}}$,
dotted lines (\linedotted,  {\color{gray}\linedotted}) are	$g_{\bld{u} \rightarrow \mathsfbi{T}}$, and
dash-dot lines (\linedshdot, {\color{gray} \linedshdot}) are	$g_{\mathsfbi{T} \rightarrow \mathsfbi{T}}$.
Black lines are plane Couette flow and grey lines are plane Poiseuille flow.  
} 
\label{fig:f3_ratio}
\end{figure} 
The forms (\ref{f1f2_components}) and (\ref{f3_components}) are particularly convenient because one can clearly delineate and independently study various input-output pairs.
We can numerically estimate the exact value of $f_{q \rightarrow r}$ or $g_{q \rightarrow r}$ for some $r,q \in \{ \bld{u},\mathsfbi{T} \}$ without resorting to the theoretical results by setting the appropriate coefficients in (\ref{Bdefns}) and (\ref{Hdefns}) equal to zero and then numerically solving the Lyapunov equations (\ref{WLyap}) and (\ref{XLyap}) or (\ref{ZLyap}). 
The component $f_{q \rightarrow r}$ or $g_{q \rightarrow r}$ can then be found using the appropriate equations from (\ref{f1defn})--(\ref{f3defn}).
For example to compute $f_{\mathsfbi{T} \rightarrow \bld{u}}$ in the strongly elastic regime, i.e. the contribution to $f$ from the kinetic energy due to stochastic forcing in the polymer stresses, we set $\mathcal{B}_{v,2}$, $\mathcal{B}_{v,3}$, $\mathcal{B}_{\eta}$ and $\bld{\mathcal{H}}_{\mathsfbi{T}}$ equal to zero while leaving the remaining coefficients in (\ref{Bdefns}) and (\ref{Hdefns}) as defined in the appendix. 
Solving (\ref{WLyap}) and (\ref{XLyap}) and using the solutions in (\ref{f1defn}) and (\ref{f2defn}) gives us the value of $\Elas f_{\mathsfbi{T} \rightarrow \bld{u}}$. 

The numerical results are presented in Figs. \ref{fig:f1f2_weak} through \ref{fig:f3_strong}.  
A weaker dependence of $f_{q \rightarrow r}$ and $g_{q \rightarrow r}$ on $\Elas$ indicates a stronger agreement with the theoretical prediction based on the lowest-order term, i.e., as shown in (\ref{f1f2_components}) and (\ref{f3_components}). We note that the results scale with variance of the forcing, here arbitrarily set to unity for both velocity and polymer stresses.

Previous authors produced some of the results we show in this section. 
\citet{Hoda2009} were solely concerned with steady-state kinetic energy due to stochastic forcing in the velocity field and thus their results correspond to the plots in panel (a) of Figs. \ref{fig:f1f2_weak} through \ref{fig:f3_strong}. 
\citet{Jovanovic2011} on the other hand considered kinetic energy and polymer stress energy due to stochastic forcing in the velocity field in the strongly elastic regime $\Elas \rightarrow \infty$. 
The authors used numerical simulation of the linearized equations to supplement their analytical computations of the output variance.
Their results are approximately equivalent to panels (a) and (c) of Figs. \ref{fig:f1f2_strong} and \ref{fig:f3_strong}.

\subsection{Scaling in the weakly elastic regime}

\begin{figure}
\centering
\includegraphics[scale=1.0]{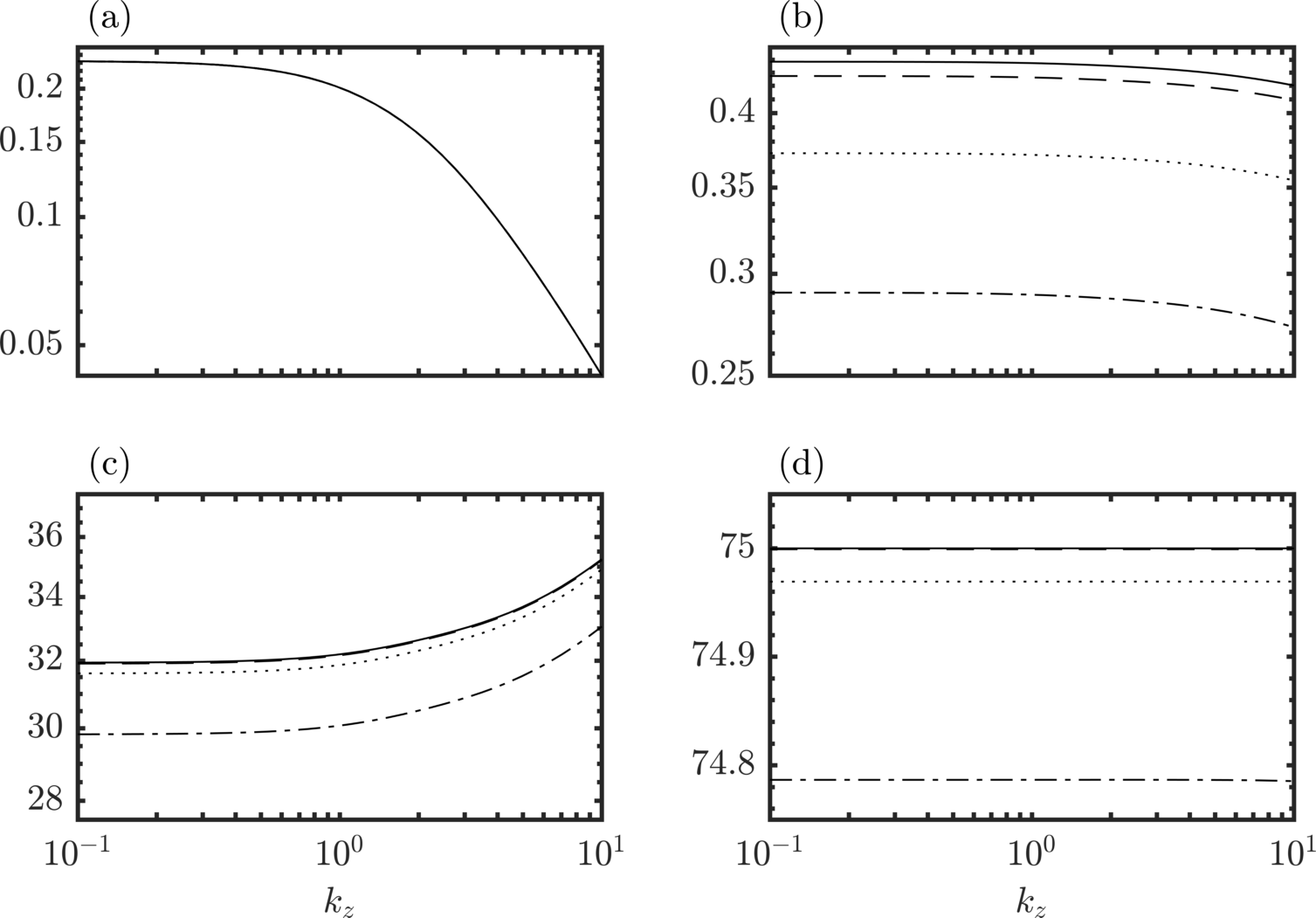}
\caption{Components of base-flow independent part $f_1+f_2$ of $\mathcal{E}$ at $\beta = 0.9$ in the weakly elastic regime ($\Elas \ll 1$) scaled by powers of $\Elas$ according to theoretical prediction. Collapse of lines indicates good agreement. (a) and (b) are the kinetic energy components and (c) and (d) are the polymer stress energy components. In addition, (a) and (c) are the responses when the forcing is only in the velocity field and (b) and (d) are the responses when the forcing is only in the polymer stress field. 
Lines $\{$\linesolids, \linedashed, \linedotted, \linedshdot$\}$ correspond to $\Elas = \{ 10^{-8}, 10^{-7}, 10^{-6}, 10^{-5}\}$.
Results are for both Couette and channel flow since $f_1 + f_2$ does not depend on the base flow.}
\label{fig:f1f2_weak}
\end{figure}
\begin{figure}
\centering
\includegraphics[scale=1.0]{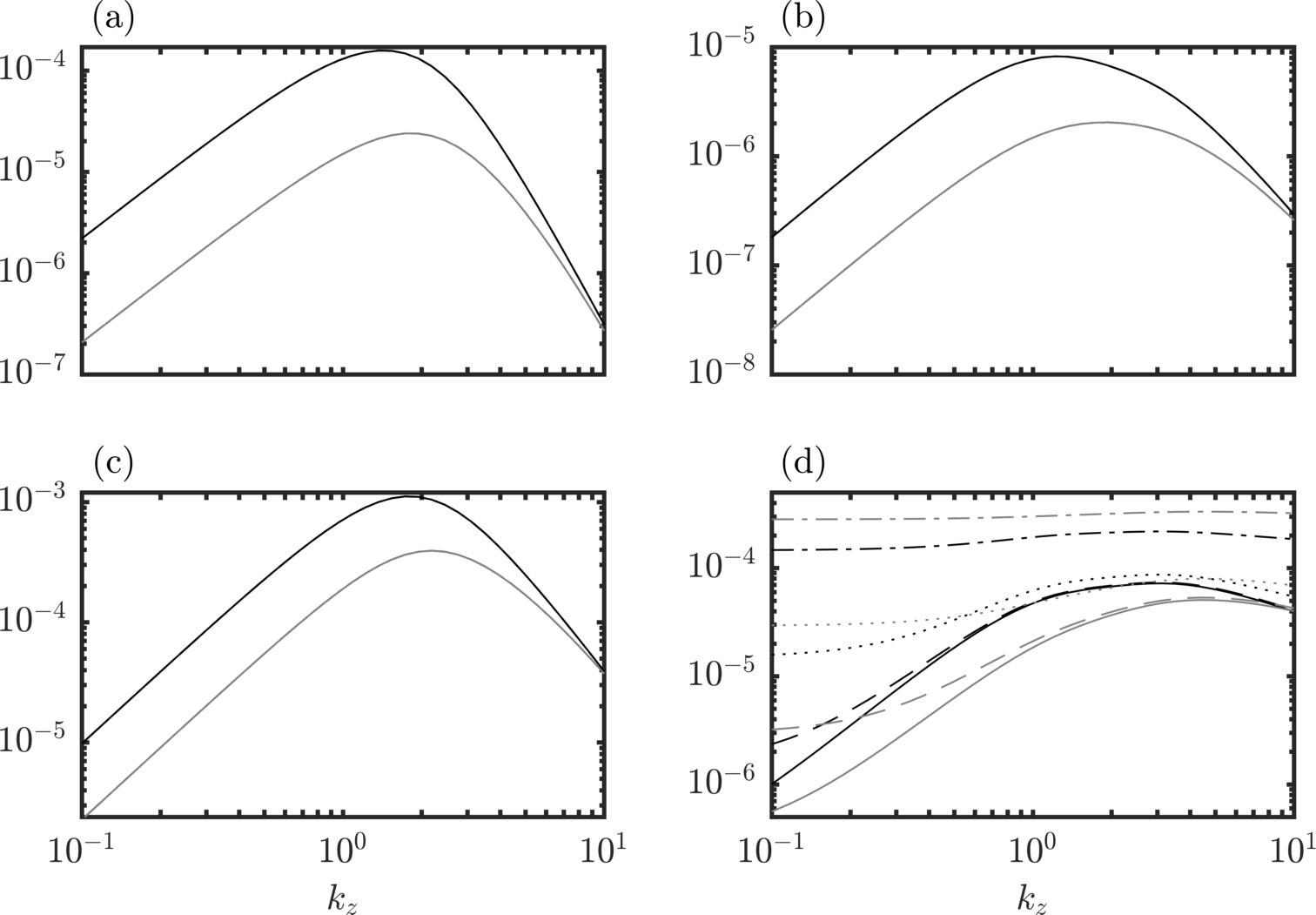}
\caption{Components of base-flow dependent part $g$ of $\mathcal{E}$ at $\beta = 0.9$ in the weakly elastic regime ($\Elas \ll 1$) scaled by powers of $\Elas$ according to theoretical prediction. Collapse of lines indicates good agreement. Refer to caption of Fig. \ref{fig:f1f2_weak}  for descriptions of figures (a)--(d).  Lines $\{$\linesolids, \linedashed, \linedotted, \linedshdot$\}$ correspond to $\Elas = \{ 10^{-8}, 10^{-7}, 10^{-6}, 10^{-5}\}$.
Black lines  are plane Couette flow and grey lines are plane Poiseuille flow. 
}
\label{fig:f3_weak}
\end{figure}

Using (\ref{f1f2_components}) and (\ref{f3_components}) into (\ref{Eexpr}), we have for $\Elas =  \Wie/\Rey \rightarrow 0$ the following expression for $\mathcal{E}$
\begin{multline}
\mathcal{E} = 
(f_{\bld{u} \rightarrow \bld{u}}+
f_{\bld{u} \rightarrow \mathsfbi{T}} )\Rey +
(g_{\bld{u} \rightarrow \bld{u}} +
g_{\bld{u} \rightarrow \mathsfbi{T}} )\Rey^3 \\
+  f_{\mathsfbi{T} \rightarrow \mathsfbi{T}} \Wie  +
 f_{\mathsfbi{T} \rightarrow \bld{u}} \frac{\Wie^2}{\Rey}  +
 \left(g_{\mathsfbi{T} \rightarrow \bld{u}}+
  g_{\mathsfbi{T} \rightarrow \mathsfbi{T}} \right)\Rey\Wie^2 
  \label{E_weak_final}
\end{multline}
where we recall $\Wie = \lambda_{\text{p}} U/L$ is the Weissenberg number. 
The coefficients $f_{a \rightarrow b}$ and $g_{a \rightarrow b}$ ($a,b\in\{\bld{u},\mathsfbi{T}\}$) are shown in Figs. \ref{fig:f1f2_weak} and \ref{fig:f3_weak} for $k_z \in [10^{-1}, 10^1]$ and for $\Elas \in \{10^{-5},10^{-6},10^{-7},10^{-8}\}$. The range of magnitudes taken on by  $f_{a \rightarrow b}$ and $g_{a \rightarrow b}$ for this $k_z$ and $\Elas$ range are listed in Table \ref{tab:coeff_mags}.
A key point to note is that even though $f_{a\rightarrow b}$ and $g_{a\rightarrow b}$ are both correct up to $\mathcal{O}(\Elas)$, the final scaling has a truncation error whose dominant term scales as $\mathcal{O}(\Wie,\Rey^2\Wie)$.
The latter arises from the base-flow dependent part of the scaling.
Thus, when considering amplification that depends on the base-flow, it is not sufficient that $\Elas \rightarrow 0$ but also that $\Wie \rightarrow 0$.

Since $\Wie/\Rey \rightarrow 0$, the $\mathcal{O}(\Rey^3)$ part of the scaling will typically dominate $\mathcal{E}$. 
The first two terms, $\mathcal{O}(\Rey)$  and $\mathcal{O}(\Rey^3)$  respectively, are analogous to the well-known scalings reported in \citep{Bamieh2001} and \citep{Hoda2009} with the difference that they now also include the contribution from the polymer stress energy due to forcing in the velocity field.
The remaining part of the scaling of $\mathcal{E}$ is an elastic correction to the $\mathcal{O}(\Rey^3)$ component and arises solely due to disturbances in the polymer stress. 

\begin{table}
\centering
\def\arraystretch{1.25}
\setlength\tabcolsep{4pt}
\begin{tabular}{c| c  r r r }
   Regime       & $a \rightarrow b$                    
& \parbox{2.5cm}{\centering $ f_{a \rightarrow b} $} 
&  \parbox{2.5cm}{$ g_{a \rightarrow b} $  (Couette)   }
&  \parbox{2.5cm}{$ g_{a \rightarrow b} $  (Poiseuille) }       \\  
\hline \\
Weak    Elasticity & $\bld{u}\rightarrow \bld{u}$           &  $[4^\dagger, 23]\times 10^{-2}$      & $ [2.4^{\dagger},160]\times10^{-6}$    & $ [2.2^{\dagger} ,240]\times10^{-7}$\\  
$(\Elas \in [10^{-8},10^{-5}])$ 		  & $\mathsfbi{T}\rightarrow \bld{u}$      &  $[2.7^\dagger,4.4]\times 10^{-1}$       & $[2.0 ^{ \dagger},82]\times10^{-7}$   & $ [2.8 ^{ \dagger},200]\times 10^{-8}$   \\  
			  & $\bld{u}\rightarrow \mathsfbi{T}$      &  $[3 ,3.5^\dagger]\times 10 \,\,\,\,\,\,\,$       & $[1.0 ^{ \dagger},1100]\times10^{-5}$  & $[ 2.5 ^{ \dagger} ,400]\times 10^{-6}$\\   
 			 & $\mathsfbi{T}\rightarrow \mathsfbi{T}$  &  $\sim 75 $                       & $[1.0 ^{ \dagger},220] \times10^{-6}$ & $ [5.8^{\dagger} ,3300]\times 10^{-7}$    \\  
\hline \\
Strong  Elasticity& $\bld{u}\rightarrow \bld{u}$           &  $[ 5, 25^\dagger] \times 10^{-2}$            & $ [4.7 ^{ \dagger}, 21000]\times10^{-8}$    & $ [1.7  ^{ \dagger},3500]\times10^{-8}$  \\  
$(\Elas \in [1,10^{4}])$ 				  & $\mathsfbi{T}\rightarrow \bld{u}$      &  $[4 ^\dagger,20 ]\times 10^{-3}$       & $[1 ^{ \dagger},6.7]\times10^{-4}$    & $ [1.3 ^{ \dagger},10]\times 10^{-4}$ \\  
  			  & $\bld{u}\rightarrow \mathsfbi{T}$      &  $[6^{\dagger}, 39]\times 10^{-2} $         & $[4^{\dagger},23] \times 10^{-2}$    & $ [6^{\dagger},48]\times 10^{-2}$          \\  
 			 & $\mathsfbi{T}\rightarrow \mathsfbi{T}$  &  $\sim 7.2 \times 10\quad$                       & $\sim 1.2 \times 10 \quad$          & $ \sim 2.3 \times 10 \quad$         \\
\hline
\end{tabular} 
\caption{Representative bounds on magnitudes of coefficients $f_{a \rightarrow b}$ and $g_{a \rightarrow b}$  for spanwise wavenumbers $k_z \in [10^{-1},10^1]$ in the asymptotic regimes of $\Elas$.
 $^\dagger$Results indicate that this bound changes monotonically if we consider $k_z \in [ 10^{-1} - \Delta k_z, 10^1 + \Delta k_z]$ for $\Delta k_z > 0$.
 }
\label{tab:coeff_mags}
\end{table}

Figures \ref{fig:f1f2_weak} and \ref{fig:f3_weak} show good agreement of the theoretical predictions of (\ref{E_weak_final}) with numerically computed quantities in the range of $\Elas$ considered ($10^{-8}$ to $10^{-5}$).
In case of perfect agreement the curves for various $\Elas$ should be indistinguishable, i.e., the ratio of any two curves should be 1.
Figures \ref{fig:f1f2_ratio} and \ref{fig:f3_ratio} show the ratio of the curves for lowest and highest $\Elas$ considered in each regime.
Both Fig. \ref{fig:f3_weak} and \ref{fig:f3_ratio} show that the base-flow dependent contribution to the polymer stress energy due to disturbances in the polymer stresses, $g_{\mathsfbi{T} \rightarrow \mathsfbi{T}}$ shows the largest discrepancy and especially at low $k_z$. 
This discrepancy can be resolved by considering higher-order terms in the series expression for $g_{\mathsfbi{T} \rightarrow \mathsfbi{T}}$
\begin{multline}
g_{\mathsfbi{T} \rightarrow \mathsfbi{T}} =
\text{tr}_{\mathcal{V}}\left(\bld{\mathscr{H}}_{\mathsfbi{T},\bld{\varphi}_2}\bld{\mathscr{Z}}_{\mathsfbi{T}\rightarrow \mathsfbi{T}}^{(0)}\bld{\mathscr{D}}_{\mathsfbi{T},\bld{\varphi}_1}  \right)+ 
\frac{\Rey}{\Wie}\text{tr}_{\mathcal{V}}\left(\bld{\mathscr{H}}_{\mathsfbi{T},\bld{\varphi}_2}\bld{\mathscr{Z}}_{\mathsfbi{T}\rightarrow \mathsfbi{T}}^{(1)} \bld{\mathscr{D}}_{\mathsfbi{T},\bld{\varphi}_1}  \right) \\
+ \left(\frac{\Rey}{\Wie}\right)^2\text{tr}_{\mathcal{V}}\left(\bld{\mathscr{H}}_{\mathsfbi{T},\bld{\varphi}_2}\bld{\mathscr{Z}}_{\mathsfbi{T}\rightarrow \mathsfbi{T}}^{(2)} \bld{\mathscr{D}}_{\mathsfbi{T},\bld{\varphi}_1}  \right) 
 + \hdots \label{rho_tautau_long}
\end{multline}
where $ \bld{\mathscr{Z}}_{\bld{u}\rightarrow \bld{u}}^{(n)} $ for $n = 0,1,2,\hdots$ are $\Elas$-independent elements of the series expression of $\bld{\mathscr{Z}}_{\bld{u}\rightarrow \bld{u}} $ .
The coefficients of the higher-order terms in the expression (\ref{E_weak_final}) for $\mathcal{E}$ are $\Wie^{n+2}\Rey^{1-n}$.
We note that $g_{\mathsfbi{T} \rightarrow \mathsfbi{T}}$ appears in the elastic correction. The difficulty of obtaining a good agreement for $g_{\mathsfbi{T} \rightarrow \mathsfbi{T}}$ may be because of the timescale separation implicit in the $\Elas \rightarrow 0$ assumption. The fast dynamics are parameterized by $\Wie$ and thus should only be weakly dependent on $\Rey$. Since $g_{\mathsfbi{T} \rightarrow \mathsfbi{T}}$ is determined by the fast dynamics, we expect its expansion in a series in $\Elas = \Wie/\Rey$ to be slow to converge.

\subsection{Scaling in the strongly elastic regime}
\begin{figure} 
\centering
\includegraphics[scale=1.0]{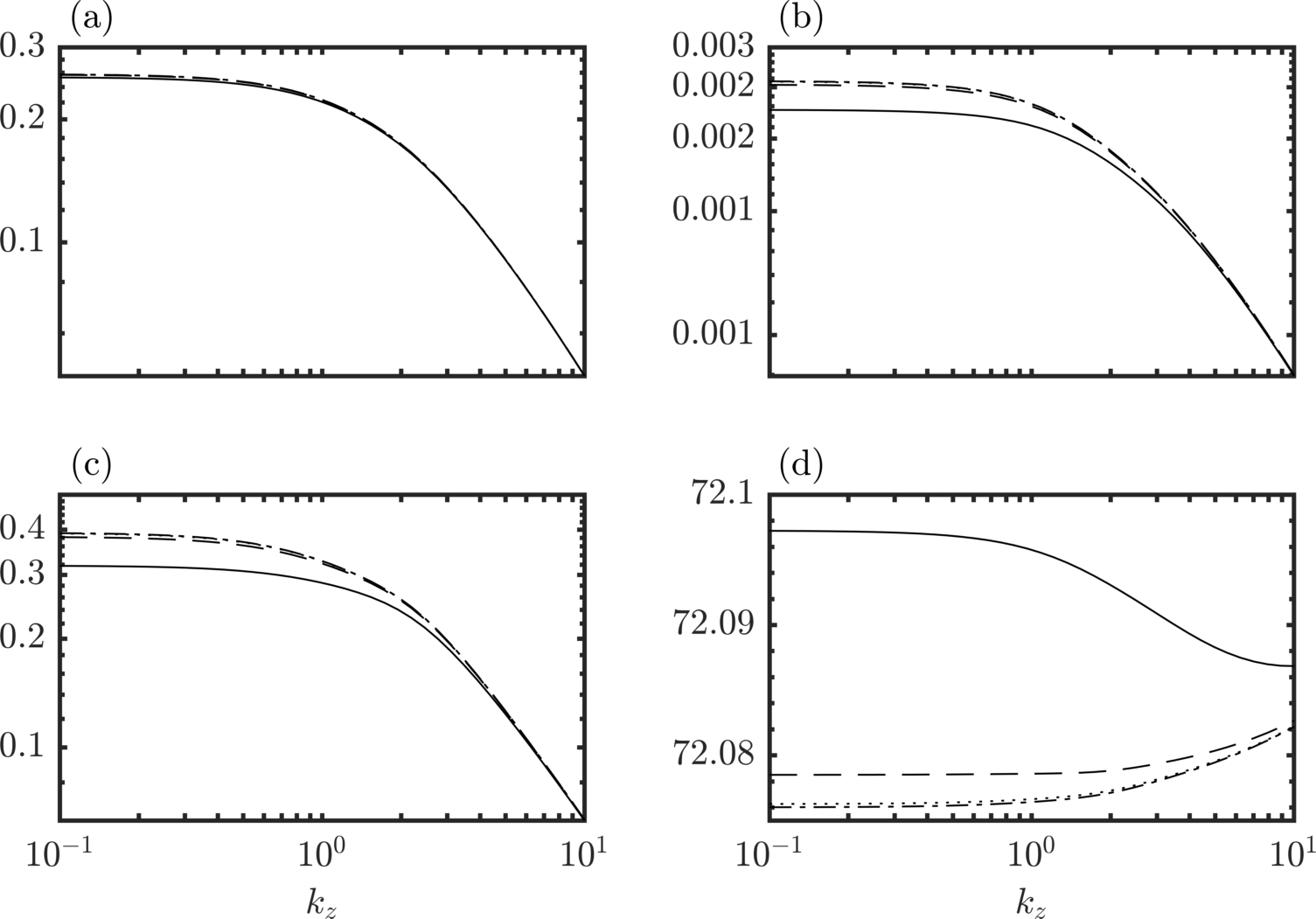}
\caption{Components of base-flow independent part $f_1+f_2$  of $\mathcal{E}$ at $\beta = 0.9$ in the strongly elastic regime ($\Elas \gg 1$) scaled by powers of 	$\Elas$ according to theoretical prediction. Collapse of lines indicates good agreement.  Refer to caption of Fig. \ref{fig:f1f2_weak} for descriptions of figures (a)--(d).  
Lines $\{$\linesolids, \linedashed, \linedotted, \linedshdot$\}$ correspond to $\Elas = \{ 1, 10, 100, 10^{4}\}$.
Results are for both Couette and channel flow since $f_1 + f_2$ does not depend on the base flow.
}
\label{fig:f1f2_strong}
\end{figure}
\begin{figure}
\centering
\includegraphics[scale=1.0]{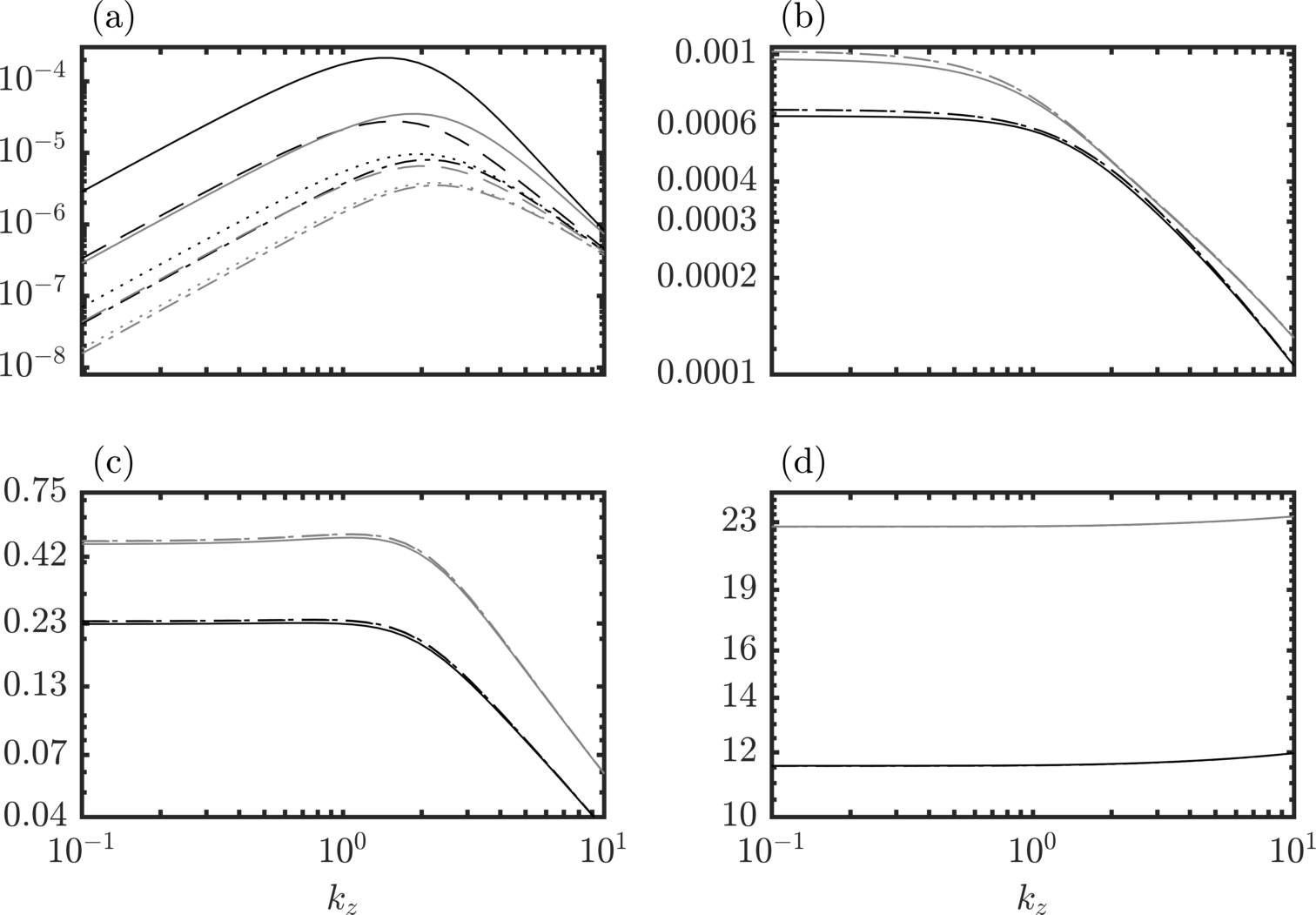}
\caption{Components of base-flow dependent part $g$ of $\mathcal{E}$ at $\beta = 0.9$ in the strongly elastic regime ($\Elas \gg 1$) scaled by powers of $\Elas$ according to theoretical prediction. Collapse of lines indicates good agreement. Refer to caption of Fig. \ref{fig:f1f2_weak}  for descriptions of figures (a)--(d). 
Lines $\{$\linesolids, \linedashed, \linedotted, \linedshdot$\}$ correspond to $\Elas = \{ 1, 10, 100, 10^{4}\}$.
Black lines (\linesolids) are plane Couette flow and grey lines ({\color{gray}\linesolids}) are plane Poiseuille flow. 
}
\label{fig:f3_strong}
\end{figure}
Substituting (\ref{f1f2_components})--(\ref{f3_components}) into (\ref{Eexpr}) we obtain the following expression for $\mathcal{E}$ when $\Elas \rightarrow \infty$
 \begin{multline}
\mathcal{E} =  
(f_{\mathsfbi{T} \rightarrow \bld{u}}  +
f_{\mathsfbi{T} \rightarrow \mathsfbi{T}})\Wie   +
(g_{\mathsfbi{T} \rightarrow \bld{u}}+
g_{\mathsfbi{T} \rightarrow \mathsfbi{T}})\Wie^3\\
 + f_{\bld{u} \rightarrow \bld{u}}\Rey +
f_{\bld{u} \rightarrow \mathsfbi{T}}   \frac{\Rey^2}{\Wie}+
\left(g_{\bld{u} \rightarrow \bld{u}} +
g_{\bld{u} \rightarrow \mathsfbi{T}} \right)\Wie\Rey^2 
\label{E_strong_final}
\end{multline}
where the coefficients $f_{a \rightarrow b}$ and $g_{a \rightarrow b}$ ($a,b\in\{\bld{u},\mathsfbi{T}\}$) are shown in Figs. \ref{fig:f1f2_strong} and \ref{fig:f3_strong} for $k_z \in [10^{-1}, 10^1]$ and for $\Elas \in \{10^{0},10^{1},10^{2},10^{4}\}$. The range of magnitudes taken on by  $f_{a \rightarrow b}$ and $g_{a \rightarrow b}$ for this $k_z$ and $\Elas$ range are listed in Table \ref{tab:coeff_mags}. Unlike in the case of weak elasticity, the primary condition ($\Elas \rightarrow \infty$) is sufficient to control the truncation error in (\ref{E_strong_final}), $\mathcal{O}\left(\Rey^2/\Wie,\Rey^4/\Wie\right)$.

The expression (\ref{E_strong_final}) is analogous to the expression (\ref{E_weak_final}) retrieved for when $\Elas \rightarrow 0$. 
Instead of being dominated by $\mathcal{O}(\Rey^3)$ terms, $\mathcal{E}$ is now dominated by $\mathcal{O}(\Wie^3)$ terms that arise solely due to disturbances in the polymer stresses. 
With vanishing forcing in the velocity field, $\bld{d}_{u} \rightarrow 0$,  we have $\mathcal{E} =  
(f_{\mathsfbi{T} \rightarrow \bld{u}}  +
f_{\mathsfbi{T} \rightarrow \mathsfbi{T}})\Wie   +
(g_{\mathsfbi{T} \rightarrow \bld{u}}+
g_{\mathsfbi{T} \rightarrow \mathsfbi{T}})\Wie^3$. 
This parallels the scaling $\mathcal{E}= (f_{\bld{u} \rightarrow \bld{u}}+
f_{\bld{u} \rightarrow \mathsfbi{T}} )\Rey +
(g_{\bld{u} \rightarrow \bld{u}} +
g_{\bld{u} \rightarrow \mathsfbi{T}} )\Rey^3$  when the disturbances in the polymer stresses $\bld{d}_{\mathsfbi{T}}$ vanish and $\Elas \rightarrow 0$.
In both cases the dominant, cubic scaling is due to coupling via the base flow.
Since $\Wie/\Rey \rightarrow \infty$, the remaining terms in (\ref{E_strong_final}) which are now $\mathcal{O}(\Rey)$ act as a viscous correction to the primary $\mathcal{O}(\Wie^3)$ scaling. 
The viscous correction of (\ref{E_strong_final}) parallels the form of the elastic correction of (\ref{E_weak_final}).

The part of the scaling that arises due to forcing in the velocity field, $f_{\bld{u} \rightarrow \bld{u}}\Rey +
f_{\bld{u} \rightarrow \mathsfbi{T}}    \Rey^2 \Wie^{-1}+
\left(g_{\bld{u} \rightarrow \bld{u}} +
g_{\bld{u} \rightarrow \mathsfbi{T}} \right)\Wie\Rey^2 $, is consistent with the singular perturbation based derivation of \citet{Jovanovic2011}.
\citet{Jovanovic2011} also reported on the existence of a `viscoelastic lift-up' effect when $\Elas \rightarrow \infty$  without considering disturbances in the polymer stresses, $\bld{d}_{\mathsfbi{T}}$. 
The direct correspondence between our results, (\ref{E_weak_final}) and (\ref{E_strong_final}), and the high $\mathcal{O}(\Wie^3)$ amplification possible shown even for creeping flows is suggestive of such a mechanism.
However, in the present case, the dominant amplification arises from disturbances in the polymer stresses.

In most flows of interest we have $\Wie \sim \mathcal{O}(10^1)$, implying a creeping flow regime, i.e., $\Rey \sim \mathcal{O}(\varepsilon)$ with $\varepsilon \ll 10$ to justify the assumption $\Elas \rightarrow \infty$. 
Larger $\Wie$ due to high shear rates may be achieved locally in some flows, thereby relaxing the assumptions needed on $\Rey$. 
Our results, in the context of a local $\Wie$, provide insight into the underlying mechanisms in more complicated flow. 
For example, it has been suggested that linear mechanisms may play an important part in organizing fully turbulent flows \citep{Jimenez2013,Page2015}. We do not pursue this line of inquiry here. 

In practice we find that $\Elas \sim \mathcal{O}(1)$ is usually sufficient to provide strong agreement between the theoretical scaling (\ref{E_strong_final}) and the numerical results. This can be verified in Figs. \ref{fig:f1f2_strong} and \ref{fig:f3_strong} and further in the ratio plots in Figs. \ref{fig:f1f2_ratio} and \ref{fig:f3_ratio}. 
The exception is the base-flow dependent contribution to the kinetic energy due to  forcing in the velocity field, $g_{\bld{u} \rightarrow \bld{u}}$, that appears in the viscous correction. 
As with $g_{\mathsfbi{T} \rightarrow \mathsfbi{T}}$ when $\Elas \rightarrow 0$, the term $g_{\bld{u} \rightarrow \bld{u}}$ when $\Elas \rightarrow \infty$ shows a non-negligible dependence on $\Elas$ at low $k_z$, implying that higher-order terms are significant at these wavenumbers
\begin{align}
g_{\bld{u} \rightarrow \bld{u}} =
 \text{tr}_{\mathcal{V}}\left(\bld{\mathscr{H}}_{\bld{u},\bld{\varphi}_2} \bld{\mathscr{Z}}_{\bld{u}\rightarrow \bld{u}}^{(0)} \bld{\mathscr{D}}_{\bld{u},\bld{\varphi}_1}  \right)+ 
\frac{\Rey}{\Wie} \text{tr}_{\mathcal{V}}\left(\bld{\mathscr{H}}_{\bld{u},\bld{\varphi}_2} \bld{\mathscr{Z}}_{\bld{u}\rightarrow \bld{u}}^{(1)} \bld{\mathscr{D}}_{\bld{u},\bld{\varphi}_1}  \right) 
 + \hdots \label{rho_uu_long}
\end{align}
where $ \bld{\mathscr{Z}}_{\bld{u}\rightarrow \bld{u}}^{(n)} $ for $n = 0,1,2,\hdots$ are $\Elas$-independent elements of the series expression of $\bld{\mathscr{Z}}_{\bld{u}\rightarrow \bld{u}} $ .
The coefficients of the higher-order terms in the expression (\ref{E_strong_final}) for $\mathcal{E}$ are thus $\Rey^{n+2}\Wie^{1-n}$.
As suggested for $g_{\mathsfbi{T} \rightarrow \mathsfbi{T}}$, the difficulty of obtaining a good agreement for $g_{\bld{u} \rightarrow \bld{u}}$ may be because of the timescale separation implicit in the $\Elas \rightarrow \infty$ assumption. The fast dynamics are now parameterized by $\Rey$ and since $g_{\bld{u} \rightarrow \bld{u}}$ is determined by the fast dynamics, it should only be weakly dependent on $\Wie$. Hence the slow convergence of a series expansion in $\Elas^{-1} =  \Rey/\Wie$.

\subsection{Contrasts between the weakly and strongly elastic regimes}
In addition to the $\Rey$ to $\Wie$ `switch' seen going from the weakly elastic form of $\mathcal{E}$ in (\ref{E_weak_final}) to the strongly elastic form in (\ref{E_strong_final}), we see several other interesting contrasts between these two regimes that can be identified due to the scaling. We highlight changes in (a) the high-wavenumber damping of disturbances (b) the spanwise selectivity of the flow  and finally (c) the difference between Poiseuille and Couette flow. 

\subsubsection{High wavenumber damping}
In viscous Newtonian flows, diffusion always leads to a damping of high-wavenumber (short wavelength) disturbances. 
This behavior may be modified in viscoelastic flow since the governing equations of the polymer stresses are hyperbolic.

The base-flow dependent part of $\mathcal{E}$, shown component-wise in Figs. \ref{fig:f3_weak} and \ref{fig:f3_strong} shows high-wavenumber damping in both regimes ($\Elas \rightarrow 0$ and $\Elas \rightarrow \infty$). The exception is $g_{\mathsfbi{T} \rightarrow \mathsfbi{T}}$, the polymer stress energy due to disturbances in the polymer stresses. As shown in Figs. \ref{fig:f3_weak}(d) and \ref{fig:f3_strong}(d), this component switches from high-wavenumber damping in the weakly elastic regime to no discernable damping in the strongly elastic regime. As discussed below, most of the components $f_{a \rightarrow b}$ of the base-flow independent part  of $\mathcal{E}$, however, show the opposite behavior (switching to high wavenumber damping when switching to the strongly elastic regime).

Figures \ref{fig:f1f2_weak}(a) and (c) show  $f_{\bld{u} \rightarrow \bld{u}}$ and  $f_{\bld{u} \rightarrow \mathsfbi{T}}$, the base-flow independent kinetic and polymer stress energy due to  the forcing $\bld{d}_{\bld{u}}$, in the weakly elastic regime.
The polymer stress energy $f_{\bld{u} \rightarrow \mathsfbi{T}}$, unlike the kinetic energy $f_{\bld{u} \rightarrow \bld{u}}$ which monotonically decays with $k_z$, shows a monotonic increase with $k_z$ indicating the importance of small wavelength disturbances in viscoelastic flow.
We expect the kinetic energy to drop off with $k_z$ due to momentum diffusion but no such significant diffusion mechanism exists in viscoelastic flows and therefore we expect large energy amplification at small wavelengths.
This behavior seems to be modified whenever the polymer relaxation time is at least as slow as the diffusion time ($\Elas \sim 1$) and can be seen in the results for the strongly elastic regime as shown in Figs. \ref{fig:f1f2_strong}(c).
The polymer stress energy due to forcing in the velocity field, $f_{\bld{u} \rightarrow \mathsfbi{T}}$, now also begins to show high-wavenumber damping. It can be seen in Fig. \ref{fig:f1f2_strong}(a) that the kinetic energy due to forcing in the velocity field, $f_{\bld{u} \rightarrow \bld{u}}$, shows surprisingly little change between the two regimes.

The lack of high-wavenumber damping in the weakly elastic regime is also evident in $f_{\mathsfbi{T} \rightarrow \bld{u}}$ and $f_{\mathsfbi{T} \rightarrow \mathsfbi{T}}$, the kinetic and polymer stress energy due to disturbances in the polymer stresses. This is shown in Figs. \ref{fig:f1f2_weak}(b) and \ref{fig:f1f2_weak}(d). 
The kinetic energy $f_{\mathsfbi{T} \rightarrow \bld{u}}$ due to the polymer stress forcing, shown scaled in Fig. \ref{fig:f1f2_weak}(b), shows only very modest decrease with increasing $k_z$.
This seems to imply that the diffusion operator is more weakly damping for disturbances that originate in the polymer stresses and propogate through to the velocity field than those that originate as forcing in the velocity field. This behavior is modified in the strongly elastic regime. As shown in Fig. \ref{fig:f1f2_strong}(b), the slower polymer relaxation time leads to high-wavenumber damping in $f_{\mathsfbi{T} \rightarrow \bld{u}}$.

The switch in the high-wavenumber damping behavior of the cross terms $f_{\mathsfbi{T} \rightarrow \bld{u}}$ and $f_{\bld{u} \rightarrow \mathsfbi{T}}$ that we have already mentioned and that can be seen in Figs. \ref{fig:f1f2_weak}(b)--(c) and \ref{fig:f1f2_strong}(b)--(c) can be attributed to the interplay between polymer relaxation and damping due to viscous diffusion. 
A higher elasticity number $\Elas$ indicates a larger polymer relaxation time compared to viscous diffusion.
Therefore the amplitudes of the polymer stresses take longer to adjust to the high $k_z$ input forcing than the time-scale on which we observe damping in the velocity field due to viscous diffusion.
The forcing in the polymer stresses can only propagate to the velocity field via these slowly evolving polymer stresses and therefore we see high wavenumber damping in $f_{\mathsfbi{T} \rightarrow \bld{u}}$ for the strongly elastic case as shown in Fig. \ref{fig:f1f2_strong}(b). 
Conversely, if the high $k_z$ forcing entering the fluid equations is damped by viscous diffusion faster than the polymer relaxation time, the forcing becomes less effective in achieving high amplitude growth in the polymer stresses. 
This is the high $k_z$ damping in $f_{\bld{u} \rightarrow \mathsfbi{T}}$ shown in Fig. \ref{fig:f1f2_strong}(c).
This explanation is tenable due to the lack of convective effects in the base-flow independent components $f_{q \rightarrow r}$ for $r,q \in \{\bld{u},\mathsfbi{T} \}$, i.e., polymer relaxation and viscous diffusion are the dominant mechanisms. The behavior of the base-flow dependent $g_{\mathsfbi{T} \rightarrow \mathsfbi{T}}$, however, involves the interplay of viscous diffusion, convective effects and polymer relaxation. 
The switching behavior in $g_{\mathsfbi{T} \rightarrow \mathsfbi{T}}$ is opposite to what we see in the base-flow independent components, i.e., $g_{\mathsfbi{T} \rightarrow \mathsfbi{T}}$ is damped at high-wavenumbers in the weakly elastic regime and broadband in the strongly elastic regime.
This seems to imply that convective effects play a role in relation to polymer relaxation that is contrary to the role played by viscous diffusion in relation to the same.

The high-wavenumber damping behavior of $f_{\mathsfbi{T} \rightarrow \mathsfbi{T}}$, the base-flow independent  polymer stress energy due to disturbances in the polymer stresses shows a more complicated switching behavior. We see from Fig. \ref{fig:f1f2_weak}(d) that as $\Elas \rightarrow 0$ the polymer stress energy $f_{\mathsfbi{T} \rightarrow \mathsfbi{T}}$ is broadband in $k_z$ with no high-wavenumber damping, reflecting the lack of diffusion in the polymer stress equations. 
As we increase $\Elas$, the polymer stress energy $f_{\mathsfbi{T} \rightarrow \mathsfbi{T}}$ now shows high-wavenumber damping.
This can be seen in the $\Elas = 1$ case shown in Fig. \ref{fig:f1f2_strong}(d).
As we further increase $\Elas$, we completely lose high-wavenumber damping and instead begin to see high-wavenumber amplification. The cases $\Elas = \{10,100,10^4\}$ in Fig. \ref{fig:f1f2_strong}(d) illustrate this behavior.

\subsubsection{Spanwise selectivity}
It is well-known that the spanwise selectivity of the flow arises due to the base-flow dependent part of the scaling. This can be seen by comparing Figs. \ref{fig:f1f2_weak} and \ref{fig:f1f2_strong} with Figs. \ref{fig:f3_weak} and \ref{fig:f3_strong}; only the base-flow dependent part of the scaling shows distinct stationary points (maxima) with respect to $k_z$. 

From Fig. \ref{fig:f3_weak} we see that all components $g_{q \rightarrow r}$ for $r,q \in \{\bld{u}, \mathsfbi{T} \}$ show spanwise selectivity with optimal $k_z \sim \mathcal{O}(1)$ in the weakly elastic regime ($\Elas \rightarrow 0$). This behavior changes drastically in the strongly elastic regime. This can be observed in Fig. \ref{fig:f3_strong} where all the components $g_{q \rightarrow r}$ lose their spanwise selectivity when $\Elas \rightarrow \infty$ except the forcing in the velocity field to kinetic energy component,  $g_{\bld{u} \rightarrow \bld{u}}$.  The latter always shows an $\mathcal{O}(1)$ optimal $k_z$ but, as seen in Fig. \ref{fig:f3_strong}(a),  the optimal $k_z$ shows a dependence on $\Elas$ when $\Elas \rightarrow \infty$ -- increasing $\Elas$ increases optimal $k_z$.

\citet{Jovanovic2011} produced Figs. \ref{fig:f3_strong}(a) and (b) and also observed the lack of selectivity in the cross component $g_{\bld{u} \rightarrow \mathsfbi{T}}$ (forcing in velocity field to polymer stress energy) when $\Elas \rightarrow \infty$ and attributed the lack of selectivity to the hyperbolic/non-diffusive nature of the polymer stress equations.
However, as noted earlier, we see selectivity in the component $g_{\bld{u} \rightarrow \mathsfbi{T}}$ in the weakly elastic regime as shown in Fig. \ref{fig:f3_weak}(c) inspite of the hyperbolic nature of the polymer stress equations.

The switch from selective to non-selective behavior occurs at some $\Elas \in [10^{-5}, 1]$ and is ostensibly due to a stronger influence of the polymer dynamics.

\subsubsection{Base-flow dependence}
The difference due the flow configurations can be seen in the results for (base-flow dependent) $g$ in Figs. \ref{fig:f3_weak} and \ref{fig:f3_strong}. 
In the weakly elastic regime (Fig. \ref{fig:f3_weak}) Couette flow typically shows a higher amplification than Poiseuille flow. 
This trend is reversed in the strongly elastic regime as seen in Fig. \ref{fig:f3_strong} and which can also be seen in Fig. \ref{fig:f3_weak}(d) for $\Elas = 10^{-5}$.  
These results indicate that the interplay of $\ddtwo{\overline{u}}{y} \neq 0$ and the elasticity in the problem leads to a lower amplification at lower $\Elas$ and a higher amplification at higher $\Elas$.
The exception to this trend is $g_{\bld{u} \rightarrow \bld{u}}$, the kinetic energy due to the forcing in the velocity field, which shows no reversal going from $\Elas \rightarrow 0$ to $\Elas \rightarrow \infty$, implying that the change in behavior may arise due to a direct effect of $\ddtwo{\overline{u}}{y}$ on the polymer stresses.
We note that in the expressions (\ref{swconstsys})--(\ref{OpsOpsdefn}), the only term that contains $\ddtwo{\overline{u}}{y}$ is in the governing equation for $\hat{T}_{xy}$ and is given by $-\ddtwo{\overline{u}}{y}v$. 
In the Newtonian limit there is no term in the streamwise constant equation that contains $\ddtwo{\overline{u}}{y}$. In either case, however, the change in base-flow has a wider effect due to the change in shear rate $\ddo{\overline{u}}{y}$.

Finally, Figs. \ref{fig:f3_weak} and \ref{fig:f3_strong} also show that whenever an optimal $k_z$ exists, it occurs at a lower $k_z$ for Couette flow when compared to Poiseuille flow but all are $\mathcal{O}(1)$.

\section{Conclusions and Future Work} \label{sec:conclusions}
The problem of determining the parametric behavior of non-Newtonian fluids is important and difficult at the same time. The large multi-dimensional parameter space and associated timescales leads to a complicated amalgam of regimes. An understanding of the dynamics in these regimes and the corresponding dependence on the parameters of the problem is a fundamental open problem in the field. 

We have approached this problem by considering an Oldroyd-B fluid operating in a regime where the nonlinearities are negligible. 
We derive the scaling of $\mathcal{E}$ with shear rate, $\Rey$ and $\Wie$.
In doing so we found it convenient to partition the parameter space such that we obtain the two regimes of weak and strong elasticity, $\Elas \rightarrow 0$ and $\Elas \rightarrow \infty$. 
The scaling of amplification $\mathcal{E}$ with $\Wie$ and $\Rey$ shows a somewhat surprising symmetry between the two regimes considered.  
Numerical results point to the validity of our analytical derivations and also bring to the fore several interesting behaviors, such as the switching of high-wavenumber damping between the two regimes. These interesting behaviors may provide a starting point for future work.

One of the novelties of the current work is to apply stochastic forcing in the polymer equations. This is the key ingredient that reveals interesting aspects of the problem not previously seen, such as the symmetry between low $\Elas$ and high $\Elas$ regimes. 
Modeling of non-Newtonian flows and deriving the associated constitutive relations is a large and intensive area of research. The rich tapestry of constitutive relations available allows one to replicate most experimentally observed macroscale phenomena but do not model several phenomena that are not necessarily insignificant from a dynamical point of view, for e.g., molecular aggregation, stress diffusion and/or polymer degradation. 
In addition, the physical setup may lead to not insignificant sources of noise in polymer stress equations, for e.g., via thermal fluctuations.
An unaddressed challenge in the field is how to account for these unmodeled effects in the constitutive relations. 
Our modest contribution in this work has been to parameterize these effects as additive stochastic white noise. The drawbacks of such a parameterization are evident; it ignores all multiplicative uncertainties and does not incorporate uncertainty in the parameters. 
However, even with such a simplistic assumption, we find that uncertainty can have drastic effects on polymer stresses and velocity fields.  
These effects may provide an explanation for the finding by \citet{Warholic1999} that the level of turbulent drag reduction due to polymers was sensitive to how the polymers were introduced into the flow.  

The significant energy amplification due to polymer stress disturbances that we show in the present work also opens up new possibilities for flow control, specifically by systematically perturbing the dissolved polymers. Such perturbations could be achieved by appropriate temperatures variations (with the use of temperature-sensitive polymers) or, as  \citet{Warholic1999} noticed in their drag reduction experiments, by modifying the injection flow rates.

A gap in the present work is the consideration of disturbance amplification in the moderately elastic regime. Many phenomena of interest do not fall in the $\Elas \rightarrow 0$ or the $\Elas \rightarrow \infty$ regime and it is of interest to understand the behavior in this regime. 
Unfortunately, not much analytical work has been done in this area owing to the difficulty of not having a small parameter to reduce the problem to one of a small deviation from a simpler problem. 
Finally we must mention that we did not consider $\hat{T}_{xx}$ in the current work and made the streamwise-constant assumption. The streamwise normal stress $\hat{T}_{xx}$ can be amplified greatly due to noise in the velocity or remaining polymer stresses as was demonstrated in \citep{Jovanovic2011}. However, $\hat{T}_{xx}$ does not affect the remaining system states due to a decoupling brought about due to the streamwise-constant assumption. The latter removes coupling with the mean streamwise normal stress, proportional to $\left(\ddo{\overline{u}}{y}\right)^2$. This mean streamwise normal stress may prove important when deriving shear rate scaling of $\mathcal{E}$ at $k_x \neq 0$.


\appendix
\section{Reynolds number scaling}
The system we are considering is given by
\begin{align}
\p_t\begin{bmatrix}
\bld{\varphi}_1  \\
 \bld{\varphi}_2  \\
\bld{\varphi}_3
\end{bmatrix} &=
\begin{bmatrix}
\frac{1}{Re}\bld{\mathcal{L}} & 0  & 0 \\
\bld{\mathcal{C}} & \frac{1}{Re}\bld{\mathcal{S}} & 0 \\
0 & \bld{\mathcal{A}} & -\frac{1}{{\Elas}\Rey}
\end{bmatrix}\begin{bmatrix}
\bld{\varphi}_1  \\
\bld{\varphi}_2  \\
\bld{\varphi}_3
\end{bmatrix} + 
\begin{bmatrix}
\bld{\mathcal{B}}_{11}
& \bld{\mathcal{B}}_{12} \\
\bld{\mathcal{B}}_{21}
& \bld{\mathcal{B}}_{22} \\
\bld{0}
& \bld{\mathcal{B}}_{32}
\end{bmatrix}\begin{bmatrix}
\bld{\hat{d}}_\bld{u} \\  \bld{\hat{d}}_\mathsfbi{T}
\end{bmatrix}  
\end{align}
where $\bld{\varphi}_1 = \begin{bmatrix} \hat{v }& \hat{T}_{yy} & \hat{T}_{yz} & \hat{T}_{zz}\end{bmatrix}^{\mathsf{T}}$, 
$\bld{\varphi}_2 = \begin{bmatrix} \hat{\eta} & \hat{T}_{xy} & \hat{T}_{xz} \end{bmatrix}$ and $\bld{\varphi}_3 =  \hat{T}_{xx} $.  
\begin{align}\bld{\mathcal{P}}=
\begin{bmatrix}
\bld{\mathcal{P}}_{11} & \bld{\mathcal{P}}_{12} & \bld{\mathcal{P}}_{13}  \\
\bld{\mathcal{P}}_{21} & \bld{\mathcal{P}}_{22} & \bld{\mathcal{P}}_{23} \\
\bld{\mathcal{P}}_{31} & \bld{\mathcal{P}}_{32} & \bld{\mathcal{P}}_{33}
\end{bmatrix}
\end{align}
is the solution to the Lyapunov equation
\begin{align}
\begin{bmatrix}
\frac{1}{\Rey}\bld{\mathcal{L}} & 0  & 0 \\
\bld{\mathcal{C}} & \frac{1}{\Rey}\bld{\mathcal{S}} & 0 \\
0 & \bld{\mathcal{A}} & -\left({\Elas}\Rey\right)^{-1}
\end{bmatrix}\bld{\mathcal{P}}
+
\bld{\mathcal{P}}
\begin{bmatrix}
\frac{1}{\Rey} \bld{\mathcal{L}}^*  & \bld{\mathcal{C}}^*  & 0  \\
0 & \frac{1}{\Rey} \bld{\mathcal{S}}^* & \bld{\mathcal{A}}^* \\
0 & 0 & -({\Elas}\Rey)^{-1}
\end{bmatrix}  = - \bld{\mathcal{B}} \bld{\mathcal{B}}^*
%
 \end{align} 
 Using the fact that $\bld{\mathcal{P}}^* = \bld{\mathcal{P}}$, we find the following by equating the elements on the right and left-hand sides  \begin{align}
 \bld{\mathcal{L}}\bld{\mathcal{P}}_{11}
+
\bld{\mathcal{P}}_{11}\bld{\mathcal{L}}^*  &= -\Rey\,\bld{\mathcal{B}}_{1}\bld{\mathcal{B}}_{1}^*
 \label{P11Lyap}  \\
 \bld{\mathcal{S}}\bld{\mathcal{P}}_{12}^* + \bld{\mathcal{P}}_{12}^*\bld{\mathcal{L}}^* 
 &= - \Rey\,\bld{\mathcal{C}} \bld{\mathcal{P}}_{11}
 \label{P12Sylv} \\
 \bld{\mathcal{S}}\bld{\mathcal{P}}_{22}
 +\bld{\mathcal{P}}_{22}\bld{\mathcal{S}}^*    &= 
 -\Rey\left( \bld{\mathcal{C}}\bld{\mathcal{P}}_{12}
+ \bld{\mathcal{P}}_{12}^*\bld{\mathcal{C}}^*
 +\bld{\mathcal{B}}_{2}\bld{\mathcal{B}}_{2}^* \right)  
 \label{P22Lyap} \\
 \left(\bld{\mathcal{L}} -\frac{1}{{\Elas}} \right)\bld{\mathcal{P}}_{13} &=-\Rey\bld{\mathcal{P}}_{12}\bld{\mathcal{A}}^* 
 \label{P13Eqn}  \\
\left( \bld{\mathcal{S}}-\frac{1}{{\Elas}}  \right)\bld{\mathcal{P}}_{23} &= - \Rey\left(\bld{\mathcal{C}}\bld{\mathcal{P}}_{13}  - \bld{\mathcal{P}}_{22}\bld{\mathcal{A}}^* \right)   
 \label{P23Eqn} \\
\bld{\mathcal{P}}_{33} &=  \frac{{\Elas}\Rey}{2}\left(\bld{\mathcal{A}}\bld{\mathcal{P}}_{23} +\bld{\mathcal{P}}_{23}^*\bld{\mathcal{A}}^*   + \bld{\mathcal{B}}_{3}\bld{\mathcal{B}}_{3}^*\right) 
 \label{P33Eqn} 
 \end{align}
 We can rewrite (\ref{P11Lyap}) and (\ref{P12Sylv}) in terms of Reynolds numbers independent operators as
  \begin{align}
  \bld{\mathcal{L}} \bld{\mathcal{X}} 
+
\bld{\mathcal{X}}  \bld{\mathcal{L}} ^*  &= - \bld{\mathcal{B}}_{1}\bld{\mathcal{B}}_{1}^*,
\quad \Rightarrow \quad
 \bld{\mathcal{P}}_{11} = \Rey\,\bld{\mathcal{X}} \\
\Rightarrow \quad
  \bld{\mathcal{S}} \bld{\mathcal{Y}}
  +
   \bld{\mathcal{Y}} \bld{\mathcal{L}}^* 
 &= -  \,\bld{\mathcal{C}} \bld{\mathcal{X}} 
 \quad
 \Rightarrow
 \quad
  \bld{\mathcal{P}}_{12}^* =  \Rey^2\,\bld{\mathcal{Y}}
 %
 \end{align}
Since the equations are linear and $ \bld{\mathcal{S}} $ is independent of $Re$ we can define, (\ref{P22Lyap}) implies the following form for $\bld{\mathcal{P}}_{22}$
\begin{align}
\bld{\mathcal{P}}_{22} = \Rey\,\bld{\mathcal{W}}  + \Rey^3\,\bld{\mathcal{Z}}
\end{align}
 where $\bld{\mathcal{Z}}$ and $\bld{\mathcal{W}}$ are independent of $\Rey$ and hence solutions of 
\begin{align}
 \bld{\mathcal{S}} \bld{\mathcal{W}}
 +\bld{\mathcal{W}} \bld{\mathcal{S}}^*    &= 
 - \bld{\mathcal{B}}_{2}\bld{\mathcal{B}}_{2}^* \\
 \bld{\mathcal{S}}\bld{\mathcal{Z}}
 + \bld{\mathcal{Z}}\bld{\mathcal{S}}^*    &= 
 - \left( \bld{\mathcal{C}} \bld{\mathcal{Y}}^*
+   \bld{\mathcal{Y}}\bld{\mathcal{C}}^* \right).
\end{align}
This so far constitutes the derivation for (\ref{Eexpr}) that was given in \citep{Hoda2009}. We proceed further with the remaining equations. Using the definition of $\bld{\mathcal{Y}}$ we have
\begin{align} 
 \left(\bld{\mathcal{L}} -\frac{1}{{\Elas}} \right)\tilde{ \bld{\mathcal{P}}}_{13} ^{(3)}     &=- \,\bld{\mathcal{Y}}^*\bld{\mathcal{A}}^* 
 \quad
 \Rightarrow
 \quad
 \bld{\mathcal{P}}_{13} = \Rey^3 \tilde{ \bld{\mathcal{P}}}_{13}^{(3)}     \label{calJeqn}
\end{align} 
and therefore, again, since the equations are linear and $ \bld{\mathcal{S}} $ is independent of $\Rey$, the form of (\ref{P23Eqn}) implies
\begin{align}
\bld{\mathcal{P}}_{23} = \Rey^2\,\tilde{\bld{\mathcal{P}}}_{23}^{(2)}   + \Rey^4\,\tilde{\bld{\mathcal{P}}}_{23}^{(4)} 
\end{align}
where $\tilde{\bld{\mathcal{P}}}_{23}^{(2)}$ and $\tilde{\bld{\mathcal{P}}}_{23}^{(4)}$ are independent of $\Rey$ and hence solutions of 
\begin{align}
\left( \bld{\mathcal{S}}-\frac{1}{{\Elas}}  \right)\tilde{\bld{\mathcal{P}}}_{23}^{(2)}&= 
-    \bld{\mathcal{W}}\bld{\mathcal{A}}^*    \label{calMeqn}\\
\left( \bld{\mathcal{S}}-\frac{1}{{\Elas}}  \right)\tilde{\bld{\mathcal{P}}}_{23}^{(4)} &= 
-   \left(\bld{\mathcal{C}} \tilde{ \bld{\mathcal{P}}}_{13}^{(3)}    + \bld{\mathcal{Z}}\bld{\mathcal{A}}^* \right) \label{calEeqn}  
\end{align}
Finally, substituting these results in (\ref{P33Eqn}) we deduce in a similar fashion that
\begin{align}
\bld{\mathcal{P}}_{33} =\Rey\,\tilde{\bld{\mathcal{P}}}_{33}^{(1)} + \Rey^3\, \tilde{\bld{\mathcal{P}}}_{33}^{(3)}  + \Rey^5\,\tilde{\bld{\mathcal{P}}}_{33}^{(5)}
\end{align}
 where $\tilde{\bld{\mathcal{P}}}_{33}^{(1)}$, $\tilde{\bld{\mathcal{P}}}_{33}^{(3)}$ and $\tilde{\bld{\mathcal{P}}}_{33}^{(5)}$ are independent of $\Rey$ and hence given by
 \begin{align}
\tilde{\bld{\mathcal{P}}}_{33}^{(1)} &= \frac{{\Elas}}{2}\bld{\mathcal{B}}_{3}\bld{\mathcal{B}}_{3}^*\\
\tilde{\bld{\mathcal{P}}}_{33}^{(3)} &= \frac{{\Elas}}{2}\left(\bld{\mathcal{A}}\tilde{\bld{\mathcal{P}}}_{23}^{(2)}   +  (\tilde{\bld{\mathcal{P}}}_{23}^{(2)})^*\bld{\mathcal{A}}^*\right)  \\
\tilde{\bld{\mathcal{P}}}_{33}^{(5)}&=\frac{{\Elas}}{2}
\left(\bld{\mathcal{A}}\tilde{\bld{\mathcal{P}}}_{23}^{(4)}   +  (\tilde{\bld{\mathcal{P}}}_{23}^{(4)})^*\bld{\mathcal{A}}^* \right).
 \end{align}
Now redefining $\bld{\mathcal{H}}$ to include $\bld{\varphi}_3$ in the output we have
\begin{align}
\bld{\mathcal{H}}^* \bld{\mathcal{H}} 
&= 
(\bld{\mathcal{H}}_{\bld{u}}+\bld{\mathcal{H}}_{\mathsfbi{T}})^* (\bld{\mathcal{H}}_{\bld{u}}+\bld{\mathcal{H}}_{\mathsfbi{T}}) \\
&= 
\text{diag}\left(
\bld{\mathcal{H}}_{\bld{u},\bld{\varphi}_1} + \bld{\mathcal{H}}_{\mathsfbi{T},\bld{\varphi}_1},
\bld{\mathcal{H}}_{\bld{u},\bld{\varphi}_2} + \bld{\mathcal{H}}_{\mathsfbi{T},\bld{\varphi}_2},
\bld{\mathcal{H}}_{\mathsfbi{T},\bld{\varphi}_3} \right)  
\end{align}
where $\bld{\mathcal{H}}_{\mathsfbi{T},\bld{\varphi}_3} =  \mathcal{H}_{xx}^* \mathcal{H}_{xx} $.
Then from $E(k_z; \Rey,\beta,{\Elas})  = \text{tr}(\bld{\mathcal{H}}^* \bld{\mathcal{H}} \bld{\mathcal{P}})$ we obtain the following Reynolds number scaling
\begin{align}
E(k_z; \Rey,\beta,{\Elas}) &= \Rey\,\mathfrak{f}_{1}(k_z;\beta,{\Elas}) + \Rey^3\,\mathfrak{f}_{2}(k_z;\beta,{\Elas}) + \Rey^5\,\mathfrak{f}_{3}(k_z;\beta,{\Elas}) 
\end{align}
where
\begin{align}
\mathfrak{f}_1(k_z;\beta,{\Elas}) &= \text{tr}
 \left[ (\bld{\mathcal{H}}_{\bld{u},\bld{\varphi}_1} + \bld{\mathcal{H}}_{\mathsfbi{T},\bld{\varphi}_1} )\bld{\mathcal{X}}
 + (\bld{\mathcal{H}}_{\bld{u},\bld{\varphi}_2} + \bld{\mathcal{H}}_{\mathsfbi{T},\bld{\varphi}_2})\bld{\mathcal{W}}  
+ \frac{{\Elas}}{2}   \bld{\mathcal{H}}_{\mathsfbi{T},\bld{\varphi}_3}\bld{\mathcal{B}}_{3}\bld{\mathcal{B}}_{3}^* \right]\\
\mathfrak{f}_2(k_z;\beta,{\Elas}) &= \text{tr} \left[(\bld{\mathcal{H}}_{\bld{u},\bld{\varphi}_2} 
+ \bld{\mathcal{H}}_{\mathsfbi{T},\bld{\varphi}_2})\bld{\mathcal{Z}} \right]
+ \frac{{\Elas}}{2}\text{tr}\left[\bld{\mathcal{H}}_{\mathsfbi{T},\bld{\varphi}_3} 
 \left(\bld{\mathcal{A}}\tilde{\bld{\mathcal{P}}}_{23}^{(2)}   +  (\tilde{\bld{\mathcal{P}}}_{23}^{(2)})^*\bld{\mathcal{A}}^*\right)  \right] \\
\mathfrak{f}_3(k_z;\beta,{\Elas}) &=  \frac{{\Elas}}{2}\text{tr} 
\left[\bld{\mathcal{H}}_{\mathsfbi{T},\bld{\varphi}_3}  \left(\bld{\mathcal{A}}\tilde{\bld{\mathcal{P}}}_{23}^{(4)}   +  (\tilde{\bld{\mathcal{P}}}_{23}^{(4)})^*\bld{\mathcal{A}}^* \right) \right].
\end{align}
The operators $\bld{\mathcal{X}}$, $\bld{\mathcal{W}}$ and  $\bld{\mathcal{Z}}$ can be determined from (\ref{WLyap})--(\ref{ZLyap}) as before while $\tilde{\bld{\mathcal{P}}}_{23}^{(2)}$ and $\tilde{\bld{\mathcal{P}}}_{23}^{(4)}$ can be determined from $\bld{\mathcal{X}}$, $\bld{\mathcal{W}}$ and  $\bld{\mathcal{Z}}$  and the solutions of (\ref{calJeqn}), (\ref{calMeqn}) and (\ref{calEeqn}).

\section{General Solution}\label{app_gs}
In this appendix we consider solutions of the generic system (\ref{presysA}). Before we proceed with the solution, we first introduce some notation to make the notation compact. In order to make the dependence of a quantity on a matrix explicit without confusing tensor products (block-wise matrix multiplication) with matrix multiplication we first define the `vertical stacking' operation by a superscript enclosed in brackets and defined as
\begin{align}
 \bld{a}^{[k]}  \equiv \bld{1}_{k\times 1}\otimes \bld{a} \label{stackdefn}
\end{align}
where we note that $\bld{a}^{[1]} = \bld{a}$. Following this we define a `restacking' operator $\mathsfbi{I}_{[k]}^{[m]}$ such that
\begin{align}
\bld{a}^{[m]} \equiv
\mathsfbi{I}_{[k]}^{[m]}\bld{a}^{[k]}.  \label{restackdefn}
\end{align} 
The actual matrix form of $\mathsfbi{I}_{[k]}^{[m]}$ can be easily derived but is not necessary in the subsequent development. This notation allows us be unambiguous while writing for a given vector composed of matrices $\mathbf{A}_i$ and $\mathbf{X}$
\begin{align}
\begin{bmatrix}
\mathbf{A}_1 \mathbf{X} \\ \hdots \\ \mathbf{A}_m \mathbf{X}
\end{bmatrix} &= \text{diag}\left(\mathbf{A}_1,\hdots,\mathbf{A}_m\right)\mathbf{X}^{[m]} 
= \text{diag}\left(\mathbf{A}_1,\hdots,\mathbf{A}_m\right)\mathsfbi{I}_{[1]}^{[m]}\bld{X} \nonumber
\end{align}
as long as we interpret $\mathsfbi{I}_{[1]}^{[m]}\mathbf{X}$ as $\mathsfbi{I}_{[1]}^{[m]}\mathbf{X}^{[1]}$.
Using this notation, we are then interested in solutions  $\bld{a}_{\bld{u}}  \in \mathbb{C}^{{\Nzero}\times 1}$, $\bld{a}_{\bld{1}}  \in \mathbb{C}^{{\Nzero}M\times 1}$ of the system
\begin{align}
\begin{bmatrix}
\beta \bld{\Gamma}_{\bld{a}}  & \alpha\bld{\Gamma}_{\bld{b}}    \\
{\Elas}^{-1}  \bld{\Gamma}_{\bld{c}} \mathsfbi{I}_{[1]}^{[M]}        
& \bld{\Gamma}_{\bld{d}}  (\alpha,\beta,{\Elas}) 
\end{bmatrix}\begin{bmatrix} \bld{a}_{\bld{u}} \\ \bld{a}_{\bld{1}}\end{bmatrix} =
- \begin{bmatrix}\bld{b}_{\bld{u}} \\ \bld{b}_{\mathsfbi{T}} \end{bmatrix}\label{gs_sysA}
\end{align}
    where $\bld{\Gamma}_{\bld{a}} \in \mathbb{C}^{{\Nzero}\times{\Nzero}}$, $\bld{\Gamma}_{\bld{b}} \in \mathbb{C}^{N\times {\Nzero}M}$, and $\bld{\Gamma}_{\bld{c}}  \in \mathbb{C}^{{\Nzero}M\times {\Nzero}M}$  are matrices independent of $\beta$ and ${\Elas}$. The vectors $\bld{b}_{\bld{u}} \in \mathbb{C}^{{\Nzero}\times 1}$ and $\bld{b}_{\mathsfbi{T}} \in \mathbb{C}^{{\Nzero}M\times 1}$ are known vectors and we define $\bld{\Gamma}_{\bld{d}} : [0,1]\times[0,1]\times (0,\infty)\mapsto  \mathbb{C}^{{\Nzero}M\times {\Nzero}M}$ by
\begin{align}
\bld{\Gamma}_{\bld{d}}  (\alpha,\beta,{\Elas}) &\equiv  \bld{\gamma}(\alpha,\beta) - \frac{1}{{\Elas}}\mathsfbi{I}_{\Nzero M}\label{gs_Gamma22defn} \\ 
\bld{\gamma}(\alpha,\beta)  &\equiv \alpha \bld{\gamma}_\alpha + \beta \bld{\gamma}_\beta \label{gs_gammadefn} 
\end{align}
where $\bld{\gamma}_\alpha,\bld{\gamma}_\beta \in \mathbb{C}^{{\Nzero}M\times {\Nzero}M}$ are independent of $\beta$ and ${\Elas}$ and
$\bld{\gamma}: [0,1]\times[0,1]\mapsto \mathbb{C}^{{\Nzero}M\times {\Nzero}M}$. Furthermore $\bld{\Gamma}_{\bld{a}}$, $\bld{\Gamma}_{\bld{d}}$ and $\bld{\gamma}$ are nonsingular, guaranteeing existence and uniqueness of $\bld{a}_{\bld{u}}$ and $\bld{a}_{\bld{1}}$. The system (\ref{gs_sysA}) is equivalent to the system (\ref{presysA}).
We will derive series solutions $\bld{a}_{\bld{u}}  \in \mathbb{C}^{{\Nzero}\times 1}$, $\bld{a}_{\bld{1}}  \in \mathbb{C}^{{\Nzero}M\times 1}$ of the system (\ref{gs_sysA}) for the two asymptotic limits  ${\Elas} \gg 1$ and ${\Elas} \ll 1$, and in both cases also derive convergence criteria for the solutions in terms of lower/upper bounds on ${\Elas}$. Note that $\bld{\Gamma}_{r}$ is analogous to $\bld{L}_{r}$, $\bld{S}_{r}$ and $\bld{G}_{r}$ for $r \in \{ \bld{a},\bld{b},\bld{c},\bld{d} \}$. Similarly, $\bld{\gamma}$ is analogous to $\bld{\ell}$, $\bld{\mathfrak{s}}$ and $\bld{\mathfrak{g}}$.

Solving (\ref{gs_sysA}) for $\bld{a}_{\bld{u}}$ and $\bld{a}_{\bld{1}}$ immediately gives
\begin{align}
  \bld{a}_{\bld{u}} &= \bld{\Gamma}_{\bld{0}}^{-1} \left( -\bld{b}_{\bld{u}}  + \alpha \bld{\Gamma}_{\bld{b}}  \bld{\Gamma}_{\bld{d}}^{-1} \bld{b}_{\mathsfbi{T}}  \right)
  \label{gs_afsol1}\\
\bld{a}_{\bld{1}}
&=  \bld{\Gamma}_{\bld{d}}^{-1}\Big[ 
  \frac{1}{{\Elas}}   \bld{\Gamma}_{\bld{c}}\mathsfbi{I}_{[1]}^{[M]}
 \bld{\Gamma}_{\bld{0}}^{-1}
   \bld{b}_{\bld{u}}   
-  \Big( \mathsfbi{I}_{N^2M} + \frac{\alpha}{{\Elas}}   \bld{\Gamma}_{\bld{c}}\mathsfbi{I}_{[1]}^{[M]}
 \bld{\Gamma}_{\bld{0}}^{-1}    \bld{\Gamma}_{\bld{b}}  \bld{\Gamma}_{\bld{d}}^{-1}    \Big) \bld{b}_{\mathsfbi{T}}\Big].  \label{gs_apsol1}
\end{align}
where we defined
\begin{align}
\bld{\Gamma}_{\bld{0}} \equiv
\beta \bld{\Gamma}_{\bld{a}}   
-\frac{\alpha}{{\Elas}}  \bld{\Gamma}_{\bld{b}}
  \bld{\Gamma}_{\bld{d}}^{-1}
 \bld{\Gamma}_{\bld{c}}  \mathsfbi{I}_{[1]}^{[M]}  \label{gs_Gamma0defn}
\end{align}
for notational convenience. Note $\mathsfbi{I}_{[1]}^{[M]} = \bld{1}_{M\times1} \otimes \mathsfbi{I}$. This is a sufficient solution but does not make the dependence on ${\Elas}$ explicit. In order to make this dependence on ${\Elas}$ explicit, one needs to make the dependence of $\bld{\Gamma}_{\bld{d}}^{-1}$ and $\bld{\Gamma}_{\bld{0}}^{-1}$ on ${\Elas}$ explicit.
Since  $\bld{\Gamma}_{\bld{d}}^{-1}(\alpha,\beta,{\Elas}) =  
 \bld{\gamma}^{-1} \left( \mathsfbi{I} - \frac{\bld{\gamma}^{-1}}{{\Elas}}\right) ^{-1}
= - {\Elas} \left( \mathsfbi{I}- {\Elas} \bld{\gamma} \right) ^{-1}$, 
we can use ${\Elas}$ to control the spectral radius of the deviation of the quantities in the parentheses  from $\mathsfbi{I}$. It follows therefore that we can use the Neumann series to obtain \emph{convergent} series solutions for $\bld{\Gamma}_{\bld{d}}^{-1}$ as
\begin{align}
\bld{\Gamma}_{\bld{d}}^{-1}(\alpha,\beta,{\Elas}) 
&=\begin{cases}
\sum_{n=0}^{\infty} \bld{\gamma}_n {\Elas}^{-n} & {\Elas} \gg 1\quad \text{Strong elasticity} \\
\sum_{n=0}^{\infty} (-1) \bld{\gamma}_n{\Elas}^{n+1}   & {\Elas} \ll 1\quad \text{Weak elasticity}
\end{cases}  \label{gs_Gamma22inv}
\end{align}
where the conditions $ {\Elas} \gg 1$  and  ${\Elas} \ll 1$ are defined such that the series are convergent (conditions on ${\Elas}$ guaranteeing this are described in a later subsection) and
 $\bld{\gamma}_n: [0,1]\times[0,1]\mapsto \mathbb{C}^{{\Nzero}M\times {\Nzero}M}$ is given by
\begin{align}
\bld{\gamma}_n(\alpha,\beta)   
&=\begin{cases}
 \bld{\gamma}^{-n-1}  & {\Elas} \gg 1\quad \text{Strong elasticity} \\
  \bld{\gamma}^n	& {\Elas} \ll 1\quad \text{Weak elasticity}.
\end{cases}  \label{gs_gammandefn}
\end{align}
Note that the ${\Nzero} \times {\Nzero}$ block elements of $\bld{\gamma}_n$ are referred to as $\gamma_{n,ij}$.
Using (\ref{gs_Gamma22inv}) in the definition (\ref{gs_Gamma0defn}), we immediately obtain a series expression for $\bld{\Gamma}_{\bld{0}}$
\begin{align}
\bld{\Gamma}_{\bld{0}} 
    =
\begin{cases} 
 \beta \bld{\Gamma}_{\bld{a}}-\frac{\alpha}{{\Elas}}\sum_{n=0}^{\infty}\tilde{\gamma}_n{\Elas}^{-n} 
  & {\Elas} \gg 1\quad \text{Strong elasticity} \\
 \beta \bld{\Gamma}_{\bld{a}} +   \alpha\sum_{n=0}^{\infty}\tilde{\gamma}_{n}{\Elas}^{n} 
& {\Elas} \ll 1\quad \text{Weak elasticity} \label{gs_Gamma0defn2}
\end{cases} 
 \end{align} 
where $\tilde{\gamma}_n : [0,1]\times[0,1]\mapsto \mathbb{C}^{{\Nzero}\times{\Nzero}}$ is obtained from the contraction $ \bld{\Gamma}_{\bld{b}}
  \bld{\Gamma}_{\bld{d}}^{-1}
 \bld{\Gamma}_{\bld{c}} \mathsfbi{I}_{[1]}^{[M]}  \bld{a}_{\bld{u}}  $ 
   \begin{align}
 \tilde{\gamma}_n(\alpha,\beta) 
  &= \bld{\Gamma}_{\bld{b}}\bld{\gamma}_{n}\bld{\Gamma}_{\bld{c}} \mathsfbi{I}_{[1]}^{[M]}    \label{gs_tildegamman0} \\
 &= \sum_{i,j,k = 1}^M \Gamma_{b,i}\gamma_{n,ij}\Gamma_{c,jk}
 \label{gs_tildegamman}  
  \end{align}
where we denote the $N\times N$ block elements of $\bld{\Gamma}_{\bld{b}}$ and $\bld{\Gamma}_{\bld{c}}$ as $\Gamma_{b,i}$ and $ \Gamma_{c,ij}$, respectively.  
Invoking the Neumann series again to invert $\bld{\Gamma}_{\bld{0}}$ in the expressions in (\ref{gs_Gamma0defn2}),  we find   
   \begin{align}
\bld{\Gamma}_{\bld{0}}^{-1} 
=\begin{cases}
  \beta^{-1} \bld{\tilde{\Gamma}}_{\bld{0}}^{-1} 
  &  {\Elas} \gg 1,\,\beta \neq 0
\\
%
  %
  \bld{\tilde{\Gamma}}_{\bld{0}}^{-1}
  &   {\Elas} \ll 1
  \end{cases} \label{gs_Gamma0inv}
 \end{align}
 where
  \begin{align}
\bld{\tilde{\Gamma}}_{\bld{0}}^{-1} 
\equiv
\begin{cases}
 \sum_{m=0}^\infty  \left(\frac{\alpha}{{\Elas}\beta}\right)^m\left(
  \bld{\Gamma}_{\bld{a}}^{-1} \sum_{n=0}^{\infty}\tilde{\gamma}_n{\Elas}^{-n}  \right)^m\bld{\Gamma}_{\bld{a}}^{-1}  
  &  {\Elas} \gg 1, \, \beta \neq 0
\\
%
  %
 \sum_{m=0}^\infty
 \left[-\alpha(\beta \bld{\Gamma}_{\bld{a}} + \alpha\tilde{\gamma}_0)^{-1}\sum_{n=1}^{\infty}\tilde{\gamma}_{n}{\Elas}^{n}\right]^m \times & \\
\hspace{1.75in} (\beta \bld{\Gamma}_{\bld{a}} + \alpha\tilde{\gamma}_0)^{-1}
  & {\Elas} \ll 1.
  \end{cases} \label{gs_Gamma0lineinv1}
 \end{align}
Lower/upper bounds on ${\Elas}$ for convergence are discussed in the next subsection.
 Finally, substituting (\ref{gs_Gamma22inv}) and (\ref{gs_Gamma0inv}) into (\ref{gs_afsol1}) and (\ref{gs_apsol1}), we obtain series solutions for $\bld{a}_{\bld{u}}$ and $\bld{a}_{\bld{1}}$ where the dependence on ${\Elas}$ is explicit
\begin{align}
\bld{a}_{\bld{u}} &=
\begin{cases}
\frac{1}{\beta}\bld{\Gamma}_{\bld{u} \rightarrow \bld{u}} \bld{b}_{\bld{u}}
+ 
\frac{\alpha}{\beta}\bld{\Gamma}_{\mathsfbi{T} \rightarrow \bld{u}}\bld{b}_{\mathsfbi{T}} & {\Elas} \gg 1, \, \beta \neq 0 \\
%
%
\bld{\Gamma}_{\bld{u} \rightarrow \bld{u}}\bld{b}_{\bld{u}} + {\Elas}\alpha\bld{\Gamma}_{\mathsfbi{T} \rightarrow \bld{u}}\bld{b}_{\mathsfbi{T}}  & {\Elas} \ll 1
\end{cases} \label{gs_afsol3}\\
\bld{a}_{\bld{1}} &=
\begin{cases}
\bld{\Gamma}_{\mathsfbi{T} \rightarrow \bld{1a}}\bld{b}_{\mathsfbi{T}}
+ \frac{1}{{\Elas}\beta}\left(\bld{\Gamma}_{\bld{u} \rightarrow \bld{1}}\bld{b}_{\bld{u}} + \alpha \bld{\Gamma}_{\mathsfbi{T} \rightarrow \bld{1b}}\bld{b}_{\mathsfbi{T}}\right) & {\Elas} \gg 1, \, \beta \neq 0 \\
%
%
\bld{\Gamma}_{\bld{u} \rightarrow \bld{1}}\bld{b}_{\bld{u}}
+{\Elas}\left(\bld{\Gamma}_{\mathsfbi{T} \rightarrow \bld{1a}} 
+\alpha\bld{\Gamma}_{\mathsfbi{T} \rightarrow \bld{1b}}\right)\bld{b}_{\mathsfbi{T}} & {\Elas} \ll 1
\end{cases} \label{gs_apsol3}
\end{align}
where
  \begin{align}
\bld{\Gamma}_{\bld{u} \rightarrow \bld{u}} &=  -\bld{\tilde{\Gamma}}_{\bld{0}}^{-1} \label{gs_Gammaff}\\
\bld{\Gamma}_{\mathsfbi{T} \rightarrow \bld{u}} &=  
\begin{cases}
   \bld{\tilde{\Gamma}}_{\bld{0}}^{-1} \bld{\Gamma}_{\bld{b}} \sum_{n=0}^{\infty}  \bld{\gamma}_n{\Elas}^{-n}    & {\Elas} \gg 1 \\
 -  \bld{\tilde{\Gamma}}_{\bld{0}}^{-1} \bld{\Gamma}_{\bld{b}} \sum_{n=0}^{\infty}  \bld{\gamma}_n{\Elas}^{n}   & {\Elas} \ll 1
\end{cases}\label{gs_Gammapf}\\
\bld{\Gamma}_{\bld{u} \rightarrow \bld{1}} &=
\begin{cases}
 \sum_{n=0}^{\infty}  \bld{\gamma}_n{\Elas}^{-n} \bld{\Gamma}_{\bld{c}}\mathsfbi{I}_{[1]}^{[M]} \bld{\tilde{\Gamma}}_{\bld{0}}^{-1}& {\Elas} \gg 1 \\
 -\sum_{n=0}^{\infty}  \bld{\gamma}_n{\Elas}^{n} \bld{\Gamma}_{\bld{c}}\mathsfbi{I}_{[1]}^{[M]} \bld{\tilde{\Gamma}}_{\bld{0}}^{-1} & {\Elas} \ll 1
\end{cases} \label{gs_Gammafp}\\
\bld{\Gamma}_{\mathsfbi{T} \rightarrow \bld{1a}} &= 
\begin{cases}
 -\sum_{n=0}^{\infty} \bld{\gamma}_n{\Elas}^{-n} & {\Elas} \gg 1 \\
\sum_{n=0}^{\infty} \bld{\gamma}_n{\Elas}^{n} 
 & {\Elas} \ll 1
\end{cases}\label{gs_Gammapp1} \\
\bld{\Gamma}_{\mathsfbi{T} \rightarrow \bld{1b}} &= 
\begin{cases}
-\sum_{n=0}^{\infty} \bld{\gamma}_n {\Elas}^{-n}\bld{\Gamma}_{\bld{c}}\mathsfbi{I}_{[1]}^{[M]}\bld{\tilde{\Gamma}}_{\bld{0}}^{-1}\bld{\Gamma}_{\bld{b}}  \sum_{p=0}^{\infty} \bld{\gamma}_p {\Elas}^{-p}
   & {\Elas} \gg 1 \\
-\sum_{n=0}^{\infty}  \bld{\gamma}_n{\Elas}^{n} \bld{\Gamma}_{\bld{c}}\mathsfbi{I}_{[1]}^{[M]}\bld{\tilde{\Gamma}}_{\bld{0}}^{-1}\bld{\Gamma}_{\bld{b}}  \sum_{p=0}^{\infty} \bld{\gamma}_p{\Elas}^{p}
& {\Elas} \ll 1
\end{cases} \label{gs_Gammapp2}
\end{align}

\subsection{Convergence criteria} \label{app_gscc}
We first consider (\ref{gs_Gamma22inv}), which is in the form of a matrix geometric series. 
A convergence criterion for this can be straightforwardly obtained by restricting the spectral radii of the terms in the series to be less than unity, i.e.,   $\lambda_{\max}(\bld{\gamma}_n {\Elas}^{\pm n}) < 1$ for the weakly and strongly elastic cases, respectively. Since $\sigma_{\max}(A) \geq |\lambda_{\max}(A)|$ and $\sigma_{\max}(A^{-1}) = 1/\sigma_{\min}(A) $ for nonsingular $A$, sufficient conditions for the convergence of (\ref{gs_Gamma22inv}) for the cases ${\Elas} \gg 1$ and ${\Elas} \ll 1$ are
\begin{align}
{\Elas} > \frac{1}{\sigma_{\min}(\bld{\gamma})}, \quad 
{\Elas} < \frac{1}{\sigma_{\max}(\bld{\gamma})} \label{gscc_cond1}
\end{align}
respectively. 
\begin{figure}
\centering
\includegraphics{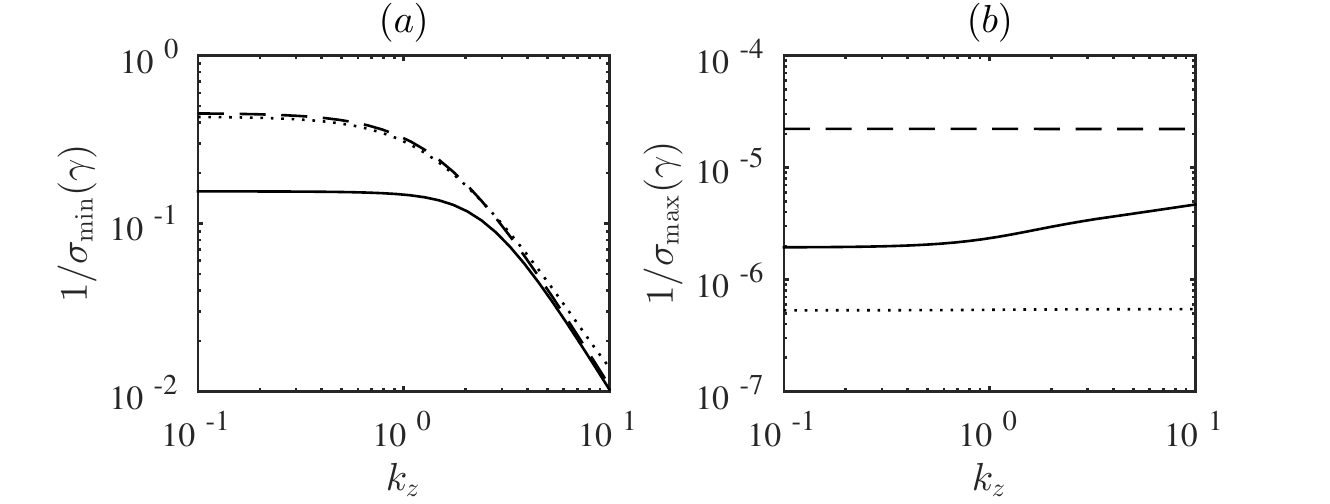}
\caption{Bounds on $\Elas$ indicated in (\ref{gscc_cond1})  at $\beta = 0.9$. Solid lines (\lline) are for $\bld{\gamma} = \bld{\ell}$, dashed lines (\dashed) are for $\bld{\gamma} = \bld{\mathfrak{s}}$, dotted lines (\dotted) are for $\bld{\gamma} = \bld{\mathfrak{g}}$. The plots in (a) show a necessary lower bound on $\Elas$ for convergence of the series in the strong elastic regime denoted $\Elas \gg 1$. Similarly, the plots in (b) show a necessary upper bound on $\Elas$ for convergence of the series in the weakly elastic regime, $\Elas \ll 1$.}
\label{fig:convg_1}
\end{figure}
Figure \ref{fig:convg_1}  shows the bounds indicated by (\ref{gscc_cond1}) plotted against $k_z$. 

\begin{figure}
\centering
\includegraphics{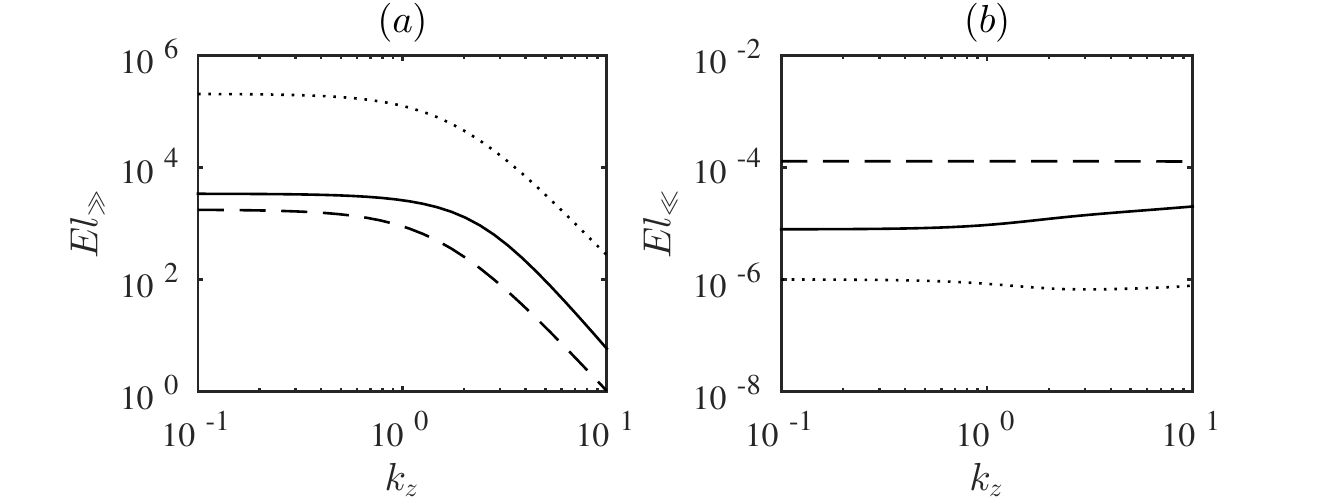}
\caption{Bounds on $\Elas$ indicated in (\ref{gscc_muboundfinal1})--(\ref{gscc_muboundfinal2})  at $\beta = 0.9$. Solid lines (\lline) are for $\bld{\gamma} = \bld{\ell}$, dashed lines (\dashed) are for $\bld{\gamma} = \bld{\mathfrak{s}}$, dotted lines (\dotted) are for $\bld{\gamma} = \bld{\mathfrak{g}}$. The plots in (a) show a sufficient lower bound on $\Elas$ for convergence of the series in the strong elastic regime, $\Elas \gg 1$. Similarly, the plots in (b) show a sufficient upper bound on $\Elas$ for convergence of the series in the weakly elastic regime, $\Elas \ll 1$.}
\label{fig:convg_2}
\end{figure}

Secondly, we are interested in the convergence of the outer series in (\ref{gs_Gamma0lineinv1})
\begin{align}
\bld{\tilde{\Gamma}}_{\bld{0}}^{-1} 
=\begin{cases}
 \sum_{m=0}^\infty  \left(\frac{\alpha}{{\Elas}\beta}
   \bld{\Gamma}_{\bld{a}}^{-1} \sum_{n=0}^{\infty}\tilde{\gamma}_n{\Elas}^{-n}  \right)^m\bld{\Gamma}_{\bld{a}}^{-1}  
  &  {\Elas} \gg 1, \, \beta \neq 0
\\ 
 \sum_{m=0}^\infty \left[-\alpha(\beta \bld{\Gamma}_{\bld{a}}+ \alpha\tilde{\gamma}_0)^{-1}\sum_{n=1}^{\infty}\tilde{\gamma}_{n}{\Elas}^{n}\right]^m(\beta \bld{\Gamma}_{\bld{a}}+ \alpha\tilde{\gamma}_0)^{-1}
  & {\Elas} \ll 1.
  \end{cases} \nonumber
   \end{align}
  From (\ref{gs_tildegamman0}), (\ref{gs_gammandefn}) and the submultiplicativity of the spectral norm, we have
      \begin{align}
 \sigma_{\max}\left(\tilde{\gamma}_n \right)
  &= 
   \sigma_{\max}\left(\bld{\Gamma}_{\bld{b}}
  \bld{\gamma}_{n}
  \bld{\Gamma}_{\bld{c}} \left( \bld{1}_{M\times1} \otimes \mathsfbi{I}   \right)    \right) 
  \leq 
  \begin{cases}
     \kappa
  \left[\sigma_{\min}\left(\bld{\gamma} \right)\right]^{-n} 
  & {\Elas} \gg 1 \\
\kappa\left[\sigma_{\max}\left(\bld{\gamma}\right)\right]^n
  & {\Elas} \ll 1  
  \end{cases} \label{gscc_gammatbound}
  \end{align}
  where we defined $\kappa$ as
  \begin{align}
  \kappa \equiv
  \begin{cases}
  \frac{\sigma_{\max}\left(\bld{\Gamma}_{\bld{b}}\right)
    \sigma_{\max}\left(\bld{\Gamma}_{\bld{c}} \left( \bld{1}_{M\times1} \otimes \mathsfbi{I}   \right)    \right) 
    }{\sigma_{\min}\left(\bld{\gamma} \right)} & {\Elas} \gg 1 \\
    \sigma_{\max}\left(\bld{\Gamma}_{\bld{b}}\right)
    \sigma_{\max}\left(\bld{\Gamma}_{\bld{c}} \left( \bld{1}_{M\times1} \otimes \mathsfbi{I}   \right)    \right)  & {\Elas} \ll 1 \\
    \end{cases}
    \label{kappadefn}
  \end{align}
  Using (\ref{gscc_gammatbound}), the triangle inequality and submultiplicative property for the spectral norm and the sum of a convergent geometric series, we then have for ${\Elas} \gg 1$ 
  \begin{align}
 \sigma_{\max}\left( \sum_{n=0}^\infty \tilde{\gamma}_n {\Elas}^{-n} \right) 
 &\leq  \sum_{n=0}^\infty {\Elas}^{-n}\sigma_{\max}\left(\tilde{\gamma}_n  \right)
 %
\leq
 \kappa
  \sum_{n=0}^\infty\left[ {\Elas}\,\sigma_{\min}\left(\bld{\gamma}\right) \right]^{-n} 
  =
    \frac{{\Elas} \kappa}{{\Elas} - [\sigma_{\min}\left(\bld{\gamma}\right)]^{-1} } 
    \label{gscc_sumbound1}
  \end{align}
and for ${\Elas} \ll 1$ we have
    \begin{align}
 \sigma_{\max}\left( \sum_{n=1}^\infty \tilde{\gamma}_n {\Elas}^{n} \right) 
 &\leq  \sum_{n=1}^\infty {\Elas}^{n}\sigma_{\max}\left(\tilde{\gamma}_n  \right)  
\leq
 \kappa
  \sum_{n=1}^\infty\left[{\Elas}\,\sigma_{\max}\left(\bld{\gamma}\right) \right]^n 
  =     \frac{ {\Elas}\kappa\,\sigma_{\max}\left(\bld{\gamma} \right)}{1 - {\Elas}\,\sigma_{\max}\left(\bld{\gamma} \right) }.  
   \label{gscc_sumbound2}
  \end{align}
 Since $\sigma_{\max}(A) \geq |\lambda_{\max}(A)| $, using (\ref{gscc_sumbound1}) in (\ref{gs_Gamma0lineinv1}) and invoking the submultiplicative property of the spectral norm, we obtain the following sufficient condition
     \begin{multline}
    {\Elas} > \frac{1}{\sigma_{\min}\left(\bld{\gamma} \right)}
     +  \frac{\kappa\alpha}{\beta} 
 \sigma_{\max}\left(
  \bld{\Gamma}_{\bld{a}}^{-1}\right) \label{gscc_muboundfinal1}   \\
=   \frac{1}{\sigma_{\min}(\bld{\gamma}) }\left(
1 + 
\frac{1 - \beta}{\beta} 
\frac{\sigma_{\max}\left(\bld{\Gamma}_{\bld{b}}\right)
    \sigma_{\max}\left[ \bld{\Gamma}_{\bld{c}}( \bld{1}_{M\times1} \otimes \mathsfbi{I}   )   \right]}
{ \sigma_{\min}\left(  \bld{\Gamma}_{\bld{a}}  \right)   }
\right) = \Elas_{\gg}
     \end{multline}
for convergence of the series in (\ref{gs_Gamma0lineinv1}) for ${\Elas} \gg 1$.
Similarly, using (\ref{gscc_sumbound2}) in (\ref{gs_Gamma0lineinv1}) we obtain the following sufficient condition
\begin{multline}
{\Elas} 
  < 
  \left(\frac{\sigma_{\min} (\beta \bld{\Gamma}_{\bld{a}}+ \alpha\tilde{\gamma}_0)}{\alpha \, \kappa   + \sigma_{\min} (\beta \bld{\Gamma}_{\bld{a}}+ \alpha\tilde{\gamma}_0) }\right)\frac{1}{\sigma_{\max}\left(\bld{\gamma}\right)} \label{gscc_muboundfinal2} \\
 =  
\frac{1}{  \sigma_{\max}\left(\bld{\gamma}\right)}
\left( 1
+ \frac{ 1-\beta }{\beta}\,     
   \frac{ \sigma_{\max}\left(\bld{\Gamma}_{\bld{b}}\right)
\sigma_{\max}\left[\bld{\Gamma}_{\bld{c}} \left( \bld{1}_{M\times1} \otimes \mathsfbi{I}   \right)    \right] }
{\sigma_{\min} [ \bld{\Gamma}_{\bld{a}}
+ \frac{ 1-\beta }{\beta}   \bld{\Gamma}_{\bld{b}} \bld{\Gamma}_{\bld{c}} \left( \bld{1}_{M\times1} \otimes \mathsfbi{I}   \right)   ]}
\right) ^{-1} \equiv \Elas_{\ll}
\end{multline}
for convergence of the series in (\ref{gs_Gamma0lineinv1}) for ${\Elas} \ll 1$. The bounds $\Elas_{\gg}$ and $\Elas_{\ll}$ at $\beta = 0.9$ are shown in Fig. \ref{fig:convg_2} plotted vs. $k_z$.
\section{Reduction to Generic Form}
\subsection{Base-flow Independent contribution to $\mathcal{E}$} \label{sec:baseIndep_math}
As discussed in section \ref{sec:sheardep}, the shear independent contribution to $\mathcal{E}$ is given by $\Rey(f_1 + f_2)$. This appendix provides details on the derivation of the generic form for Lyapunov equation (\ref{XLyap}), whose solution $\bld{\mathcal{X}}$ is used to compute $f_1$ using (\ref{f1defn}) . The procedure for $\bld{\mathcal{W}}$ is similar and so we omit it here.

Algebraically reducing  (\ref{XLyap}) and vectorizing the resulting equations to construct a linear system in the generic form (\ref{presysA}), we obtain
\begin{align} 
\beta \bld{L}_{\bld{a}}\bld{x}_{\bld{u}} + \alpha \bld{L}_{\bld{b}}   \bld{x}_{\bld{1}}   
&=
-  \bld{\mathscr{D}}_{\bld{u},\bld{\varphi}_1}   \label{sysLA} \\
{\Elas}^{-1}\bld{L}_{\bld{c}}  (\bld{1}_{6\times 1}\otimes \bld{x}_{\bld{u}} )
+ \bld{L}_{\bld{d}}(\alpha,\beta,{\Elas})\bld{x}_{\bld{1}} 
&= \nonumber\\
-   \frac{\alpha{\Elas}}{2} &
\begin{bmatrix}
\text{diag}\left(
\mathsfbi{I} \otimes \mathcal{L}_{12}, 
\mathsfbi{I} \otimes \mathcal{L}_{13}, 
\mathsfbi{I} \otimes \mathcal{L}_{14}
\right)  \\
\text{diag}\left( 
\overline{\mathcal{L}_{12}}\otimes \mathsfbi{I}, 
\overline{\mathcal{L}_{13}}\otimes \mathsfbi{I}, 
\overline{\mathcal{L}_{14}}\otimes \mathsfbi{I}
\right)
\end{bmatrix} \bld{\mathscr{D}}_{\mathsfbi{T},\bld{\varphi}_1}  \label{sysLB}
\end{align}
where  $\bld{L}_{\bld{a}}  \in \mathbb{C}^{\Nzero\times \Nzero}$, $\bld{L}_{\bld{b}}  \in \mathbb{C}^{\Nzero\times 6\Nzero}$ and $\bld{L}_{\bld{c}}  \in \mathbb{C}^{6\Nzero\times 6\Nzero}$ are  
\begin{align} 
\bld{L}_{\bld{a}} &\equiv  \overline{\mathcal{L}_{11}}\oplus \mathcal{L}_{11}  \nonumber \\
\bld{L}_{\bld{b}}  &\equiv
\begin{bmatrix} 
\overline{\mathcal{L}_{12}} \otimes \mathsfbi{I} &
\overline{\mathcal{L}_{13}} \otimes \mathsfbi{I} &
\overline{\mathcal{L}_{14}} \otimes \mathsfbi{I} &
\mathsfbi{I} \otimes \mathcal{L}_{12} &
\mathsfbi{I} \otimes \mathcal{L}_{13} &
\mathsfbi{I} \otimes \mathcal{L}_{14} 
\end{bmatrix}    \\
\bld{L}_{\bld{c}}  &\equiv
\text{diag}\left(
\overline{\mathcal{L}_{21}} \otimes \mathsfbi{I} ,
\overline{\mathcal{L}_{31}} \otimes \mathsfbi{I} ,
\overline{\mathcal{L}_{41}} \otimes \mathsfbi{I} ,
\mathsfbi{I} \otimes \mathcal{L}_{21} ,
\mathsfbi{I} \otimes \mathcal{L}_{31} ,
\mathsfbi{I} \otimes \mathcal{L}_{41} \right)   \nonumber	
\end{align} 
and  $\bld{L}_{\bld{d}}  \in \mathbb{C}^{6\Nzero\times 6\Nzero}$ is given by
\begin{align}  
\bld{L}_{\bld{d}} \equiv  \bld{\ell}(\alpha,\beta) - \frac{1}{{\Elas}}\mathsfbi{I}_{6{\Nzero}}, \quad 
\bld{\ell}(\alpha,\beta)  \equiv  \beta  \bld{\ell}_{\beta} + \alpha \bld{\ell}_{\alpha}  .
\end{align} 
The symbol $\oplus$ indicates the Kronecker sum, i.e., $\bld{A} \oplus \bld{B} = \bld{A} \otimes \mathsfbi{I} + \mathsfbi{I} \otimes \bld{B}$ for $\bld{A}$, $\bld{B}$ with compatible dimensions. All the matrix operators in the previous definitions are independent of $\Elas$ except for $\bld{L}_{\bld{d}}$. In addition only $\bld{L}_{\bld{d}}$ and $\bld{\ell}$ depend on $\beta$.
The matrix operators  $\bld{\ell}_{\alpha}, \bld{\ell}_{\beta} \in \mathbb{C}^{\tilde{6N}\times 6{\Nzero}}$ are defined in Appendix \ref{sec:app_opDefns}. The operator $\bld{\ell}_{\beta}$ is a diagonal matrix only consisting of terms involving $\mathcal{L}_{11}$ -- the diffusion-like operator in the $v$ equation in (\ref{swconstsys}). The operator $\bld{\ell}_{\alpha}$ consists of $\mathcal{L}_{1i}$ and $\mathcal{L}_{i1}$ for $i \in \{2,3,4\}$ which couple $\hat{v}$ to $\hat{T}_{yy}$, $\hat{T}_{yz}$ and $\hat{T}_{zz}$ and vice versa in (\ref{swconstsys}). 

Since the  system (\ref{sysLA}) and (\ref{sysLB}) is in the generic form (\ref{presysA}), we can utilize the solution to (\ref{presysA}) provided in the appendix to obtain solutions for $\bld{x}_{\bld{u}}$ and $\bld{x}_{\bld{1}}$ that have an explicit dependence on $\Elas$. The remaining block elements $\mathcal{X}_{ij}$ in $\bld{\mathcal{X}}$ can be reconstructed using $\mathcal{X}_{1i}$ (for $i =  1,\hdots, 4$) corresponding to the elements of $\bld{x}_1$ and the following expressions derived from (\ref{XLyap})
\begin{align}
 \mathcal{X}_{ii}  &= \frac{1}{2} \left(\mathcal{L}_{i1}\mathcal{X}_{1i} + \mathcal{X}_{1i}^* \mathcal{L}_{i1}^* \right) + \frac{{\Elas}}{2}\mathcal{B}_{k_i} \mathcal{B}_{k_i}^*, \quad 
 \text{for } \{ k_i \} = \{ 0,yy, yz, zz \} \label{Xiiexpr}\\
 \mathcal{X}_{ij}  &= \frac{1}{2} \left(\mathcal{X}_{1i}^* \mathcal{L}_{j1}^*  + \mathcal{L}_{i1}\mathcal{X}_{1j}\right) \label{Xijexpr}
\end{align}
for $i,j = 2,\hdots,4$. In particular, we use the vectorized form of (\ref{Xiiexpr}) to construct $\bld{x}_{\mathsfbi{T}}$  in terms of $\bld{x}_{\bld{1}}$. We can then use  $\bld{x}_{\bld{u}}$ and $\bld{x}_{\mathsfbi{T}}$  to compute $f_1$ from the alternative expression  (\ref{vecf1}). Invoking the weak and strong elasticity solutions to (\ref{presysA}) provided in the appendix, we thus find expressions for $f_1$.

\subsection{Shear dependent contribution to $\mathcal{E}$} \label{sec:shearDep_math}
The shear dependent contribution to $\mathcal{E}$ as discussion section \ref{sec:sheardep} is given by $\Rey^3 g(k_z; \beta,{\Elas})$. We can determine $g$ from $\bld{\mathcal{Z}}$ which requires one to solve the Lyapunov equation (\ref{ZLyap}). The Lyapunov equation (\ref{ZLyap}) implicitly contains $\bld{\mathcal{X}}$ via $\bld{\mathcal{Y}}$ and thus must be solved in multiple steps. We proceed with the following three steps:
\begin{enumerate}
\item We first solve for $\bld{\mathcal{X}}$ in terms $\bld{\mathscr{D}}_{\bld{u},\bld{\varphi}_1}$ and $\bld{\mathscr{D}}_{\mathsfbi{T},\bld{\varphi}_1}$.
\item Using the expressions for $\bld{\mathcal{X}}$ we solve the Sylvester equation (\ref{YSylv}) for $\bld{\mathcal{Y}}$ in terms of $\bld{\mathcal{X}}$ via $\bld{\mathcal{Q}} = \bld{\mathcal{C}}\bld{\mathcal{X}}$.
\item Finally we solve (\ref{ZLyap}) for $\bld{\mathcal{Z}}$ using the right-hand side $\bld{\mathcal{R}} = \bld{\mathcal{C}}\,\bld{\mathcal{Y}} ^*  + \bld{\mathcal{Y}} \,\bld{\mathcal{C}}^* $ and the expressions for $\bld{\mathcal{Y}}$ derived in (\textit{b}).  
\end{enumerate} 

The Sylvester equation (\ref{YSylv}) and the Lyapunov equation (\ref{ZLyap}) can both be reduced to the form (\ref{presysA}). 
The procedure to find $g$ in terms of $\bld{\mathscr{D}}_{\bld{u},\bld{\varphi}_1}$ and $\bld{\mathscr{D}}_{\mathsfbi{T},\bld{\varphi}_1}$ then simply becomes a problem of successively solving three linear systems of the form (\ref{presysA}).  

\begin{enumerate} 
\item Solving for $\bld{\mathcal{X}}$ is immediate from the previous section. Casting the relevant Lyapunov equation (\ref{XLyap}) in the reduced form (\ref{sysLA})--(\ref{sysLB}) and invoking the generic solution to (\ref{presysA}) given in the appendix, we obtain expressions for $\bld{x}_{\bld{u}}$ and $\bld{x}_{\bld{1}}$. The remaining vectorized elements of $\bld{\mathcal{X}}$ can be constructed using  $\bld{x}_{\bld{u}}$, $\bld{x}_{\bld{1}}$ and the vectorized form of the expressions (\ref{Xiiexpr})--(\ref{Xijexpr}).

\item Using a procedure analogous to that used to obtain the reduced systems (\ref{sysLA})--(\ref{sysLB}), we can reduce the Sylvester equation (\ref{YSylv}) for $\bld{\mathcal{Y}}$ to a system for $\bld{y}_{\bld{0}}$ and $\bld{y}_{\bld{1}}$ 
\begin{align} 
\beta \bld{G}_{\bld{a}}\bld{y}_{\bld{0}}   +  \alpha \bld{G}_{\bld{b}}\bld{y}_{\bld{1}} 
&=- 
\bld{Q}_{\bld{f0}}\bld{x}_{\bld{u}}  \label{sysGA}
\\
\frac{1}{{\Elas}}  \bld{G}_{\bld{c}}    (\bld{1}_{5\times 1} \otimes \bld{y}_{\bld{0}} )
+ \bld{G}_{\bld{d}}(\alpha,\beta,{\Elas})\bld{y}_{\bld{1}} 
&=
- \bld{Q}_{\bld{f1}}\bld{x}_{\bld{u}}-(\bld{Q}_{\bld{11}}^{(0)} 
+ \alpha{\Elas}\bld{Q}_{\bld{11}}^{(1)} )\bld{x}_{\bld{1}}   \nonumber \\
 &\hspace{1.25in}
- \alpha {\Elas}^2   \bld{Q}_{\bld{p1}}\bld{\mathscr{D}}_{\mathsfbi{T},\bld{\varphi}_1} 
  \label{sysGB}
\end{align} 
where the operators $\bld{G}_{\bld{a}}\in \mathbb{C}^{{\Nzero}\times {\Nzero}}$ , $\bld{G}_{\bld{b}}\in \mathbb{C}^{{\Nzero}\times 5{\Nzero}}$ and $\bld{G}_{\bld{c}}\in \mathbb{C}^{5{\Nzero}\times 5{\Nzero}}$ are  given by
\begin{align}
\bld{G}_{\bld{a}} &\equiv \overline{\mathcal{L}_{11}} \oplus \mathcal{S}_{11}\nonumber\\
\bld{G}_{\bld{b}} &\equiv
\begin{bmatrix} 
\overline{\mathcal{L}_{12}} \otimes \mathsfbi{I} &
\overline{\mathcal{L}_{13}} \otimes \mathsfbi{I} &
\overline{\mathcal{L}_{14}} \otimes \mathsfbi{I} &
\mathsfbi{I} \otimes \mathcal{S}_{12}  &
\mathsfbi{I} \otimes \mathcal{S}_{13}
 \end{bmatrix}\\
\bld{G}_{\bld{c}} &\equiv 
\text{diag}\left( 
\overline{\mathcal{L}_{21}} \otimes \mathsfbi{I} ,
\overline{\mathcal{L}_{31}} \otimes \mathsfbi{I} ,
\overline{\mathcal{L}_{41}} \otimes \mathsfbi{I} ,
\mathsfbi{I} \otimes \mathcal{S}_{21}, 
\mathsfbi{I} \otimes \mathcal{S}_{31}
\right)   \nonumber  
\end{align} 
and $\bld{G}_{\bld{d}} \in \mathbb{C}^{5{\Nzero}\times 5{\Nzero}}$ is defined by
\begin{align}
\bld{G}_{\bld{d}}  \equiv \bld{\mathfrak{g}} - \frac{1}{{\Elas}}\mathsfbi{I}_{5{\Nzero}}, \quad
\bld{\mathfrak{g}} \equiv \alpha \bld{\mathfrak{g}}_\alpha + \beta \bld{\mathfrak{g}}_\beta  
\end{align}
The operators $\bld{\mathfrak{g}}_\alpha$, $\bld{\mathfrak{g}}_\beta$ are defined in appendix \ref{sec:app_opDefns}.  The operator $\bld{\mathfrak{g}}_\beta$ is a diagonal matrix consisting of $\mathcal{L}_{11}$ and $\mathcal{S}_{11}$ which are the diffusion operators in  the $v$ and $\eta$ equations. All the operators are independent of  $\Elas$ except $\bld{G}_{\bld{d}}$. In addition only $\bld{G}_{\bld{d}}$ and $\bld{\mathfrak{g}}$ depend on $\beta$. The matrix operators in the right-hand side $\bld{Q}_{\bld{f0}}$, $\bld{Q}_{\bld{p1}}$ , $\bld{Q}_{\bld{f1}}$, $\bld{Q}_{\bld{11}}^{(0)}$  and $\bld{Q}_{\bld{11}}^{(1)}$ are defined in appendix \ref{sec:app_opDefns} and originate from the expression $\bld{\mathcal{Q}} = \bld{\mathcal{C}}\bld{\mathcal{X}}$ that appears in the right-hand side of the Sylvester equation (\ref{YSylv}) and the expressions (\ref{Xiiexpr})--(\ref{Xijexpr}) that allow us to write all the vectorized block elements of $\bld{\mathcal{X}}$ in terms of $\bld{x}_{\bld{u}}$ and $\bld{x}_{\bld{1}}$. The latter expressions lead to the appearance of $\bld{\mathscr{D}}_{\mathsfbi{T},\bld{\varphi}_1}$ in the right-hand side of (\ref{sysGB}).

As with the linear system (\ref{sysLA})--(\ref{sysLB}), the system (\ref{sysGA})--(\ref{sysGB}) is in the generic form (\ref{presysA}) and therefore we can utilize the solutions in the appendix to immediately find expressions for $\bld{y}_{\bld{0}}$ and $\bld{y}_{\bld{1}}$  in the two limits ${\Elas} \gg 1$ and ${\Elas} \ll 1$. 
These solutions require expressions  for $\bld{x}_{\bld{u}}$ and $\bld{x}_{\bld{1}}$ computed in the previous step in order to be explicit in $\bld{\mathscr{D}}_{\bld{u},\bld{\varphi}_1}$ and $\bld{\mathscr{D}}_{\mathsfbi{T},\bld{\varphi}_1}$. The remaining vectorized elements of $\bld{\mathcal{Y}}$ in $\bld{y}_{\bld{2}}$ and $\bld{y}_{\bld{3}}$, defined by
\begin{align}
\bld{y}_{\bld{2}} &\equiv  
\mathcal{V}\big(\begin{bmatrix}  
\mathcal{V}(\mathcal{Y}_{22}) & 
\mathcal{V}(\mathcal{Y}_{33})  
\end{bmatrix}\big) , \quad
\bld{y}_{\bld{3}} \equiv 
\mathcal{V}\big(\begin{bmatrix}  
\mathcal{V}(\mathcal{Y}_{32}) & 
\mathcal{V}(\mathcal{Y}_{23}) & 
\mathcal{V}(\mathcal{Y}_{24}) & 
\mathcal{V}(\mathcal{Y}_{34})  
\end{bmatrix}\big) \label{y2y3defn}
\end{align} 
can be written in terms of $\bld{y}_{\bld{1}}$, $\bld{x}_{\bld{1}}$ and the polymer disturbance $\bld{\mathscr{D}}_{\mathsfbi{T},\bld{\varphi}_1}$ using the vectorized form of following expression obtained directly from (\ref{YSylv})
\begin{align}
\mathcal{Y}_{ij}^* &= \frac{1}{2}\left({\Elas} \mathcal{Q}_{ij}^* + \mathcal{Y}_{1j}^*\mathcal{S}_{i1}^*  + \mathcal{L}_{j1} \mathcal{Y}_{i1}^*  \right)  , \quad i = 2,3;\,j = 2,3,4 .
\end{align}
Explicit expressions for $\bld{y}_{\bld{2}}$ and $\bld{y}_{\bld{3}}$ in terms of $\bld{y}_{\bld{0}}$, $\bld{y}_{\bld{1}}$ and the vectorized elements of $\bld{\mathcal{X}}$ can then be easily deduced.

\item Substituting the expressions for $\bld{\mathcal{S}}$ and $\bld{\mathcal{C}}$ given in (\ref{Opsdefn}) in the Lyapunov equation (\ref{ZLyap}) for $\bld{\mathcal{Z}}$, we obtain a reduced system for $\bld{z}_{\bld{u}}$ and $\bld{z}_{\bld{1}}$ analogous to (\ref{sysLA})--(\ref{sysLB}) and (\ref{sysGA})--(\ref{sysGB}) and given by
\begin{align} 
\beta \bld{S}_{\bld{a}}\bld{z}_{\bld{u}} + \alpha \bld{S}_{\bld{b}}\bld{z}_{\bld{1}}  
&= 
- \bld{R}_{\bld{0f}}^{(0)} \begin{bmatrix}
 \bld{y}_{\bld{0}} \\ \overline{\bld{y}_{\bld{0}}}
\end{bmatrix}
\label{sysSA} \\ 
{\Elas}^{-1} \bld{S}_{\bld{c}} (\bld{1}_{4\times 1}\otimes\bld{z}_{\bld{u}}) + \bld{S}_{\bld{d}}\bld{z}_{\bld{1}} 
&= 
-
\bld{R}_{\bld{01}}^{(0)}
\begin{bmatrix}
 \bld{y}_{\bld{0}} \\ \overline{\bld{y}_{\bld{0}}}
\end{bmatrix}
-
\left(\bld{R}_{\bld{11}}^{(0)}
+ \alpha  {\Elas} \bld{R}_{\bld{11}}^{(1)} \right)
\begin{bmatrix}
 \bld{y}_{\bld{1}} \\ \overline{\bld{y}_{\bld{1}}}
\end{bmatrix} \nonumber \\
&\hspace{0.75in}-\alpha  {\Elas}
\left( 
\bld{R}_{\bld{21}}^{(1)} 
\begin{bmatrix}
 \bld{y}_{\bld{2}} \\ \overline{\bld{y}_{\bld{2}}}
\end{bmatrix}
+
\bld{R}_{\bld{31}}^{(1)}
\begin{bmatrix}
 \bld{y}_{\bld{3}} \\ \overline{\bld{y}_{\bld{3}}}
\end{bmatrix}\right)  
\label{sysSB}
\end{align} 
where $\bld{S}_{\bld{a}}\in \mathbb{C}^{{\Nzero}\times {\Nzero}}$, $\bld{S}_{\bld{b}}\in \mathbb{C}^{{\Nzero}\times 4{\Nzero}}$, $\bld{S}_{\bld{c}}  \in \mathbb{C}^{4{\Nzero}\times 4{\Nzero}}$ are given by
\begin{align}   
\bld{S}_{\bld{a}} &\equiv
 \overline{\mathcal{S}_{11}} \oplus \mathcal{S}_{11} , \quad
\nonumber\\ 
\bld{S}_{\bld{b}} &\equiv
\begin{bmatrix} 
\overline{\mathcal{S}_{12}} \otimes \mathsfbi{I} &
\overline{\mathcal{S}_{13}} \otimes \mathsfbi{I} & 
\mathsfbi{I} \otimes \mathcal{S}_{12}  &
\mathsfbi{I} \otimes \mathcal{S}_{13}
 \end{bmatrix} \\
\bld{S}_{\bld{c}} &\equiv 
\text{diag}\left( 
\overline{\mathcal{S}_{21}} \otimes \mathsfbi{I},
\overline{\mathcal{S}_{31}} \otimes \mathsfbi{I}, 
\mathsfbi{I} \otimes \mathcal{S}_{21}, 
\mathsfbi{I} \otimes \mathcal{S}_{31}
\right)  \nonumber
\end{align}
and $\bld{S}_{\bld{d}}\in \mathbb{C}^{{4\Nzero}\times {4\Nzero}}$ is given by
\begin{align}
\bld{S}_{\bld{d}} \equiv \bld{\mathfrak{s}} - \frac{1}{{\Elas}}\mathsfbi{I}_{4{\Nzero}}, \quad
\bld{\mathfrak{s}} \equiv \alpha \bld{\mathfrak{s}}_\alpha + \beta \bld{\mathfrak{s}}_\beta 
\end{align}
where $\bld{\mathfrak{s}}_\alpha$, $\bld{\mathfrak{s}}_\beta$ are  defined in the appendix. The matrix operators in the right-hand side $\bld{R}_{\bld{ij}}^{(0)}$, $\bld{R}_{\bld{ij}}^{(1)}$ are defined appendix \ref{sec:app_opDefns} and originate from the expression $\bld{\mathcal{R}} = \bld{\mathcal{C}}\,\bld{\mathcal{Y}} ^*  + \bld{\mathcal{Y}} \,\bld{\mathcal{C}}^* $ which appears in the right-hand side of (\ref{ZLyap}). All the operators are independent of $\Elas$ except $\bld{G}_{\bld{22}}$. In addition none of operators depend on $\beta$ except $\bld{G}_{\bld{22}}$  and $\bld{\mathfrak{s}}$. We note that in the derivation of the scaling of $f_2(k_z;\beta,\Elas)$, omitted in the current work, we obtain a system similar to (\ref{sysSA})--(\ref{sysSB}) with the only difference being in the right-hand side.

 The system (\ref{sysSA})--(\ref{sysSB})  is in the generic form (\ref{presysA}) and thus as before we can utilize the solutions of the generic form to find expressions for $\bld{z}_{\bld{u}}$ and $\bld{z}_{\bld{1}}$. The elements of $\bld{z}_{\mathsfbi{T}}$ are the relevant remaining vectorized elements of $\bld{\mathcal{Z}}$ and can be written in terms of $\bld{z}_{\bld{u}}$ and $\bld{z}_{\bld{1}}$ using the expression below derived directly from the diagonal elements of (\ref{ZLyap})
\begin{align}
\mathcal{Z}_{ii} = \frac{1}{2}\left( {\Elas} \mathcal{R}_{ii} + \mathcal{S}_{i1}\mathcal{Z}_{1i} + \mathcal{Z}_{1i}^* \mathcal{S}_{i1}^* \right), \quad i = 2,3.
\end{align} 
\end{enumerate}


\section{Operator definitions} \label{sec:app_opDefns} 
\subsection{Input/Output coefficients}
The input coefficients in (\ref{Bdefns}) are, if not zero, given by
\begin{align}
\mathcal{B}_{v,2} &=  -k_z^2\Delta^{-1} ,\quad
\mathcal{B}_{v,3} = -\text{i}k_z \Delta^{-1}\p_y, \quad
\mathcal{B}_{\eta} = \text{i}k_z \\
\mathcal{B}_{xx} &= \mathcal{B}_{xy} = \mathcal{B}_{xz} = \mathcal{B}_{yy} = \mathcal{B}_{yz} = \mathcal{B}_{zz}  = 1
\end{align} 
Similarly the output coefficients in (\ref{Hdefns}) are, if not zero, given by
\begin{align}
\mathcal{H}_{u} &= \frac{1}{\text{i}k_z},\quad
\mathcal{H}_{v} = 1 ,\quad
\mathcal{H}_{w} = -\frac{1}{\text{i}k_z} \p_y \\
\mathcal{H}_{xx} &= \mathcal{H}_{xy} = \mathcal{H}_{xz} = \mathcal{H}_{yy} = \mathcal{H}_{yz} = \mathcal{H}_{zz}  = 1
\end{align}
\subsection{Operators in reduced  equations}
We first define $\bld{\ell}_{(\cdot)}$ as first used in (\ref{sysLA})--(\ref{sysLB}). Recall $\bld{L}_{\bld{d}}  = \bld{\ell} - \frac{1}{{\Elas}}\mathsfbi{I}_{6{\Nzero}} =  \beta  \bld{\ell}_{\beta} + \alpha \bld{\ell}_{\alpha}   - \frac{1}{{\Elas}}\mathsfbi{I}_{6{\Nzero}}$. Here $\bld{\ell}_{\alpha}$ and $\bld{\ell}_{\beta}$ are defined as
\begin{align} 
\bld{\ell}_\alpha(i,j) & \equiv
\frac{1}{2}
\begin{cases}
  \sum_{k=2}^4 \mathsfbi{I}\otimes   \mathcal{L}_{1k}\mathcal{L}_{k1}     & i = j = 1,3 \\ 
\sum_{k=2}^4  \overline{\mathcal{L}_{1k}} \overline{\mathcal{L}_{k1}}   \otimes \mathsfbi{I}  & i = j = 4,6 \\
\overline{\mathcal{L}_{(i+1)1}} \otimes \mathcal{L}_{1(j-2)} & i = 1,3;\, j = 4,6 \\
 \overline{\mathcal{L}_{1(j+1)}}\otimes \mathcal{L}_{(i-2)1} & i = 4,6;\, j = 1,3 \\
 0 & \text{otherwise}
\end{cases}  \\ 
\bld{\ell}_\beta &\equiv
\text{diag}\left( 
\mathsfbi{I}\otimes \mathcal{L}_{11}, \mathsfbi{I}\otimes \mathcal{L}_{11}, \mathsfbi{I}\otimes \mathcal{L}_{11},
\overline{\mathcal{L}_{11}} \otimes \mathsfbi{I}, 
\overline{\mathcal{L}_{11}} \otimes \mathsfbi{I}, 
\overline{\mathcal{L}_{11}} \otimes \mathsfbi{I} 
\right).
\end{align}
Similarly recall (\ref{sysGA})--(\ref{sysGB}) where we invoke $\bld{G}_{\bld{d}} = \bld{\mathfrak{g}} - \frac{1}{{\Elas}}\mathsfbi{I}_{5{\Nzero}}= \alpha \bld{\mathfrak{g}}_\alpha + \beta \bld{\mathfrak{g}}_\beta- \frac{1}{{\Elas}}\mathsfbi{I}_{5{\Nzero}}$. Here we define $\bld{\mathfrak{g}}_{\alpha}$ and $\bld{\mathfrak{g}}_{\beta}$ as
\begin{align} 
\bld{\mathfrak{g}}_\alpha(i,j) &\equiv \frac{1}{2}
\begin{cases} 
\sum_{k = 2}^3 \mathsfbi{I} \otimes \mathcal{S}_{1k}\mathcal{S}_{k1}  & i = j = 1,2,3 \\
\sum_{k = 2}^4 \overline{\mathcal{L}_{1k}} \overline{\mathcal{L}_{k1}} \otimes \mathsfbi{I} & i = j = 4,5\\
\overline{\mathcal{L}_{(i+1)1}}  \otimes \mathcal{S}_{1(j-2)}
 & i = 1,2,3; \, j = 4,5 \\
\overline{\mathcal{L}_{1(j+1)}} \otimes \mathcal{S}_{(i-2)1}
 & i = 4,5; \, j = 1,2,3 \\
 0 & \text{otherwise}
\end{cases} \\
\bld{\mathfrak{g}}_\beta &\equiv
\text{diag}\left(\mathsfbi{I}\otimes \mathcal{S}_{11},\mathsfbi{I}\otimes \mathcal{S}_{11},\mathsfbi{I}\otimes \mathcal{S}_{11},\overline{\mathcal{L}_{11}} \otimes \mathsfbi{I},\overline{\mathcal{L}_{11}} \otimes \mathsfbi{I}\right).
\end{align}
Finally we have the expression $\bld{S}_{\bld{d}}  = \bld{\mathfrak{s}} - \frac{1}{{\Elas}}\mathsfbi{I}_{4{\Nzero}} = \alpha \bld{\mathfrak{s}}_\alpha + \beta \bld{\mathfrak{s}}_\beta - \frac{1}{{\Elas}}\mathsfbi{I}_{4{\Nzero}}$ in (\ref{sysSA})--(\ref{sysSB}) and we define $\bld{\mathfrak{s}}_{\alpha}$ and $\bld{\mathfrak{s}}_{\beta}$ as
\begin{align} 
\bld{\mathfrak{s}}_\alpha(i,j) &\equiv
\frac{1}{2}
\begin{cases}
  \sum_{k=2}^3 \mathsfbi{I}\otimes   \mathcal{S}_{1k}\mathcal{S}_{k1}     & i = j = 1,2 \\ 
\sum_{k=2}^3  \overline{\mathcal{S}_{1k}} \overline{\mathcal{S}_{k1}}   \otimes \mathsfbi{I}  & i = j = 3,4 \\
\overline{\mathcal{S}_{(i+1)1}} \otimes \mathcal{S}_{1(j-1)} & i = 1,2;\, j = 3,4 \\
 \overline{\mathcal{S}_{1(j+1)}}\otimes \mathcal{S}_{(i-1)1} & i = 3,4;\, j = 1,2 \\
 0 & \text{otherwise}
\end{cases}  \\ 
\bld{\mathfrak{s}}_\beta &\equiv 
\text{diag}\left( 
\mathsfbi{I}\otimes \mathcal{S}_{11}, \mathsfbi{I}\otimes \mathcal{S}_{11},
\overline{\mathcal{S}_{11}} \otimes \mathsfbi{I}, \overline{\mathcal{S}_{11}} \otimes \mathsfbi{I} 
\right).
\end{align}

In order to define the remaining operators in a concise manner, we introduce some operators and notation. We first introduce the vectorized transpose operator $\mathsf{T}_{\mathcal{V}}: \mathbb{C}^{nm \times 1} \rightarrow \mathbb{C}^{nm \times 1}$ for $n,m \in \mathbb{N}$ that rearranges the elements of the vectorized matrix such that it represents the vectorized transposed matrix. Thus, for example given $\bld{X} \in \mathbb{C}^{N\times N}$, we have
 \begin{align}
 \mathsf{T}_{\mathcal{V}}\,\mathcal{V}\left(  \bld{X}\right) =  \mathcal{V}\left( \bld{X}^{\mathsf{T}}\right).
 \end{align} 
It is an elementary exercise to derive the specific form of $\mathsf{T}_{\mathcal{V}}$ for a given dimension. We will often need to perform this vectorized transpose operation on a series of vectorized matrices organized in a larger vector.   For conciseness we then use the notation
 \begin{align}
 \mathsf{T}_{\mathcal{V},M} \equiv 
 \text{diag}\underbrace{\left( \mathsf{T}_{\mathcal{V}}, \mathsf{T}_{\mathcal{V}},\hdots, \mathsf{T}_{\mathcal{V}} \right)}_{M} 
 \end{align}
that allows us to conveniently represent a vector of vectorized transposed matrices. In the current work, vectorization is only ever applied to matrices of dimension $N \times N$ and therefore all the individual entries $ \mathsf{T}_{\mathcal{V},M}$ maybe taken to be the vectorized transpose operation of the same dimension.

Since we use vectorized matrices, it becomes convenient to defining a quantity $\bld{e}_{m}^{(k)}\in \mathbb{R}^{{\Nzero} \times m {\Nzero}}$ given by
\begin{align}
\bld{e}_{m}^{(k)} \equiv 
\begin{bmatrix}
\underbrace{\bld{0}_{{\Nzero} \times {\Nzero}} \quad \hdots \quad \bld{0}_{{\Nzero} \times {\Nzero}}}_{k-1} 
\quad \mathsfbi{I}_{{\Nzero} \times {\Nzero}} \quad 
\underbrace{\bld{0}_{{\Nzero} \times {\Nzero}} \quad \hdots \quad \bld{0}_{{\Nzero} \times {\Nzero}}}_{m-k}
\end{bmatrix}
\label{erowdefn}
\end{align}
that allows us to extract the $k$-th ${\Nzero}\times 1$ vector from a larger vector $\in \mathbb{C}^{m{\Nzero} \times 1}$, and where we fix ${\Nzero} = N^2$ within the definition.
 Using this, we can define the $\bld{e}_{m}^{(k_1,\hdots,k_p)}$ given by
\begin{align}
\bld{e}_{m}^{(k_1,\hdots,k_p)} \equiv 
\text{diag}\left(
\bld{e}_{m}^{(k_1)}, \bld{e}_{m}^{(k_2)},\hdots,\bld{e}_{m}^{(k_p)}
\right)
\end{align}
for multi-index $(k_1,\hdots,k_p)$, that allows us to extract $p$ vectors of dimension ${\Nzero}\times 1$ from a larger vector. With this notation, we first define the operators appearing on the right-hand side of (\ref{sysGA})--(\ref{sysGB})  that originate from $\bld{\mathcal{Q}} = \bld{\mathcal{C}}\bld{\mathcal{X}}$
\begin{align}
\bld{Q}_{\bld{p2}} 
&= \text{diag}\left(
\tilde{\mathcal{C}}_{22} ,
\tilde{\mathcal{C}}_{33}  
\right) \bld{e}_{3}^{(1,2)}\mathsfbi{I}_{[1]}^{[2]},\quad
\bld{Q}_{\bld{12}} 
= \bld{Q}_{\bld{p2}}\bld{K}_{L,p}  + \text{diag}\left( 
\tilde{\mathcal{C}}_{21}, 
\tilde{\mathcal{C}}_{31}  
\right)\bld{e}_{6}^{(1,2)} \mathsfbi{I}_{[1]}^{[2]}  \nonumber\\
\bld{Q}_{\bld{13}} 
&= \text{diag}\left(
\tilde{\mathcal{C}}_{33} , 
\tilde{\mathcal{C}}_{22} ,
\tilde{\mathcal{C}}_{22} ,
\tilde{\mathcal{C}}_{33}  
\right)\bld{e}_{6}^{(4,1,3,2)} \mathsfbi{I}_{[1]}^{[4]}  \bld{K}_{L,2}  
\nonumber \\
&\hspace{0.75in}
+ \text{diag}\left(
\tilde{\mathcal{C}}_{31}  ,
\tilde{\mathcal{C}}_{21}  ,
\tilde{\mathcal{C}}_{21}  ,
\tilde{\mathcal{C}}_{31}  
\right)\bld{e}_{6}^{(1,2,3,3)}\mathsfbi{I}_{[1]}^{[4]}\\
\bld{Q}_{\bld{f0}} 
&= \tilde{\mathcal{C}}_{11} , \quad
\bld{Q}_{\bld{f1}} 
= \text{diag}\left( \bld{0}, \bld{0},\bld{0},
\tilde{\mathcal{C}}_{21},
\tilde{\mathcal{C}}_{31}
\right) \mathsfbi{I}_{[1]}^{[5]}, \quad
\bld{Q}_{\bld{p1}} 
  = \frac{1}{2}\bld{M}_{H,d}\bld{Q}_{\bld{p2}} \nonumber\\
\bld{Q}_{\bld{11}}^{(0)} 
&= \text{diag}\big(  
 \tilde{\mathcal{C}}_{11},
\tilde{\mathcal{C}}_{11},
\tilde{\mathcal{C}}_{11} ,
 \tilde{\mathcal{C}}_{22},
\tilde{\mathcal{C}}_{33} 
\big)\bld{e}_{6}^{(1,2,3,4,5)} \mathsfbi{I}_{[1]}^{[5]}, \quad
\bld{Q}_{\bld{11}}^{(1)} 
  =   \bld{M}_{H,d}\bld{Q}_{\bld{12}}  + \bld{M}_{H,3}\bld{Q}_{\bld{13}}  \nonumber
\end{align}
where we defined the following
\begin{align} 
\bld{K}_{L,p} &\equiv \frac{1}{2}
\begin{bmatrix} 
 \text{diag}\left(\tilde{\mathcal{L}}_{21},\tilde{\mathcal{L}}_{31},\tilde{\mathcal{L}}_{41}\right),
 \text{diag}\left(\tilde{\mathcal{L}}_{21*},\tilde{\mathcal{L}}_{31*},\tilde{\mathcal{L}}_{41*}\right)
\end{bmatrix}  \nonumber\\
\bld{K}_{L,2} &\equiv 
\frac{1}{2}
\begin{bmatrix}
\begin{matrix}
\bld{0}_{2{\Nzero}\times {\Nzero}}
& \text{diag}\left(\tilde{\mathcal{L}}_{21},\tilde{\mathcal{L}}_{31}\right)
& \text{diag}\left(
\tilde{\mathcal{L}}_{31*} , 
\tilde{\mathcal{L}}_{41*}\right)
& \bld{0}_{2{\Nzero}\times {\Nzero}} \\
\bld{0}_{{\Nzero}\times 2{\Nzero}}
&  \tilde{\mathcal{L}}_{21} 
& \tilde{\mathcal{L}}_{41*}
& \bld{0}_{{\Nzero}\times 2{\Nzero}} 
\end{matrix} \\
\begin{matrix}
\text{diag}\left(\tilde{\mathcal{L}}_{31}, \tilde{\mathcal{L}}_{41}\right)
& \bld{0}_{2{\Nzero}\times \tilde{2N}} 
& \text{diag}\left( 
\tilde{\mathcal{L}}_{21*} , 
\tilde{\mathcal{L}}_{31*} \right) \\
\tilde{\mathcal{L}}_{41} 
& \bld{0}_{{\Nzero}\times 4{\Nzero}}
& \tilde{\mathcal{L}}_{21*} 
\end{matrix}
\end{bmatrix} \\
\bld{M}_{H,d} &=
\frac{1}{2}
\begin{bmatrix}
\text{diag}\left( \tilde{\mathcal{S}}_{12},  \tilde{\mathcal{S}}_{13}\right) \\
0 \qquad 0 \\
\text{diag}\left( \tilde{\mathcal{L}}_{12*},  \tilde{\mathcal{L}}_{13*}\right) 
\end{bmatrix} ,\quad
\bld{M}_{H,3}  =
\frac{1}{2}
\begin{bmatrix}
\tilde{\mathcal{S}}_{13} & 0 & 0 & 0 \\
  0 & \tilde{\mathcal{S}}_{12} & 0 & 0 \\
  0 & 0 & \tilde{\mathcal{S}}_{12} & \tilde{\mathcal{S}}_{13} \\
  0 & \tilde{\mathcal{L}}_{13*} & \tilde{\mathcal{L}}_{14*} & 0 \\
 \tilde{\mathcal{L}}_{12*}  & 0 & 0 & \tilde{\mathcal{L}}_{14*} 
\end{bmatrix} \nonumber
\end{align}
and
\begin{align}
\tilde{\mathcal{C}}_{ij}  &\equiv \mathsfbi{I} \otimes \mathcal{C}_{ij} ,\quad \tilde{\mathcal{C}}_{ij*} \equiv \overline{\mathcal{C}_{ij}} \otimes \mathsfbi{I}, \quad 
\tilde{\mathcal{L}}_{ij}  \equiv \mathsfbi{I} \otimes \mathcal{L}_{ij} ,\quad \tilde{\mathcal{L}}_{ij*} \equiv \overline{\mathcal{L}_{ij}} \otimes \mathsfbi{I} \nonumber \\
\tilde{\mathcal{S}}_{ij}  &\equiv \mathsfbi{I} \otimes \mathcal{S}_{ij} ,\quad \tilde{\mathcal{S}}_{ij*} \equiv \overline{\mathcal{S}_{ij}} \otimes \mathsfbi{I}. \nonumber
\end{align}
Note that $\mathsfbi{I}_{[k]}^{[m]}$ is the restacking operator defined in (\ref{restackdefn}).

We also define similarly, for the right-hand side of (\ref{sysSA})--(\ref{sysSB}), all the operators originating from $\bld{\mathcal{R}} = \bld{\mathcal{C}}\bld{\mathcal{Y}}^* + \bld{\mathcal{Y}}\bld{\mathcal{C}}^*$ as follows
\begin{align}
\bld{R}_{\bld{0f}}^{(0)} &= 
\begin{bmatrix} 
R_{\bld{0f}}^{(0)} & 
R_{\bld{0f}*}^{(0) } 
\end{bmatrix}, \quad
\bld{R}_{\bld{01}}^{(0)}  = 
\begin{bmatrix} 
R_{\bld{01}}^{(0)} & 
R_{\bld{01}*}^{(0)} 
\end{bmatrix} , \quad
\bld{R}_{\bld{11}}^{(0)}  = 
\begin{bmatrix} 
R_{\bld{11}}^{(0)} & 
R_{\bld{11}*}^{(0) } 
\end{bmatrix} 				\nonumber\\
\bld{R}_{\bld{11}}^{(1)} &= 
\begin{bmatrix} 
R_{\bld{11}}^{(1)} & 
R_{\bld{11}*}^{(1) } 
\end{bmatrix}  , \quad
\bld{R}_{\bld{21}}^{(1)}  = 
\begin{bmatrix} 
R_{\bld{21}}^{(1)} & 
R_{\bld{21}*}^{(1) } 
\end{bmatrix},\quad
\bld{R}_{\bld{31}}^{(1)}  = 
\begin{bmatrix} 
R_{\bld{31}}^{(1)} & 
R_{\bld{31}*}^{(1) } 
\end{bmatrix}  
\end{align}
where
\begin{align}
R_{\bld{0f}}^{(0)} &= \tilde{\mathcal{C}}_{11\ast}  ,\quad
R_{\bld{0f}*}^{(0)} = \tilde{\mathcal{C}}_{11}\mathsf{T}_{\mathcal{V}}\\
R_{\bld{01}}^{(0)} &= 
\text{diag}\left(\tilde{\mathcal{C}}_{21\ast},\tilde{\mathcal{C}}_{31\ast} ,\bld{0},\bld{0}\right)\mathsfbi{I}_{[1]}^{[4]}, \quad
R_{\bld{01}*}^{(0)}= 
\text{diag}\left(\bld{0},\bld{0},\tilde{\mathcal{C}}_{21},\tilde{\mathcal{C}}_{31} \right)\mathsfbi{I}_{[1]}^{[4]}\mathsf{T}_{\mathcal{V}} \\
R_{\bld{11}}^{(0)} &= 
\text{diag}\left(
\tilde{\mathcal{C}}_{22\ast},
\tilde{\mathcal{C}}_{33\ast},
\tilde{\mathcal{C}}_{11\ast},
\tilde{\mathcal{C}}_{11\ast} 
\right)\bld{e}_{5}^{(1,2,4,5)}\mathsfbi{I}_{[1]}^{[4]} \\
R_{\bld{11}*}^{(0)}  &= 
\text{diag}\left(
\tilde{\mathcal{C}}_{11},
\tilde{\mathcal{C}}_{11},
\tilde{\mathcal{C}}_{22},
\tilde{\mathcal{C}}_{33}
\right)\bld{e}_{5}^{(4,5,1,2)}\mathsfbi{I}_{[1]}^{[4]}\mathsf{T}_{\mathcal{V},5} \\
R_{\bld{11}}^{(1)} &=  
\frac{1}{2}
\begin{bmatrix}
\text{diag}\left( \overline{\mathcal{C}_{21}}\otimes \mathcal{S}_{12}, \overline{\mathcal{C}_{31}}\otimes \mathcal{S}_{13}\right) \\
\text{diag}\left( 
\overline{\mathcal{S}_{12}}\overline{\mathcal{C}_{21}} \otimes \mathsfbi{I} , 
\overline{\mathcal{S}_{13}}\overline{\mathcal{C}_{31}} \otimes \mathsfbi{I} \right)
\end{bmatrix} \bld{e}_{5}^{(4,5)}\mathsfbi{I}_{[1]}^{[2]}\nonumber \\
&\hspace{0.75in}
 +
\frac{1}{2}\begin{bmatrix}
\text{diag}\left( \overline{\mathcal{C}_{21}}\otimes \mathcal{S}_{13}, \overline{\mathcal{C}_{31}}\otimes \mathcal{S}_{12} \right) \\
\text{diag}\left( \overline{\mathcal{S}_{13}}\overline{\mathcal{C}_{21}}\otimes \mathsfbi{I}, \overline{\mathcal{S}_{12}}\overline{\mathcal{C}_{31}}\otimes \mathsfbi{I} \right)\end{bmatrix} \bld{e}_{5}^{(5,4)} \mathsfbi{I}_{[1]}^{[2]}, \\
R_{\bld{11}*}^{(1)} &= 
\frac{1}{2} \begin{bmatrix}
\text{diag}\left( \mathsfbi{I}\otimes \mathcal{S}_{12}\mathcal{C}_{21}, \mathsfbi{I}\otimes \mathcal{S}_{13}\mathcal{C}_{31}\right) \\
\text{diag}\left( \overline{\mathcal{S}_{12}} \otimes \mathcal{C}_{21} , \overline{\mathcal{S}_{13}} \otimes  \mathcal{C}_{31}\right)
\end{bmatrix} \bld{e}_{5}^{(4,5)} \mathsfbi{I}_{[1]}^{[2]}\mathsf{T}_{\mathcal{V},5} \nonumber \\
&\hspace{0.75in}
+ \frac{1}{2}\begin{bmatrix}
\text{diag}\left( \mathsfbi{I}\otimes \mathcal{S}_{13}\mathcal{C}_{31}, \mathsfbi{I}\otimes \mathcal{S}_{12}\mathcal{C}_{21} \right) \\
\text{diag}\left( \overline{\mathcal{S}_{13}}\otimes \mathcal{C}_{31}, \overline{\mathcal{S}_{12}}\otimes \mathcal{C}_{21} \right)\end{bmatrix} \bld{e}_{5}^{(4,5)} 
\mathsfbi{I}_{[1]}^{[2]}\mathsf{T}_{\mathcal{V},5} \\ 
R_{\bld{21}}^{(1)} &=  
\frac{1}{2}\begin{bmatrix}
\text{diag}\left( \overline{\mathcal{C}_{22}} \otimes \mathcal{S}_{12}, \overline{\mathcal{C}_{33}} \otimes \mathcal{S}_{13}\right) \\
\text{diag}\left( \overline{\mathcal{S}_{12}} \overline{\mathcal{C}_{22}} \otimes \mathsfbi{I} , \overline{\mathcal{S}_{13}}\overline{\mathcal{C}_{33}}  \otimes \mathsfbi{I} \right)
\end{bmatrix} \bld{e}_{2}^{(1,2)}\mathsfbi{I}_{[1]}^{[2]} \nonumber \\
R_{\bld{21}*}^{(1)}  &= 
\frac{1}{2}\begin{bmatrix}
\text{diag}\left( \mathsfbi{I}\otimes \mathcal{S}_{12}\mathcal{C}_{22}, \mathsfbi{I}\otimes \mathcal{S}_{13}\mathcal{C}_{33}\right) \\
\text{diag}\left( \overline{\mathcal{S}_{12}} \otimes \mathcal{C}_{22} , \overline{\mathcal{S}_{13}} \otimes \mathcal{C}_{33}\right)
\end{bmatrix} \bld{e}_{2}^{(1,2)}\mathsfbi{I}_{[1]}^{[2]}\mathsf{T}_{\mathcal{V},2} \\
R_{\bld{31}}^{(1)} &= 
\frac{1}{2}\begin{bmatrix}
\text{diag}\left( \overline{\mathcal{C}_{22}}\otimes \mathcal{S}_{13}, \overline{\mathcal{C}_{33}}\otimes \mathcal{S}_{12} \right) \\
\text{diag}\left( \overline{\mathcal{S}_{13}}\overline{\mathcal{C}_{22}}\otimes \mathsfbi{I}, \overline{\mathcal{S}_{12}}\overline{\mathcal{C}_{33}}\otimes \mathsfbi{I} \right)\end{bmatrix} \bld{e}_{4}^{(1,2)}\mathsfbi{I}_{[1]}^{[2]}\nonumber\\
R_{\bld{31}*}^{(1)}  &= 
\frac{1}{2}\begin{bmatrix}
\text{diag}\left( \mathsfbi{I}\otimes \mathcal{S}_{13}\mathcal{C}_{33}, \mathsfbi{I}\otimes \mathcal{S}_{12} \mathcal{C}_{22}\right) \\
\text{diag}\left( \overline{\mathcal{S}_{13}}\otimes \mathcal{C}_{33}, \overline{\mathcal{S}_{12}}\otimes \mathcal{C}_{22}\right)\end{bmatrix} \bld{e}_{4}^{(2,1)} \mathsfbi{I}_{[1]}^{[2]}\mathsf{T}_{\mathcal{V},4}.
\end{align} 

\subsection{Operators in solutions of reduced equations}
We present the operators used in the solutions to the reduced equations in this appendix. The solutions are presented in Sections \ref{sec:baseIndep_math} and \ref{sec:shearDep_math} of the main text.
\subsubsection{Base flow independent component}
The operators $\bld{\mathscr{X}}_{(\cdot)\rightarrow (\cdot)}$ in the expression for $f_1 + f_2$ for $\Elas \ll 1$ in (\ref{chi_uu})--(\ref{chi_tautau}) are given by
 \begin{align}
\bld{\mathscr{X}}_{\bld{u} \rightarrow \bld{u}} &\equiv  
  -\bld{\tilde{L}}_{\bld{a}} ^{-1}  + \mathcal{O}({\Elas})     \label{scrXffdefnA}\\
\bld{\mathscr{X}}_{\mathsfbi{T} \rightarrow \bld{u}} &\equiv 
-  \frac{\alpha^2 }{2} 
  \begin{bmatrix} 
\bld{\tilde{L}}_{\bld{a}} ^{-1}(\overline{\mathcal{L}_{12}}\otimes \mathcal{L}_{12})&
\hdots & 
\bld{\tilde{L}}_{\bld{a}} ^{-1}(\overline{\mathcal{L}_{14}}\otimes \mathcal{L}_{14})  
\end{bmatrix}     
+ \mathcal{O}({\Elas})   \label{scrXpfdefnA} \\
\bld{\mathscr{X}}_{\bld{u} \rightarrow \mathsfbi{T}} &\equiv 
 -   \sum_{j=2}^4(\overline{\mathcal{L}_{j1}}\otimes \mathcal{L}_{j1}) \bld{\tilde{L}}_{\bld{a}} ^{-1}  + \mathcal{O}({\Elas})   \label{scrXfpdefnA}\\
\bld{\mathscr{X}}_{\mathsfbi{T} \rightarrow \mathsfbi{T}} &\equiv 
\frac{1}{2}\mathsfbi{I}_{3{\Nzero}} + \mathcal{O}({\Elas})   \label{scrXppdefnA}
\end{align} 
where
\begin{align}
\bld{\tilde{L}}_{\bld{a}} &\equiv -\Big[\beta \Big(\overline{\mathcal{L}_{11}} \oplus \mathcal{L}_{11}\Big)+ \alpha \sum_{j = 2}^4 
  \overline{\mathcal{L}_{1j}}\overline{\mathcal{L}_{j1}} \oplus \mathcal{L}_{1j}\mathcal{L}_{j1}  \Big]. 
\nonumber
\end{align}
Similarly, for $\Elas \gg 1$ we have
\begin{align}
\bld{\mathscr{X}}_{\bld{u} \rightarrow \bld{u}} &\equiv 
-\beta^{-1}\bld{L}_{\bld{a}}^{-1}  
 + \mathcal{O}(\frac{1}{{\Elas}})    \label{scrXffdefnB}\\
\bld{\mathscr{X}}_{\mathsfbi{T} \rightarrow \bld{u}} &\equiv 
 \frac{ \alpha^2}{2\beta} \bld{L}_{\bld{a}}^{-1}  \bld{L}_{\bld{b}}  \bld{\ell}^{-1} 
\begin{bmatrix}
\text{diag}\left(
\mathsfbi{I} \otimes  \mathcal{L}_{12}, 
\mathsfbi{I} \otimes \mathcal{L}_{13},
\mathsfbi{I} \otimes \mathcal{L}_{14}
\right)  \\
\text{diag}\left( 
\overline{\mathcal{L}_{12}} \otimes \mathsfbi{I}, 
\overline{\mathcal{L}_{13}} \otimes \mathsfbi{I}, 
\overline{\mathcal{L}_{14}} \otimes \mathsfbi{I}
\right)
\end{bmatrix}   + \mathcal{O}(\frac{1}{{\Elas}})     \label{scrXpfdefnB} \\
\bld{\mathscr{X}}_{\bld{u} \rightarrow \mathsfbi{T}} &\equiv 
\beta^{-1}  \bld{L}_{\bld{A}}\bld{L}_{\bld{c}}\mathsfbi{I}_{[1]}^{[6]} \bld{L}_{\bld{a}}^{-1}  + \mathcal{O}(\frac{1}{{\Elas}})     \label{scrXfpdefnB}\\
\bld{\mathscr{X}}_{\mathsfbi{T} \rightarrow \mathsfbi{T}} &\equiv 
\frac{1}{2}\mathsfbi{I}_{3{\Nzero}} 
-  \frac{\alpha}{2}  \bld{L}_{\bld{A}}
\begin{bmatrix}
\text{diag}\left(
\mathsfbi{I} \otimes  \mathcal{L}_{12}, 
\mathsfbi{I} \otimes  \mathcal{L}_{13},
\mathsfbi{I} \otimes  \mathcal{L}_{14}
\right)  \\
\text{diag}\left( 
\overline{\mathcal{L}_{12}} \otimes \mathsfbi{I}, 
\overline{\mathcal{L}_{13}} \otimes \mathsfbi{I}, 
\overline{\mathcal{L}_{14}} \otimes \mathsfbi{I}
\right)
\end{bmatrix} + \mathcal{O}(\frac{1}{{\Elas}})   \label{scrXppdefnB} 
\end{align} 
where
\begin{align} 
  \bld{L}_{\bld{A}} &\equiv
  \frac{1}{2}\begin{bmatrix}  
\text{diag}\left(
\mathsfbi{I} \otimes  \mathcal{L}_{21},
\mathsfbi{I} \otimes  \mathcal{L}_{31},
\mathsfbi{I} \otimes  \mathcal{L}_{41}\right) &
\text{diag}\left(
\overline{\mathcal{L}_{21}} \otimes \mathsfbi{I},
\overline{\mathcal{L}_{31}} \otimes \mathsfbi{I},
\overline{\mathcal{L}_{41}} \otimes \mathsfbi{I} \right)    
\end{bmatrix} \bld{\ell}^{-1}. \nonumber
\end{align}  
Expressions analogous to (\ref{scrXffdefnA})--(\ref{scrXppdefnB}) can be derived for the operators $\bld{\mathscr{W}}_{(\cdot)\rightarrow (\cdot)}$. These expressions are the same form as  (\ref{scrXffdefnA})--(\ref{scrXppdefnB}) but with all  matrices, vectors and summations modified by replacing $\mathcal{L}_{ij}$ with $\mathcal{S}_{ij}$  and removing the $\mathcal{L}_{14}$, $\mathcal{L}_{41}$ dependent submatrices.
\subsubsection{Shear dependent component}
The operators $\bld{\mathscr{Z}}_{(\cdot)\rightarrow(\cdot)}$ in the expressions (\ref{rho_uu})--(\ref{rho_tautau}) for the shear dependent component $g$ in the weakly elastic regime $\Elas \ll 1$ are given by
\begin{align}
\bld{\mathscr{Z}}_{\bld{u}\rightarrow \bld{u}} &=
\bld{\tilde{S}}_{\mathsfbi{T}} + \bld{\tilde{S}}_{\mathsfbi{T}\ast}
+ \mathcal{O}(\Elas) \nonumber \\
\bld{\mathscr{Z}}_{\mathsfbi{T}\rightarrow \bld{u}} &=  \alpha^2  
\begin{bmatrix}
\bld{1}_{1\times 3} \otimes \bld{\tilde{S}}_{\mathsfbi{T}} &
\bld{1}_{1\times 3} \otimes \bld{\tilde{S}}_{\mathsfbi{T}\ast} 
\end{bmatrix}
\begin{bmatrix}
\text{diag}\left( \overline{\mathcal{L}_{12}} \otimes \mathcal{L}_{12} ,\hdots, \overline{\mathcal{L}_{14}} \otimes  \mathcal{L}_{14}\right)   \\
\text{diag}\left( \mathcal{L}_{12} \otimes \overline{\mathcal{L}_{12}},\hdots, \mathcal{L}_{14} \otimes \overline{\mathcal{L}_{14}}\right) 
\end{bmatrix}
+ \mathcal{O}(\Elas) \nonumber\\   
\bld{\mathscr{Z}}_{\bld{u}\rightarrow\bld{p}} &= 
2\,\text{diag}\left( \overline{\mathcal{S}_{21}} \otimes \mathcal{S}_{21}, \overline{\mathcal{S}_{31}} \otimes \mathcal{S}_{31}\right)\mathsfbi{I}_{[1]}^{[2]}( \bld{\tilde{S}}_{\mathsfbi{T}} + \bld{\tilde{S}}_{\mathsfbi{T}\ast}  )   
 + \mathcal{O}(\Elas)
\nonumber \\
\bld{\mathscr{Z}}_{\mathsfbi{T}\rightarrow\bld{p}} &=
2\alpha^2
\text{diag}\left( \overline{\mathcal{S}_{21}} \otimes \mathcal{S}_{21}, \overline{\mathcal{S}_{31}} \otimes \mathcal{S}_{31}\right)
\Bigg( (
\bld{1}_{2\times 3} \otimes \bld{\tilde{S}}_{\mathsfbi{T}} )
\text{diag}\left( \overline{\mathcal{L}_{12}} \otimes \mathcal{L}_{12} ,\hdots, \overline{\mathcal{L}_{14}} \otimes  \mathcal{L}_{14}\right)  
\nonumber \\
&\hspace{0.5in} 
+ (
\bld{1}_{2\times 3} \otimes \bld{\tilde{S}}_{\mathsfbi{T}\ast} )
\text{diag}\left( \mathcal{L}_{12} \otimes \overline{\mathcal{L}_{12}},\hdots, \mathcal{L}_{14} \otimes \overline{\mathcal{L}_{14}}\right) \Bigg)
  + \mathcal{O}(\Elas)  \nonumber
\end{align} 
where we defined for notational convenience 
%
%
\begin{align}
\bld{\tilde{S}}_{\mathsfbi{T}}  &=\bld{\tilde{S}}_{11}^{-1}\tilde{\mathcal{C}}_{11*}  \bld{\tilde{G}}_{11}^{-1}
\tilde{\mathcal{C}}_{11}
\bld{\tilde{L}}_{\bld{a}} ^{-1}  , \qquad
\bld{\tilde{S}}_{\mathsfbi{T}*}  =\bld{\tilde{S}}_{11}^{-1}\tilde{\mathcal{C}}_{11}\mathsf{T}_{\mathcal{V}}\overline{ \bld{\tilde{G}}_{11}^{-1}
\tilde{\mathcal{C}}_{11}
\bld{\tilde{L}}_{\bld{a}} ^{-1} } \nonumber\\ 
\bld{\tilde{G}}_{11} &= -\Big[\beta \Big(\overline{\mathcal{L}_{11}} \oplus \mathcal{S}_{11} \Big)
+ \alpha\Big(   \overline{\mathcal{L}_{14}}\overline{\mathcal{L}_{41}} \otimes \mathsfbi{I}
   + \sum_{j = 2}^3 \overline{\mathcal{L}_{1j}}\overline{\mathcal{L}_{j1}} \oplus \mathcal{S}_{1j}\mathcal{S}_{j1}  \Big)\Big]  \nonumber\\
\bld{\tilde{S}}_{11} &= 
-\Big[\beta \Big(\overline{\mathcal{S}_{11}} \oplus \mathcal{S}_{11}\Big) + \alpha\sum_{j = 2}^3 
  \overline{\mathcal{S}_{1j}}\overline{\mathcal{S}_{j1}} \oplus \mathcal{S}_{1j}\mathcal{S}_{j1}  \Big]  \nonumber\\
  \tilde{\mathcal{C}}_{11} &= \mathsfbi{I} \otimes \mathcal{C}_{11}, \qquad
\tilde{\mathcal{C}}_{11*} = \overline{\mathcal{C}_{ij}} \otimes \mathsfbi{I}  \nonumber \\
\bld{K}_{H,d} &= 
\frac{1}{2}\begin{bmatrix}
\text{diag}\left( \mathsfbi{I}\otimes \mathcal{S}_{21},\mathsfbi{I}\otimes \mathcal{S}_{31}\right) 
& \bld{0}_{2\tilde{N}\times\tilde{N}} &
\text{diag}\left(\overline{\mathcal{L}_{21}} \otimes \mathsfbi{I}, \overline{\mathcal{L}_{31}} \otimes \mathsfbi{I}\right)  \end{bmatrix} \nonumber\\
\bld{K}_{H,1}&=\frac{1}{2}\begin{bmatrix}
\mathsfbi{I}\otimes \mathcal{S}_{31} & 0 & 0 & 0 & \overline{\mathcal{L}_{21}}\otimes \mathsfbi{I} \\  
0 & \mathsfbi{I}\otimes \mathcal{S}_{21} & 0 & \overline{\mathcal{L}_{31}}\otimes \mathsfbi{I} & 0 \\  
0 & 0 & \mathsfbi{I}\otimes \mathcal{S}_{21} & \overline{\mathcal{L}_{41}}\otimes \mathsfbi{I} & 0 \\  
0 & 0 & \mathsfbi{I}\otimes \mathcal{S}_{31} & 0 & \overline{\mathcal{L}_{41}}\otimes \mathsfbi{I}   
\end{bmatrix}  \nonumber
\end{align}

Similarly, the operators  in the expressions (\ref{rho_uu})--(\ref{rho_tautau}) for $g$ in the strongly elastic regime $\Elas \gg 1$ are given by
\begin{align}
\bld{\mathscr{Z}}_{\bld{u}\rightarrow \bld{u}} &=
  \left(\frac{\alpha}{\beta}\right)^2
  \begin{bmatrix}
  \bld{S}_{\bld{a}}^{-1} \bld{S}_{\bld{b}}  &
 \bld{S}_{\bld{a}}^{-1} \bld{S}_{\bld{b}} 
  \end{bmatrix}
   \bld{R}_{\bld{A}} 
   \left(
\begin{bmatrix}
\bld{Q}_{\bld{A}} \\
\overline{\bld{Q}_{\bld{A}}}  \end{bmatrix}
- \alpha\begin{bmatrix}
  \bld{Q}_{\bld{11}}^{(1)}\bld{Q}_{\bld{B}} \\
 \overline{\bld{Q}_{\bld{11}}^{(1)}\bld{Q}_{\bld{B}}} \end{bmatrix}\right)
  + \mathcal{O}\left(\frac{1}{\Elas}\right) \nonumber\\
\bld{\mathscr{Z}}_{\mathsfbi{T}\rightarrow \bld{u}} &=
\alpha^2
\left(\frac{\alpha}{\beta}\right)^2
\begin{bmatrix}
\bld{S}_{\bld{a}}^{-1} \bld{S}_{\bld{b}}
&
\bld{S}_{\bld{a}}^{-1} \bld{S}_{\bld{b}} 
\end{bmatrix}\bld{R}_{\bld{A}} \begin{bmatrix}
\bld{Q}_{\bld{C}}\\
\overline{\bld{Q}_{\bld{C}}}
\end{bmatrix} 
 + \mathcal{O}\left(\frac{1}{\Elas}\right)  \nonumber\\
\bld{\mathscr{Z}}_{\bld{u}\rightarrow \mathsfbi{T}} &=
\frac{1}{\beta}
\bld{K}_{\bld{A}}
\begin{bmatrix}
\bld{Q}_{\bld{12}} \bld{Q}_{\bld{B}} \\
\overline{\bld{Q}_{\bld{12}} \bld{Q}_{\bld{B}}}
\end{bmatrix}  
+\left( \frac{1}{\beta}  \bld{K}_{\bld{C}}  
   -\alpha\bld{K}_{\bld{B}}\bld{R}_{\bld{A}}  \right)
   \left(
\begin{bmatrix}
\bld{Q}_{\bld{A}} \\
\overline{\bld{Q}_{\bld{A}}}  \end{bmatrix}
- \alpha\begin{bmatrix}
  \bld{Q}_{\bld{11}}^{(1)}\bld{Q}_{\bld{B}} \\
 \overline{\bld{Q}_{\bld{11}}^{(1)}\bld{Q}_{\bld{B}}} \end{bmatrix}\right)
+ \mathcal{O}\left(\frac{1}{\Elas}\right) \nonumber
\\
\bld{\mathscr{Z}}_{\mathsfbi{T}\rightarrow\bld{p}} &=
\frac{\alpha^2}{\beta}\left(   
\bld{K}_{\bld{C}} 
- \alpha  
\bld{K}_{\bld{B}} \bld{R}_{\bld{A}} \right)\begin{bmatrix}
\bld{Q}_{\bld{C}}\\
\overline{\bld{Q}_{\bld{C}}}
\end{bmatrix} 
+\bld{K}_{\bld{A}}
\left( \frac{1}{2}\begin{bmatrix} \bld{Q}_{p2} \\ \overline{\bld{Q}_{p2} }\end{bmatrix}
-\alpha \begin{bmatrix}
\bld{Q}_{\bld{D}} \\
\overline{\bld{Q}_{\bld{D}}}\end{bmatrix} \right)
+ \mathcal{O}\left(\frac{1}{\Elas}\right) \nonumber
\end{align}
where we defined for notational convenience
\begin{align}
\bld{R}_{\bld{A}}  &= 
\text{diag}\left( \bld{\mathfrak{s}}^{-1}\left(R_{\bld{11}}^{(1)} 
+ R_{\bld{21}}^{(1)} \bld{K}_{H,d}
+R_{\bld{31}}^{(1)} \bld{K}_{H,1}\right),
\bld{\mathfrak{s}}^{-1}\left(R_{\bld{11}*}^{(1)} 
+ R_{\bld{21}*}^{(1)} \overline{\bld{K}_{H,d}}
+ R_{\bld{31}*}^{(1)} \overline{\bld{K}_{H,1}}\right)
\right)\nonumber \\ 
\bld{K}_{\bld{A}} &= \frac{1}{4}\begin{bmatrix}
\text{diag}(
  \overline{\mathcal{C}_{22}} \otimes \mathsfbi{I},
   \overline{\mathcal{C}_{33}} \otimes \mathsfbi{I} )  & \text{diag}(
 \left(\mathsfbi{I} \otimes \mathcal{C}_{22}\right)\mathsf{T}_{\mathcal{V}},
\left(\mathsfbi{I} \otimes \mathcal{C}_{33}\right)\mathsf{T}_{\mathcal{V}}
)
\end{bmatrix} \nonumber\\
\bld{K}_{\bld{B}}  
 &= \frac{1}{2}
\left[\begin{matrix}\begin{bmatrix} \bld{0}_{2\tilde{N}\times 3\tilde{N}} &
\text{diag}\left( \tilde{\mathcal{C}}_{21*},  \tilde{\mathcal{C}}_{31*} \right)
\end{bmatrix}  
+ \text{diag}(\tilde{\mathcal{C}}_{22*},\tilde{\mathcal{C}}_{33*} )\bld{K}_{H,d} 
\end{matrix}, \right.\nonumber \\&\hspace{0.75in} \left. \begin{matrix}
\begin{bmatrix} \bld{0}_{2\tilde{N}\times 3\tilde{N}} &
\text{diag}\left(\tilde{\mathcal{C}}_{21} \mathsf{T}_{\mathcal{V}}, \tilde{\mathcal{C}_{31}} \mathsf{T}_{\mathcal{V}}  \right)
\end{bmatrix}  + 
\text{diag}(\tilde{\mathcal{C}}_{22} \mathsf{T}_{\mathcal{V}},
\tilde{\mathcal{C}}_{33}\mathsf{T}_{\mathcal{V}})  \overline{\bld{K}_{H,d}}  
 \end{matrix} \right] \nonumber\\
\bld{Q}_{\bld{A}} &= \bld{\mathfrak{g}}^{-1} \bld{Q}_{\bld{f1}}   \bld{L}_{\bld{a}}^{-1}, \qquad
 \bld{Q}_{\bld{B}}  =  \bld{\ell}^{-1}  \bld{L}_{\bld{c}}\mathsfbi{I}_{[1]}^{[6]} \bld{L}_{\bld{a}}^{-1} \nonumber\\
\bld{Q}_{\bld{C}}  &= 
\frac{1}{2}\bld{\mathfrak{g}}^{-1}\bld{Q}_{\bld{11}}^{(1)}\bld{\ell}^{-1}
\begin{bmatrix}
\text{diag}\left(
\mathsfbi{I}\otimes \mathcal{L}_{12},
\mathsfbi{I}\otimes \mathcal{L}_{13},
\mathsfbi{I}\otimes \mathcal{L}_{14}
\right) \\
\text{diag}\left( 
 \overline{\mathcal{L}_{12}} \otimes \mathsfbi{I},
 \overline{\mathcal{L}_{13}} \otimes \mathsfbi{I},
 \overline{\mathcal{L}_{14}} \otimes \mathsfbi{I}
\right)
\end{bmatrix} \nonumber \\
\bld{Q}_{\bld{D}}  &= 
\frac{1}{2}\bld{Q}_{\bld{12}} \bld{\ell}^{-1} 
\begin{bmatrix}
\text{diag}\left(
\mathsfbi{I}\otimes \mathcal{L}_{12},
\mathsfbi{I}\otimes \mathcal{L}_{13},
\mathsfbi{I}\otimes \mathcal{L}_{14}
\right) \\
\text{diag}\left( 
 \overline{\mathcal{L}_{12}} \otimes \mathsfbi{I},
 \overline{\mathcal{L}_{13}} \otimes \mathsfbi{I},
 \overline{\mathcal{L}_{14}} \otimes \mathsfbi{I}
\right) \nonumber
\end{bmatrix}
\end{align} 
Note that $\mathsfbi{I}_{[k]}^{[m]}$ is the restacking operator defined in (\ref{restackdefn}).

\bibliographystyle{jfm}

\bibliography{manuscript}

\end{document}